\documentclass[twocolumn,prl,amsmath,amssymb, aps,superscriptaddress, nofootinbib]{revtex4-2}
 \usepackage{comment}
\usepackage[titletoc,toc,title]{appendix}
\usepackage{titletoc}
\usepackage{titlesec}
\usepackage[version=4]{mhchem} 
\usepackage[normalem]{ulem}
\usepackage{mhchem}
\usepackage{amsmath,bm}
\usepackage{amssymb}
\usepackage{amsfonts}
\usepackage{afterpage}
\usepackage{graphicx}
\usepackage{xfrac}
\usepackage{footmisc}
\usepackage{comment}
\usepackage{pifont}
\usepackage{times}
\usepackage[hidelinks]{hyperref}
\usepackage{enumitem}
\usepackage{centernot}
\usepackage{cancel}
\usepackage{slashed}
\usepackage{relsize}
\usepackage{amsmath,amssymb,mathrsfs}
\usepackage{times}
\usepackage{epsfig}
\usepackage{verbatim}
\usepackage{bm}
\usepackage[utf8]{inputenc}
\usepackage{graphics}
\usepackage{graphicx,epsfig,amssymb,amsmath,color,cancel}
\usepackage{subfigure}
\usepackage[english]{babel}
\usepackage{adjustbox}
\usepackage{physics}
\usepackage{placeins}

\usepackage{array}
\newcolumntype{P}[1]{>{\centering\arraybackslash}p{#1}}
\newcolumntype{M}[1]{>{\centering\arraybackslash}m{#1}}

\usepackage[table]{xcolor}
\usepackage{tabularray}

\UseTblrLibrary{booktabs}

\definecolor{zima_blue}{HTML}{1393C1}
\hypersetup{setpagesize=false,bookmarksnumbered=true,bookmarksopen=true,%
colorlinks=true,linkcolor=zima_blue,urlcolor=zima_blue,citecolor=zima_blue,linktocpage=false}

\usepackage{orcidlink}

\usepackage[
    starfontserif 
    ]{starfont}
\usepackage{amsmath}

\DeclareSymbolFont{starfontsym}{OT1}{sts}{m}{n}
\DeclareMathSymbol{\mathSun}{\mathord}{starfontsym}{115}
\DeclareMathSymbol{\mathMercury}{\mathord}{starfontsym}{102}
\DeclareMathSymbol{\mathVenus}{\mathord}{starfontsym}{103}
\DeclareMathSymbol{\mathTerra}{\mathord}{starfontsym}{76}
\DeclareMathSymbol{\mathvarTerra}{\mathord}{starfontsym}{108}
\DeclareMathSymbol{\mathMoon}{\mathord}{starfontsym}{100}
\DeclareMathSymbol{\mathvarMoon}{\mathord}{starfontsym}{97}
\DeclareMathSymbol{\mathMars}{\mathord}{starfontsym}{104}
\DeclareMathSymbol{\mathJupiter}{\mathord}{starfontsym}{106}
\DeclareMathSymbol{\mathSaturn}{\mathord}{starfontsym}{83}
\DeclareMathSymbol{\mathUranus}{\mathord}{starfontsym}{70}
\DeclareMathSymbol{\mathvarUranus}{\mathord}{starfontsym}{65}
\DeclareMathSymbol{\mathNeptune}{\mathord}{starfontsym}{71}
\DeclareMathSymbol{\mathPluto}{\mathord}{starfontsym}{74}
\DeclareMathSymbol{\mathvarPluto}{\mathord}{starfontsym}{72}

\titleformat{\section}
{\normalfont\fontsize{10}{12}\bfseries  \centering }{\thesection.}{1em}{}
\titleformat{\subsection}
{\normalfont\fontsize{10}{12}\bfseries \centering}{\thesubsection.}{1em}{}
\titleformat{\subsubsection}
{\normalfont\fontsize{10}{12}\bfseries \centering}{\thesubsubsection)}{1em}{}

\titleformat{\paragraph}
{\normalfont\fontsize{10}{12}\bfseries  }{\thesection:}{1em}{}

\begin{document}

\preprint{}

\title{Domain wall interpretation of the PTA signal confronting black hole overproduction}

\author{Yann Gouttenoire~\orcidlink{0000-0003-2225-6704}}
\email{yann.gouttenoire@gmail.com}
\affiliation{School of Physics and Astronomy, Tel-Aviv University, Tel-Aviv 69978, Israel}
\author{Edoardo Vitagliano~\orcidlink{0000-0001-7847-1281}}
\email{edoardo.vitagliano@mail.huji.ac.il}
\affiliation{Racah Institute of Physics, Hebrew University of Jerusalem, Jerusalem 91904, Israel}
\affiliation{Dipartimento di Fisica e Astronomia, Università degli Studi di Padova,
Via Marzolo 8, 35131 Padova, Italy}
\affiliation{Istituto Nazionale di Fisica Nucleare (INFN), Sezione di Padova,
Via Marzolo 8, 35131 Padova, Italy}

\begin{abstract}

Recently, Pulsar Timing Array (PTA) collaborations have detected a stochastic gravitational wave background (SGWB) at nano-Hz frequencies, with Domain Wall networks (DWs) proposed as potential sources. To be cosmologically viable, they must annihilate before dominating the universe energy budget, thus generating a SGWB. While sub-horizon DWs shrink and decay rapidly, causality requires DWs with super-horizon size to continue growing until they reach the Hubble horizon. Those entering the latest can be heavier than a Hubble patch and collapse into Primordial Black Holes (PBHs).
We conduct a Bayesian analysis of the PTA signal, interpreting it as an outcome of SGWB from DW networks, with a prior ensuring no PBH overproduction. 
 Our findings indicate that DWs
result in the production of solar-mass PBHs. 
The binary mergers occurring within these PBHs generate a second SGWB in the kilo-Hz domain which could be observable in on-going or planned Earth-based interferometers.

\end{abstract}
\maketitle

\textbf{\textit{Introduction---}}NANOGrav 15-year (NG15) \cite{NANOGrav:2023gor}, EPTA \cite{Antoniadis:2023rey}, PPTA \cite{Reardon:2023gzh}, CPTA \cite{Xu:2023wog}, the first three being summarized in \cite{InternationalPulsarTimingArray:2023mzf}, have recently reported the observation of a nano-Hz stochastic gravitational wave background (SGWB)~\cite{NANOGrav:2023gor}, confirming the hint observed in previous years \cite{NANOGrav:2020spf,Chen:2021rqp,Goncharov:2021oub,Antoniadis:2022pcn}.
Sources of nano-Hz GWs could be a population of supermassive black hole binaries (SMBH)~\cite{Sesana:2004sp,Burke-Spolaor:2018bvk,Antoniadis:2023zhi,NANOGrav:2023hfp,Ellis:2023dgf} or could be related to  early universe phenomena~\cite{NANOGrav:2023hvm,EPTA:2023xxk}, such as phase transitions~\cite{Caprini:2010xv,Ratzinger:2020koh,NANOGrav:2021flc,Bringmann:2023opz,Gouttenoire:2023bqy}, large inhomogeneities~\cite{Wang:2019kaf,Chen:2019xse,DeLuca:2020agl,Sugiyama:2020roc,Vaskonen:2020lbd,Kohri:2020qqd,Zhao:2022kvz}, topological defects~\cite{Ellis:2020ena,Blasi:2020mfx,Buchmuller:2020lbh,Bian:2020urb,King:2020hyd,Madge:2023dxc,Servant:2023mwt}, axion-gauge field dynamics \cite{Ratzinger:2020koh,Madge:2023dxc,Murai:2023gkv,Unal:2023srk,Geller:2023shn} and PBHs mergers \cite{Depta:2023qst,Gouttenoire:2023nzr}.
Domains Walls are topological defects forming after the spontaneous breaking of a discrete symmetry, e.g. $\mathcal{Z}_2$ \cite{Vilenkin:1981zs,Preskill:1991kd,Saikawa:2017hiv,Gelmini:2020bqg}.
DWs dilute slower than radiation in expanding cosmology \cite{Zeldovich:1974uw,Vilenkin:2000jqa}. To be viable, there must exist an energy bias between distinct vacua so that DWs are pulled toward annihilating with each other. Upon annihilating around QCD confinement epoch, DW networks can produce a SGWB~\cite{Gleiser:1998na,Hiramatsu:2010yz,Kawasaki:2011vv,Hiramatsu:2013qaa,Gelmini:2020bqg} with a nano-Hz peak frequency detectable by PTAs \cite{Ferreira:2022zzo}.
DW networks feature closed domains configuration which collapse to PBHs when they shrink below their Schwarzschild radius~\cite{Ferrer:2018uiu,Ge:2019ihf,Gouttenoire:2023gbn} in a process dubbed ``catastrogenesis''~\cite{Gelmini:2022nim,Gelmini:2023ngs}.
In this \textit{Letter}, we report important advances in our comprehension of PBH formation during the annihilation of long-lived DWs networks, with the details deferred to a companion paper \cite{Gouttenoire:2023gbn}. Our work builds on previous studies \cite{Ferrer:2018uiu,Gelmini:2022nim,Gelmini:2023ngs}. Our results indicate that the DWs which collapse into PBHs are those that are super-horizon at the beginning of the annihilation phase, and their abundance can be estimated in percolation theory.
Performing Bayesian analysis of PTA datasets NG15 \cite{NANOGrav:2023gor} and IPTA2 \cite{Antoniadis:2022pcn}, we find that DW networks, possibly combined with SMBH binaries, can not only be a plausible source of the observed PTA signal, but can also lead to the production of PBHs in quantities that can be probed using kilo-Hz interferometers, astrometric measurements, and 21-cm line observations.\footnote{The possibility for the DW interpretation of the PTA signal to produce solar-mass PBHs was also mentioned in \cite{Kitajima:2023cek}.}

\begin{figure*}[t!]
\centering
\begin{adjustbox}{max width=1\linewidth,center}
\raisebox{0cm}{\makebox{\includegraphics[ width=0.49\textwidth, scale=1]{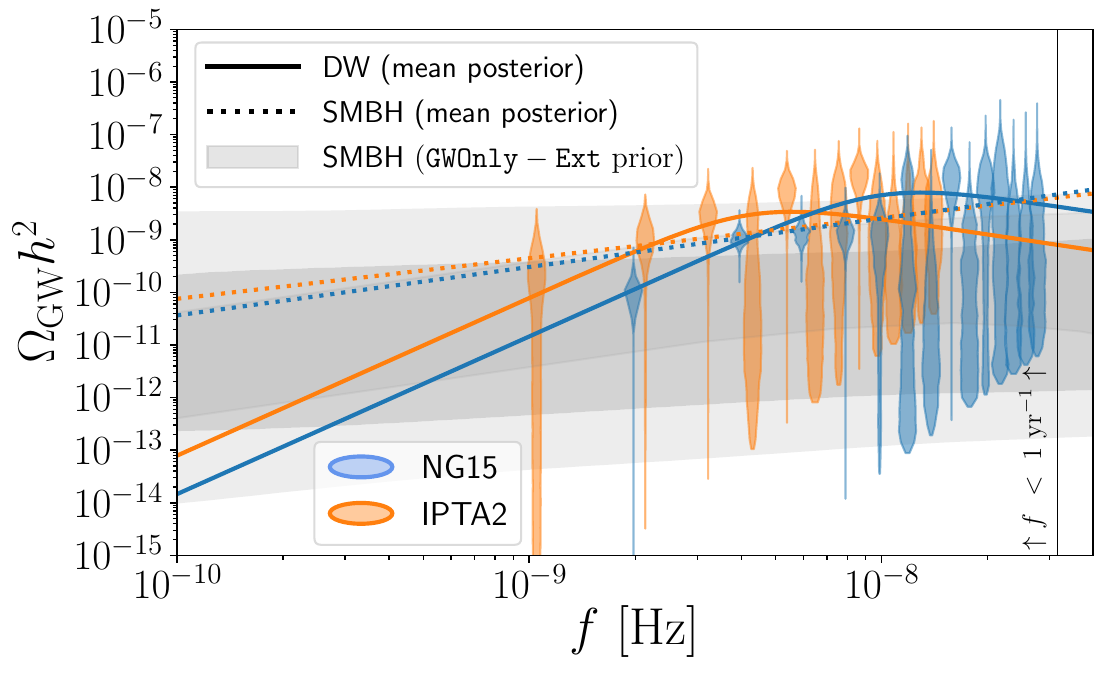}}}
\raisebox{0cm}{\makebox{\includegraphics[ width=0.49\textwidth, scale=1]{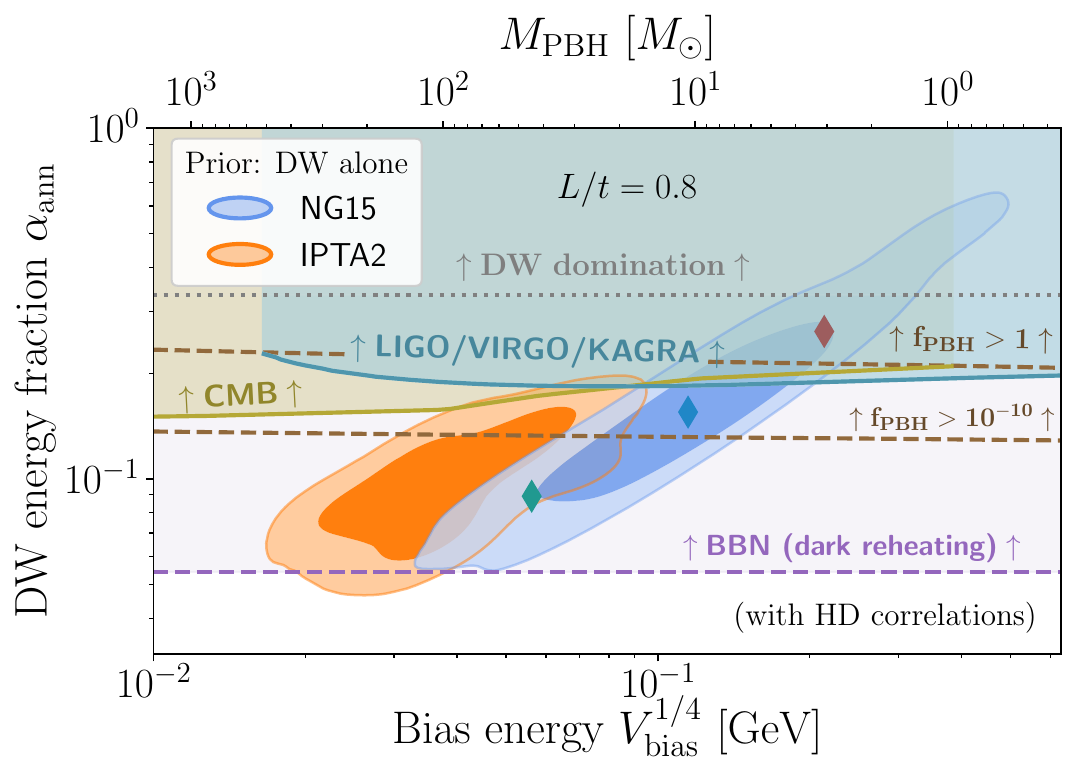}}}
\end{adjustbox}
\caption{\label{fig:DW_GWspectrum_NANO_IPTA} LEFT: SGWB from DW network annihilation (solid) preferred by PTA datasets, assuming annihilation into SM radiation and neglecting the PBH prior constraints. We show the preferred SGWB from SMBH binaries (dotted), assuming the astrophysical prior drawn from the ${\tt GWOnly-Ext}$ library \cite{NANOGrav:2023hvm}, whose $68\%$ and $90\%$ confidence levels are depicted with the gray bands.  We refer to Tab.~\ref{tab:meanPoseterior} in the SuM for the mean posterior values. RIGHT: $68\%$ (orange) and $90\%$ (blue) confidence levels of the DW interpretation of the PTA signal, assuming SM reheating and neglecting PBH prior constraints. The dashed brown region shows the PBH abundance given in Eq.~\eqref{eq:fPBH_abundance} assuming a DW network correlation length $L/t=0.8$, which might be a conservative choice, see below Eq.~\eqref{eq:rho_DW_A}. We show the associated current PBH exclusion bounds from on-going kilo-Hz GW interferometers~\cite{Nakamura:1997sm,Raidal:2018bbj,Kavanagh:2018ggo,LIGOScientific:2019kan,DeLuca:2020qqa} (blue) and CMB~\cite{Ali-Haimoud:2016mbv,Poulin:2017bwe,Serpico:2020ehh}  (yellow). Part of the DW interpretation is in tension with PBH overproduction constraints. The current BBN exclusion bound $N_{\rm eff}\lesssim 0.4$ (dashed purple) only applies if the DW network annihilate entirely into dark degrees of freedom. In dotted gray, we show the region when the bias energy density $V_{\rm bias}$ locally dominates the energy density of the universe. 
}
\end{figure*}

\textbf{\textit{Gravitational waves---}}After a phase of friction-domination whose duration is model dependent \cite{Blasi:2022ayo},
 numerical simulations have shown that the energy stored in DWs reaches the scaling regime  \cite{Vilenkin:2000jqa,Saikawa:2017hiv},
\begin{equation}
\label{eq:rho_DW_A}
\rho_{\rm DW} = \frac{\sigma}{L},\qquad \textrm{with}~L\simeq  t/\mathcal{A},
\end{equation}
where $L$ is the typical correlation length of the DW network and $\sigma$ the DW surface tension. Numerical simulations of the $\mathcal{Z}_2$-model have found $\mathcal{A} \simeq 0.8 \pm 0.1$ \cite{Hiramatsu:2013qaa}, leading to $L/t =1.25^{+0.18}_{-0.14}$. Other simulations found $L/t \simeq 1.14\pm 0.04$ \cite{Martins:2016ois}. Eq.~\eqref{eq:rho_DW_A} results in the DW energy density redshifting slower than the main radiation fluid $\rho_{\rm rad} \simeq \pi^2 g_* T^4/30 \propto 1/t^2$ such that DWs dominate the energy density of the universe after the time
\begin{equation}
     t_{\rm dom }^{\mathsmaller{\rm unbias}}~ =~ \frac{3M_{\rm pl}^2}{4\mathcal{A} \sigma},
\end{equation}
where $M_{\rm pl}\simeq 2.44 \times 10^{18}~\rm GeV$ is the reduced Planck mass.
We assume the presence of high dimensional operators that explicitly break $U(1)$ symmetry \cite{Kibble:1976sj,Vilenkin:1981zs,Sikivie:1982qv,Gelmini:1988sf}. This transforms the flat direction into a discrete collection of vacua.  DWs annihilate when the vacuum pressure $V_{\rm bias}$ between these new minimum points surpasses the pressure $C_d\, \sigma / L $ arising from their surface tension $\sigma$ at
\begin{equation}
\label{eq:t_ann}
t_{\rm ann}~ \simeq~  C_d \,\mathcal{A}\,\frac{\sigma}{V_{\rm bias}},
\end{equation}
where $C_d\simeq \rm a~few$ can be inferred from numerical simulations \cite{Kawasaki:2014sqa}. We take $C_d\simeq 3$. To be viable, DW must annihilate before the energy bias dominate the energy density $3M_{\rm pl}^2H^2\simeq V_{\rm bias}$ of the universe at
\begin{equation}
\label{eq:t_dom}
     t_{\rm dom }~ \simeq~ \frac{\sqrt{3}M_{\rm pl}}{2\sqrt{V_{\rm bias}}}.
\end{equation}
$t_{\rm dom }$ and $t_{\rm dom }^{\mathsmaller{\rm unbias}}$ are the times when the DW network dominates the universe energy density with and without the presence of the bias $V_{\rm bias}$, respectively.
We introduce the DW energy fraction of the universe at $t_{\rm ann}$,
\begin{equation}
\label{eq:alpha_ann}
    \alpha_{\rm ann} \equiv \frac{\rho_{\rm DW}}{3M_{\rm pl}^2H^2} \Big|_{t=t_{\rm ann}}
    \simeq\left( \frac{t_{\rm ann}}{t_{\rm dom }^{\mathsmaller{\rm unbias}}} \right) \simeq C_d^{-1}\left( \frac{t_{\rm ann}}{t_{\rm dom }} \right)^2,
\end{equation}
where we approximated $H\simeq 1/2t$ valid for radiation-dominated universe.
During the annihilation process, DWs are driven to relativistic speed and radiate GWs~\cite{Vilenkin:1981zs,Preskill:1991kd,Chang:1998tb,Gleiser:1998na}, leading to the power spectrum today \cite{Hiramatsu:2010yz,Kawasaki:2011vv,Hiramatsu:2013qaa}
\begin{align}
\label{eq:_Omega_GW_0_DW}
\Omega_{\rm GW}h^2 =\mathcal{D}\frac{3}{32\pi}\epsilon_{\mathsmaller{\rm GW}}\alpha_{\rm ann}^2 S(f),
\end{align}
where $\mathcal{D}\sim 10^{-5}$ accounts for redshift, $S(f)$ is a spectral function peaked at $f_{\rm peak}  \sim {\rm nHz} \left({T_{\rm ann}/10~\rm MeV} \right)$ and $\epsilon_{\mathsmaller{\rm GW}} = 0.7 \pm 0.4$ is fitted against numerical simulation \cite{Hiramatsu:2013qaa}. We defer the details concerning the SGWB in SuM.~\ref{app:DW_interp}.

We used the software tool ${\tt PTArcade}$ \cite{Mitridate:2023oar} to conduct a series of Bayesian analysis of the interpretation of PTA signal as resulting from DW networks, SMBH binaries or a combination of them. We used the first $14$ frequency bins of NG15 \cite{NANOGrav:2020spf} and the first 13 frequency bins of IPTA2 \cite{Antoniadis:2022pcn}. We use the mode ``enterprise'' of PTArcade with $5\times 10^5$ steps, including Helling-Down inter-pulsar correlations. The favored SGWB and model posteriors of the DW interpretation are visible in Fig.~\ref{fig:DW_GWspectrum_NANO_IPTA}, under the assumption of DW annihilation into Standard Model (SM) and not including PBHs constraints as a prior. Additional details are reported in the SuM.

\textbf{\textit{PBH production---}}DW networks have close configurations that are filled with the energy density $V_{\rm bias}$.
Approximating DWs as spherical objects, their mass-energy is given by \cite{Deng:2016vzb,Deng:2017uwc,Deng:2020mds}
\begin{equation}
\label{eq:M_EoM}
    M=M_{\rm bulk} + M_{\rm bdy} + M_{\rm bind},
\end{equation}
where $M_{\rm bulk} = 4\pi V_{\rm bias}R^3/3$ is a volume term, $M_{\rm bdy}= 4\pi \sigma \gamma R^2$
is a boundary term, and $M_{\rm bind}\simeq -8\pi^2 G\sigma^2 R^3$ is the repulsive surface-surface gravitational binding energy. Closes DWs collapse into PBHs when they enter their Schwarzschild radius at a time $t_{\rm PBH}$
\begin{equation}
\label{eq:Schwarzchild_radius}
    R(t_{\rm PBH}) = 2GM(t_{\rm PBH}).
\end{equation}
Assuming that DWs collapse within one light-crossing time, we can replace $R(t_{\rm PBH}) \simeq t_{\rm PBH}$ in Eq.~\eqref{eq:Schwarzchild_radius}. Approximating $M\simeq M_{\rm bulk}$, leads to $t_{\rm PBH} = 2t_{\rm dom}$. However, the bulk term only accounts for $25\%$ of the DW mass. Accounting for the  surface ($50\%$) and binding energy ($25\%$), we find that PBHs form twice lighter, half a cosmic time earlier, at \cite{Gouttenoire:2023gbn}
\begin{equation}
\label{eq:t_PBH}
    t_{\rm PBH} \simeq  t_{\rm dom}.
\end{equation}
This enhances the PBH abundance by many orders of magnitude, leading the lower limit on $\alpha_{\rm ann}$ from PBH overproduction to be stronger by a factor $\sim 4$. Such difference is crucial to investigate PBH production within the PTA posterior region. The equation of motion (EoM) of a bubble of false vacuum in the thin-shell limit in General Relativity has been studied by many authors~\cite{Berezin:1982ur,Berezin:1987bc,Blau:1986cw,Maeda:1985ye,Tanahashi:2014sma,Deng:2016vzb,Deng:2017uwc,Deng:2020mds}. By solving the EoM, we obtain that spherical DWs enter their Schwarzschild radius at $t_{\rm PBH}$ if they have a radius equal to \cite{Gouttenoire:2023gbn}
 \begin{equation}
 \label{eq:fitting_fct_Rann}
     \frac{R_{\rm ann}^{\rm PBH}}{t_{\rm ann}} \simeq 0.780\, {\rm Log_{10}^2}\left(\alpha_{\rm ann}\right) -0.618\,{\rm Log_{10}}\left(\alpha_{\rm ann}\right)  + 0.407,
 \end{equation}
at the onset of the annihilation phase at $t_{\rm ann}$. DWs larger than this threshold are called \textit{late-annihilators}. The probability to find false vacuum region large enough to fully cover a ball of radius $R_{\rm ann}^{\rm PBH}$ can be calculated using methods borrowed from percolation theory \cite{Stauffer:1978kr,Essam_1980,Lalak:1993bp}. 
We discretize the DW network on a lattice whose spacing is set by the correlation length $L$. Each site can be occupied or not with a probability $p=0.5$ according to whether the field configuration lies in the false or true vacuum. We call $s_{\rm ball}(r_{\rm ann}^{\rm PBH})$ the minimal number of lattice sites of unit length to fully cover a ball of radius $r_{\rm ann}^{\rm PBH}\equiv R_{\rm ann}^{\rm PBH}/L$. The probability that all those sites are occupied is
\begin{equation}
\label{eq:mathcal_F_percolation_th}
\mathcal{F}(r_{\rm ann}^{\rm PBH})~\simeq s_{\rm ball}(r_{\rm ann}^{\rm PBH})\times~p^{s_{\rm ball}(r_{\rm ann}^{\rm PBH})}.
\end{equation}
\begin{figure}[t]
\centering
\begin{adjustbox}{max width=1\linewidth,center}
\raisebox{0cm}{\makebox{\includegraphics[ width=0.8\textwidth, scale=1]{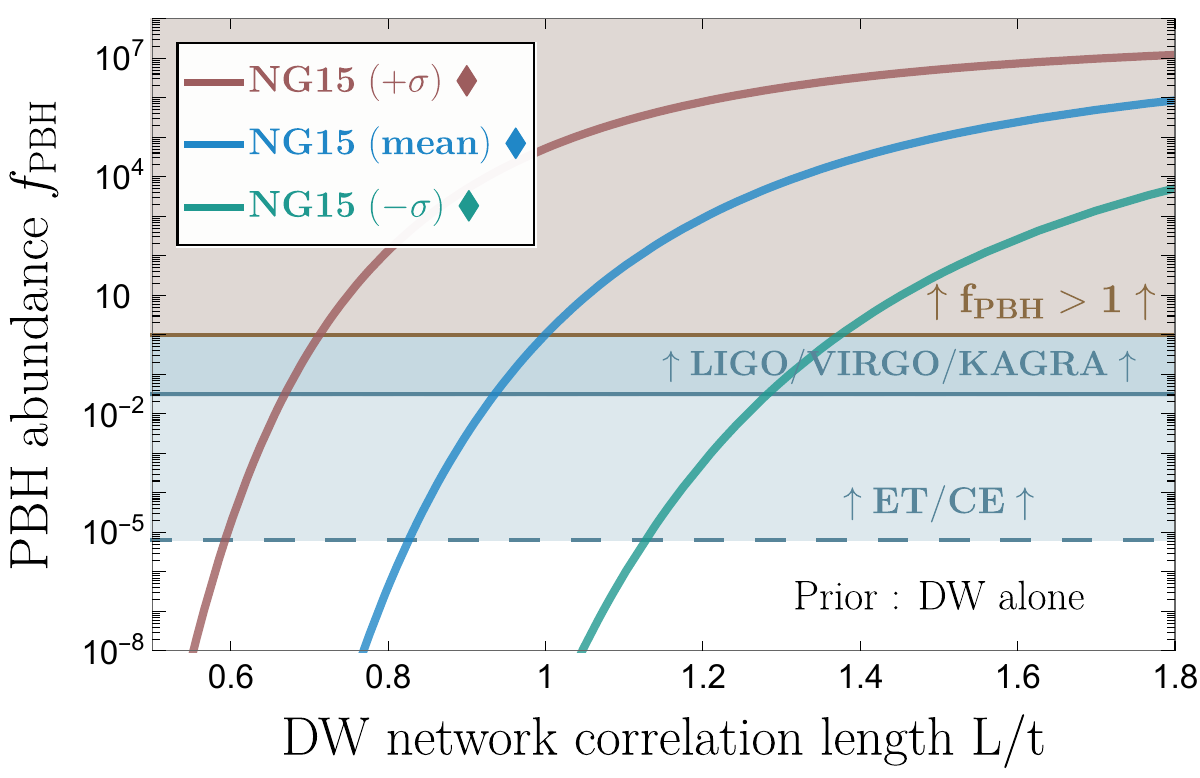}}}
\end{adjustbox}
\caption{ \label{fig:f_PBH_vs_alpha} The solid lines shows the PBH abundance from DW networks, given in Eq.~\eqref{eq:fPBH_abundance}, as a function of the correlation length $L$ of the DW network in unit of the cosmic horizon $t$. We set the DW network parameters to the mean and 68$\%$ deviation of the NG15 posteriors, represented by the three colored diamonds in Fig.~\ref{fig:DW_GWspectrum_NANO_IPTA}.  The brown region shows the overclosure bound on PBHs. The blue shows a sketch---as they slightly vary for the 3 solid curves---of the current PBH exclusion constraints from GW interferometers LVK \cite{Nakamura:1997sm,Raidal:2018bbj,Kavanagh:2018ggo,LIGOScientific:2019kan,DeLuca:2020qqa} and futuredetection prospect in CE/ET \cite{Chen:2019irf,Pujolas:2021yaw}.   }
\end{figure}
\begin{figure}[t!]
\centering
\begin{adjustbox}{max width=1\linewidth,center}
\raisebox{0cm}{\makebox{\includegraphics[ width=0.6\textwidth, scale=1]{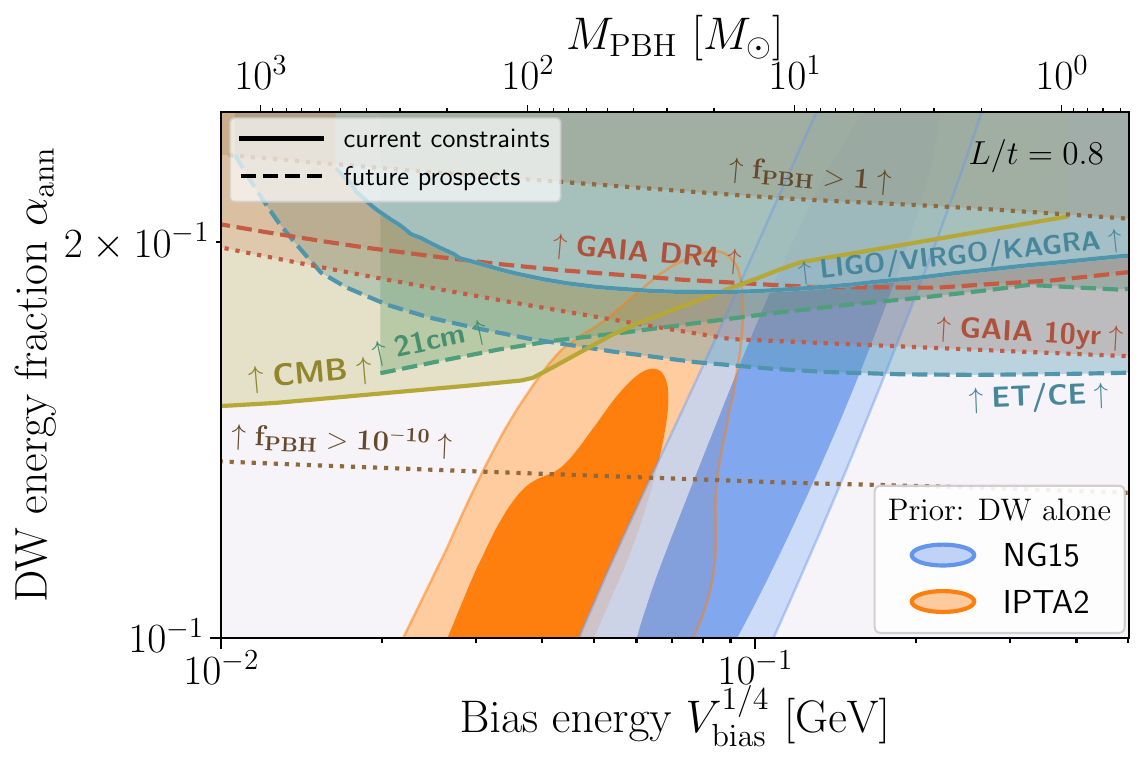}}}
\end{adjustbox}
\caption{\small \label{fig:DW_BBN_PBH_vary_alpha_ann} This is a zoom-in of Fig.~\ref{fig:DW_GWspectrum_NANO_IPTA}-right. In addition to current PBH exclusion constraints, shown with solid lines, from LVK \cite{Nakamura:1997sm,Raidal:2018bbj,Kavanagh:2018ggo,LIGOScientific:2019kan,DeLuca:2020qqa} and CMB~\cite{Ali-Haimoud:2016mbv,Poulin:2017bwe,Serpico:2020ehh}, we illustrate with dashed lines the anticipated PBH detection capabilities for CE/ET \cite{Chen:2019irf,Pujolas:2021yaw}, 21-cm survey \cite{Mena:2019nhm,Villanueva-Domingo:2021cgh,Villanueva-Domingo:2021spv}, and weak lensing analysis of GAIA data \cite{Chen:2023xyj,VanTilburg:2018ykj,Verma:2022pym}.}
\end{figure}
We calculate the function $s_{\rm ball}(r)$ using analytical summation over lattice sites and find that it is close to its upper limit $s_{\rm ball}(r)~< ~\left\lceil 2r\right\rceil^3$ given by the number of lattice sites covering a cube of radius $2R$ \cite{Gouttenoire:2023gbn}. We get rid of the unphysical discrete jumps by proposing a spline interpolation which is as close as possible as the analytical sum while being lower than its upper limit. One finds \cite{Gouttenoire:2023gbn}

\begin{equation}
\label{eq:s_sph_spline}
   \frac{s_{\rm ball}^{\rm smooth}(r)}{r^3} =8+\left(\frac{1+\tanh{(a_1\log{(r/a_2)})}}{2}\right)\left(\frac{4\pi}{3}-8 \right),
\end{equation}
where $a_1\simeq 1.15$, $a_2 \simeq 5.55$ and $r\equiv R/L$. 
The probability $\mathcal{F}(r_{\rm ann}^{\rm PBH}/L)$ in Eq.~\eqref{eq:mathcal_F_percolation_th} represents the energy fraction of the DW network that takes the form of closed DWs sufficiently large to encompass a sphere with a radius $r_{\rm ann}^{\rm PBH}$. Hence this is the energy fraction of DWs which collapse into PBHs.
We conclude that PBH abundance normalized to the observed relic DM density is
\begin{equation}
\label{eq:fPBH_abundance}
    f_{\rm PBH} \simeq \left(\frac{T_{\rm dom}}{T_{\rm eq}}\right) \mathcal{F}(r_{\rm ann}^{\rm PBH}) \simeq \left(\frac{T_{\rm dom}}{1~\rm TeV}\right)\frac{\mathcal{F}(r_{\rm ann}^{\rm PBH})}{8\times 10^{-13}},
\end{equation}
with $r_{\rm ann}^{\rm PBH}\equiv R_{\rm ann}^{\rm PBH}/L$ where $R_{\rm ann}^{\rm PBH}$ in Eq.~\eqref{eq:fitting_fct_Rann} and $\mathcal{F}(r)$ in Eq.~\eqref{eq:mathcal_F_percolation_th}. We omitted prefactors whose treatment is referred to \cite{Gouttenoire:2023gbn} since they do not lead to any visible change in the plots of this letter. The correlation length $L$ must be evaluated at $t_{\rm ann}$ when the network is still approximately in the scaling regime. In this letter, we follow two strategies. Either we set the value $L\simeq 0.8 t$, which is expected to be conservative in the $\mathcal{Z}_2$-symmetric model, see below Eq.~\eqref{eq:rho_DW_A}. Or we set $L$ as a free parameter and vary it within the range $L/t\in [0.0,1.5]$.
The PBH mass is given by the mass inside the Schwarzschild radius at horizon crossing $R_{\rm sch}(t_{\rm PBH})=t_{\rm PBH}$,
\begin{equation} \label{eq:MPBH}
    M_{\rm PBH} \simeq \frac{t_{\rm PBH}}{2G}\simeq 14~M_{\mathSun}\left( \frac{100~\rm MeV}{V_{\rm bias}^{1/4}} \right)^{2},
\end{equation}
where $t_{\rm PBH}$ is given by Eqs.~\eqref{eq:t_PBH} and \eqref{eq:t_dom}.

On the other hand, the formula in Eq.~\eqref{eq:fPBH_abundance} might overestimate the PBH abundance if friction is active during the annihilation phase, which could occur in certain scenarios \cite{Blasi:2022ayo,Blasi:2023sej}, if the number of vacua exceeds $N_{\text{DW}} > 2$, or if the energy bias $V_{\text{bias}}$ plays a role during the formation of the network \cite{Coulson:1995nv,Hindmarsh:1996xv,Pujolas:2022qvs}. These additional factors, as well as the effects of a surface tension that decreases over time~\cite{Babichev:2021uvl,Ramazanov:2021eya,Babichev:2023pbf}, are left for future studies. 

\textbf{\textit{PBH detectability---}}We have selected three points from the NG15 posterior distribution marked with diamonds in the right panel of Fig.~\ref{fig:DW_GWspectrum_NANO_IPTA}. These points include the mean and its $1\sigma$ (68$\%$) confidence interval. For each of those points, we display the PBH abundance as a function of the correlation length $L$ in Fig.~\ref{fig:f_PBH_vs_alpha}. The most stringent current constraints on these PBHs stem from the non-detection of a kilo-Hertz SGWB, which would be indicative of PBH binary mergers. They are derived from the ongoing searches by Earth-based interferometers LVK \cite{Nakamura:1997sm,Raidal:2018bbj,Kavanagh:2018ggo,LIGOScientific:2019kan,DeLuca:2020qqa}. Looking ahead, the most optimistic forecasts for detection are associated with the next generation of Earth-based interferometers Einstein Telescope and the Cosmic Explorer (ET/CE) \cite{Chen:2019irf,Pujolas:2021yaw}, assuming that all the astrophysical foregrounds would be subtracted \cite{Gouttenoire:2023gbn}.

\begin{figure}[t]
\centering
\begin{adjustbox}{max width=1\linewidth,center}
\raisebox{0cm}{\makebox{\includegraphics[ width=0.6\textwidth, scale=1]{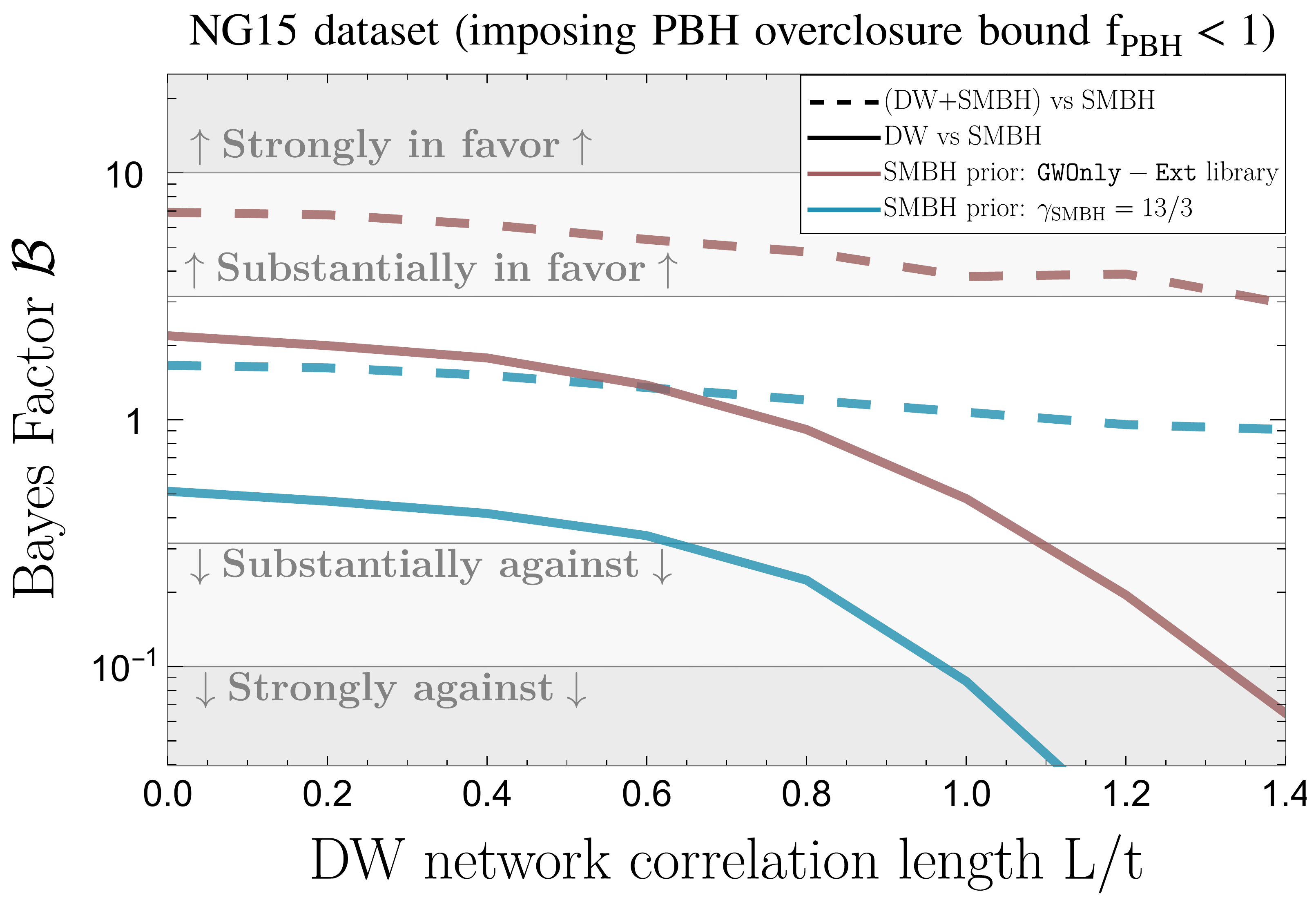}}}
\end{adjustbox}
\caption{ \label{fig:BayesFactor_NG15} Bayesian factors $\mathcal{B}_{Y,X}$, with values significantly above 1 suggest support for interpretation $Y$ over $X$. Conversely, values close to 1 indicate no clear preference between $X$ and $Y$. We evaluate the DW interpretation of the PTA signal in two scenarios: with (dashed) and without (solid) contributions from SMBH binaries, comparing these to the interpretation based solely on SMBH binaries. For the SMBH binaries, we consider two modeling approaches: a power-law with a spectral tilt of $\gamma=13/3$ (blue), and a power-law derived from the simulated $\tt{GWOnly-Ext}$ library, which incorporates astrophysical priors~\cite{NANOGrav:2023hvm} (red). Both models assume circular orbits and binary evolution driven by GWs only, see SuM.~\ref{app:DW_SMBH_interp} for the details.  DW networks with larger correlation length $L$ produce more PBHs, leading to tighter PBH overclosure bound, which in turn diminishes the quality of the PTA signal fit. The Bayes factor shows a significant decrease for the DW-only interpretation, while the reduction is more moderate in the combined DW+SMBH interpretation.  For each point of the figure, we vary the correlation length $L$ within the prior range $L\in [x,1.5]$ where $x$ is the horizontal axis of this plot.  
}
\end{figure}

We show GW interferometer reaches in Fig.~\ref{fig:DW_BBN_PBH_vary_alpha_ann} together with additional probes of PBHs that are formed under the hypothesis that the PTA signal arises from a DW network.
Firstly, accretion of PBHs leads to X-ray emissions that can modify the ionization fraction after recombination, leading to signatures in CMB \cite{Ali-Haimoud:2016mbv,Poulin:2017bwe,Serpico:2020ehh} and 21 cm
line surveys \cite{Mena:2019nhm,Villanueva-Domingo:2021cgh,Villanueva-Domingo:2021spv}. Secondly, weak lensing effects from PBHs can cause transient shifts in stars positions which could be detected in GAIA data~\cite{Chen:2023xyj,VanTilburg:2018ykj,Verma:2022pym}. We fixed the correlation length to $L=0.8t$, keeping in mind that PBH constraints and prospects would become stronger for larger $L$.

To quantify the evidence provided by the observed PTA data, denoted as $\mathcal{D}$, in favor of one model, say $X$, versus another, say $Y$, we employ the Bayesian factor
\begin{equation}
    \mathcal{B}_{Y,X} \equiv \mathcal{P}(\mathcal{D}|Y)\,/\,\mathcal{P}(\mathcal{D}/X),
\end{equation}
where $\mathcal{P}(\mathcal{D}/X)$ is the likelihood probability of observing data D given the model X.
In Fig.~\ref{fig:BayesFactor_NG15}, we show how the Bayes factor $\mathcal{B}_{Y,X}$ decreases as a function of the PBH abundance, which itself increases as a function of the correlation length $L/t$. The PBH overclosure bound can worsen the quality of both DW and DW+SMBH interpretation of NG15 signal. We model the SGWB from SMBH binaries with and without astrophysical priors, see SuM.~\ref{app:DW_SMBH_interp} for the details.  The astrophysical priors used in this work assume circular and GW-driven binaries. The inclusion of environment effects on SMBH binaries \cite{Antoniadis:2023zhi,NANOGrav:2023hfp,NANOGrav:2023hfp,Ellis:2023dgf} would improve the SMBH interpretation and lower the Bayes factor displayed in Fig.~\ref{fig:BayesFactor_NG15}. We leave it for future works.
We observe a much more limited impact of PBHs on the posterior of IPTA2.  However, IPTA2 finds no evidence for Hellings-Downs quadrupolar correlations, probably due to the incorporation of older (1994-2005) and noisier data \cite{NANOGrav:2023ctt,Antoniadis:2022pcn,EPTA:2023sfo,Zic:2023gta}.  In contrast, NG15 finds a Bayes factor of 226 in favor of HD correlation \cite{NANOGrav:2023gor}. Hence, we regard the Bayes factors derived from IPTA2 as less reliable and report them in the SuM (Fig.~\ref{fig:BayesFactor_IPTA2}). Our implementation of the PBH overclosure bound $f_{\rm PBH}<1$ as a prior in the Bayesian analysis and its limitation is discussed in Sec. II of the SuM.

\textbf{\textit{Discussion and outlook---}}We have conducted a Bayesian analysis of the NG15 and the IPTA2 datasets, examining the hypothesis that the PTA SGWB signal originates from a long-lived DW networks which annihilated in the early universe, possibly combined with a contribution from SMBH binaries which accounts for astrophysical priors.
We have reported new results concerning the modeling of PBH formation during the annihilation of long-lived DW network. The details are referred in the companion paper \cite{Gouttenoire:2023gbn}.  We find the DW interpretation of NG15 preferred to the GW-driven SMBH binary interpretation by a Bayes Factor of $\mathcal{B}=6.6$, neglecting for PBH formation. Accounting for PBHs formation, the Bayes factor falls to $\mathcal{B}=0.9$ or lower if the network correlation length is $L\gtrsim 0.8t$. Those PBHs with masses $M_{\rm PBH}= 11^{+13}_{-6} M_{\mathSun}$ for NG15, can undergo binary mergers and source kilo-Hz GW. In some region of the PTA posteriors, see Fig.~\ref{fig:DW_BBN_PBH_vary_alpha_ann}, enough PBHs can be produced to be within the detectability range of future Earth-based interferometers ET/CE, either under the form of individual events or of a SGWB.

This study enhances our comprehension of PBHs formation during the annihilation of long-lived DW networks, addressing gaps in previous literature \cite{Ferrer:2018uiu,Gelmini:2022nim,Gelmini:2023ngs}. It precisely characterizes the subset of DWs that collapse into PBHs -- termed \textit{late-annihilators} -- and develop a method to calculate their abundance using percolation theory. Although the DW interpretation of the PTA signal has been explored in numerous studies \cite{Sakharov:2021dim,Ferreira:2022zzo,ZambujalFerreira:2021cte,Madge:2023dxc,NANOGrav:2023hvm,Bai:2023cqj,Kitajima:2023cek,Blasi:2023sej,Servant:2023mwt,Lu:2023mcz,Li:2023tdx,Guo:2023hyp,King:2023cgv}, this work is pioneering in demonstrating quantitatively that such a scenario can yield a PBH population within the observable range of LVK. Furthermore, the Bayesian analysis improve previous works by incorporating uncertainties in the determination of the SGWB peak in lattice simulations \cite{Hiramatsu:2013qaa}, and modeling SMBH binaries with and without astrophysical priors. It should be noted that after the inclusion of environmental effects on SMBH binaries, which is a subject for future research, the DW interpretation might not be preferred over the SMBH interpretation anymore.


\textbf{\textit{Acknowledgments---}}YG thanks Simone Blasi, Alberto Mariotti, Andrea Mitridate, Oriol Pujol\`{a}s, Fabrizio Rompineve, Ken'ichi Saikawa, Peera Simakachorn and Ville Vaskonen for useful discussions. YG is grateful to the Azrieli Foundation for the award of an Azrieli Fellowship.
EV acknowledges support by
the European Research Council (ERC) under the European Union’s Horizon Europe research and innovation
programme (grant agreement No. 101040019). This work was conducted using the high performance computing cluster resources of Tel Aviv University. This work was supported by the Italian MUR Departments of Excellence grant 2023-2027 ``Quantum Frontiers'' and by Istituto Nazionale di Fisica Nucleare (INFN) through the Theoretical Astroparticle Physics (TAsP) project.

\appendix

\onecolumngrid

\fontsize{11}{13}\selectfont

\makeatletter
\renewcommand*{\fnum@figure}{{\normalfont \normalsize \figurename~\thefigure}}
\renewcommand*{\@caption@fignum@sep}{ : }
\makeatother

\renewcommand{\tocname}{\Large  Supplemental Material
\vspace{1 cm}}%

\titleformat{\section}
{\normalfont\fontsize{12}{14}\bfseries  \centering }{\thesection.}{1em}{}
\titleformat{\subsection}
{\normalfont\fontsize{12}{14}\bfseries \centering}{\thesubsection.}{1em}{}
\titleformat{\subsubsection}
{\normalfont\fontsize{12}{14}\bfseries \centering}{\thesubsubsection)}{1em}{}

\titleformat{\paragraph}
{\normalfont\fontsize{12}{14}\bfseries  }{\thesection:}{1em}{}

 {
 \hypersetup{linkcolor=black}
 \tableofcontents
 }

\section{Bayesian analysis of PTA signal}
\label{app:data_analysis}
\subsection{GW-induced pulsar timing residuals}

GW are linear tensor perturbations to the flat spacetime metric $g_{ab}=\eta_{ab}+h_{ab}$.
In presence of a GW propagating along unit vector $\hat{\Omega}$, the frequency $\nu$ observed at Earth of a pulse emitted by a pulsar $I$  propagating along the unit vector $-\hat{p}$ is shifted according to \cite{Allen:1997ad,Anholm:2008wy,Maggiore:2018sht}:
 \begin{equation}
 \label{eq:nu_I_over_nu}
     \frac{\delta \nu_I}{\nu} = - \Lambda^{ab}\left[ h_{ab}(t_e,\vec{x}_e) - h_{ab}(t_{I},\vec{x}_{I})\right],\qquad \textrm{with}\quad \Lambda^{ab} \equiv \frac{1}{2}\frac{\hat{p}^a\hat{p}^b}{1+\hat{\Omega}\cdot \hat{p}}.
 \end{equation}
($\vec{x}_e$, $t_e$) and ($\vec{x}_{I}$, $t_{I}=t_e-D$) are Earth and pulsar coordinates, $D$ being the earth-pulsar distance. The quantity $\Lambda^{ab}$ depends on the angle between the Earth, the pulsar and the GW source. The anomalous residual in arrival time of pulses emitted by a given pulsar is given by the integral of the accumulated frequency shift over time:
\begin{equation}
\label{eq:residual_R}
    R_I(t) \equiv - \int_0^{t} \frac{\delta \nu_I}{\nu} dt.
\end{equation}
We are searching for stochastic gravitational wave backgrounds (SGWBs). In order to extract this information and disentangle it from pure noise, we introduce the cross-power spectral
density of the time residual between two pulsars $I$ and $J$:
\begin{equation}
\label{eq:h_ab_h_ab_tau}
    S_{R,I,J}(f) \equiv  \int_{-\infty}^{+\infty} d\tau \,e^{2\pi i f \tau}\left<R_I(t)R_J(t+\tau)\right>.
\end{equation}
Plugging Eqs.~\eqref{eq:nu_I_over_nu} and \eqref{eq:residual_R} into Eq.~\eqref{eq:h_ab_h_ab_tau}, we obtain:
\begin{equation}
\label{eq:S_R_IJ_S_h}
   S_{R,IJ}(f) =  \frac{\Gamma_{IJ}}{12\pi^2}\frac{1}{f^3} h_c^2(f) =  \frac{\Gamma_{IJ}}{8\pi^4}\frac{H_0^2}{f^5}\Omega_{\rm GW},
\end{equation}
where $\Gamma_{IJ}$ are the so-called Helling-Down (HD) cross-correlation coefficients \cite{Hellings:1983fr}:
\begin{equation}
\label{eq:ORF}
    \Gamma_{IJ} =\frac{\delta_{IJ}}{2}+\frac{1}{2}+3(1-\cos{\theta_{IJ}})\left[\ln{\left(\frac{1-\cos{\theta_{IJ}}}{2} \right)-\frac{1}{6}} \right],
\end{equation}
where $\theta_{IJ}$ is the angle difference between of pulsars $I$ and $J$ seen from the Earth.
The quantity  $ h_c$ is the characteristic strain of the SGWB. It is defined from its power spectral density $S_h(f)$:
\begin{equation}
     h_c^2\equiv f S_h(f),\qquad \textrm{with} \quad S_h(f) \equiv \int_{-\infty}^{+\infty} d\tau \,e^{2\pi i f \tau}\left<h_{ab}(t)h^{ab}(t+\tau)\right> .
\end{equation}
The term $\Omega_{\rm GW}$ is the fraction of the universe energy attributed to the SGWB: 
\begin{equation}
\label{eq:Omega_GW_def}
    \Omega_{\rm GW}(f) \equiv \frac{1}{\rho_c}\frac{d\rho_{\rm GW}}{d \ln{f}} ,\qquad \textrm{with}\quad \rho_{\rm GW}= \frac{\left<\dot{h}_{ab}\dot{h}_{ab}\right>}{32\pi G},\quad \textrm{and}\quad \left<\dot{h}_{ab}\dot{h}_{ab}\right> = 8\pi^2 \int_0^\infty  df\,f^2S_h(f),
\end{equation}
where $\rho_c=3M_{\rm pl}^2H_0^2$ and $H_0$ is today Hubble constant.

\begin{figure*}[t!]
\centering
\begin{adjustbox}{max width=1\linewidth,center}
\raisebox{0cm}{\makebox{\includegraphics[ width=0.49\textwidth, scale=1]{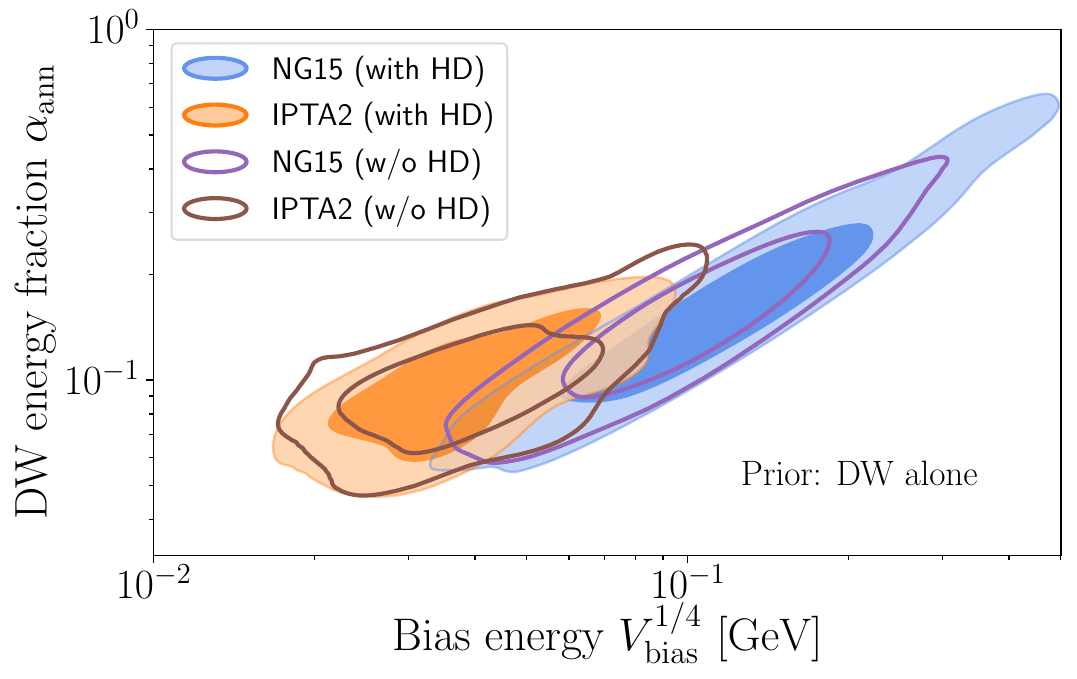}}}
\raisebox{0cm}{\makebox{\includegraphics[ width=0.49\textwidth, scale=1]{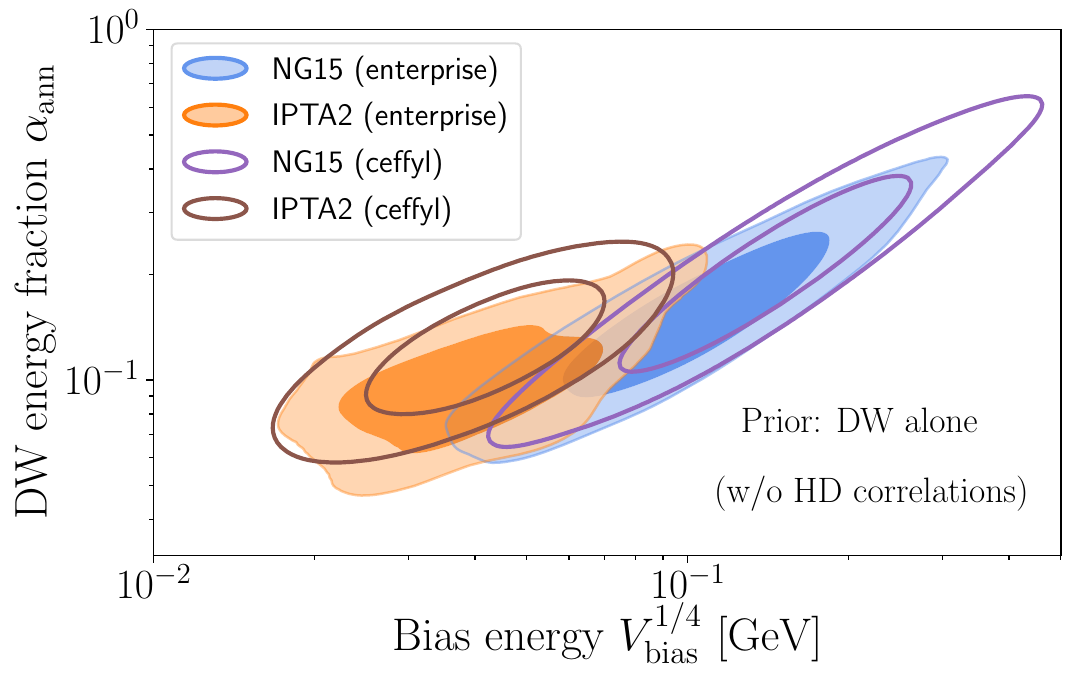}}}
\end{adjustbox}
\caption{\small \label{fig:DW_ceffyl_HD} LEFT: We compare the $68\%$ and $90\%$ preferred regions, with and without incorporating Helling-Down (HD) inter-pulsar cross-correlations. This means using Eq.~\eqref{eq:S_R_IJ_S_h} with either $\Gamma_{IJ}$ given by Eq.~\eqref{eq:ORF}, or simply setting $\Gamma_{IJ}=\delta_{IJ}$. The latter approach speeds up MCMC sampling by approximately a factor of 5. 
RIGHT: We compare the modes ``enterprise'' and ``ceffyl'' of ${\tt PTArcade}$. The former performs a MCMC sampling over the complete likelihood, including all noise parameters from each pulsar.  The latter performs a MCMC sampling on a reduced likelihood built from the free spectrum template, cf. the violins in Fig.~\ref{fig:DW_GWspectrum_NANO_IPTA}-left of the main text, speeding up the calculation by approximately a factor $10^4$. We use the mode ``enterprise'' with HD correlations for the remainder of this study. }
\end{figure*}

\subsection{Noise analysis}

\underline{Likelihood function}  ---
The theoretical prediction of pulse arrival times can achieve precision up to $100~\rm ns$ \cite{Hobbs:2006cd,Edwards:2006zg,Hobbs:2009yn} using timing models embedded in software tools such as  $\tt TEMPO2$ \cite{2012ascl.soft10015H} or $\tt PINT$ \cite{Luo:2020ksx}, both of which are integrated into ${\tt ENTERPRISE}$~\cite{enterprise}. The predicted time residuals can be described by a deterministic part $\vec{R}_{\rm det}$ which can be minimised by the timing model and a stochastic part which contains the SGWB signal along with various sources of noises:
\begin{equation}
\label{eq:R_th}
    \vec{R}_{\rm th} = \vec{R}_{\rm det}+ \vec{R}_{\rm GW} + \sum_n\vec{R}_{\rm noise}^{(n)},
\end{equation}
where vector components denote different times of observation and distinct pulsars. The $n$-sum runs over the different types of noise: red noise, dispersion measure noise, white noise and jitter noise \cite{NANOGrav:2015qfw,NANOGrav:2015aud,NANOGrav:2023ctt}.
Each source of noise can be modeled as a Gaussian process with a covariance matrix of Fourier coefficients denoted by $\left[S_n\right] $.
The likelihood for observing a time series of residuals $\vec{R}_{\rm obs}$ assuming all the model parameters contained in $\vec{R}_{\rm th}$, denoted by $\vec{\theta}$, can be written schematically as \cite{NANOGrav:2015qfw,NANOGrav:2015aud,NANOGrav:2023ctt}
 \begin{equation}
 \label{eq:likelihood}
\mathcal{P}(\vec{R}_{\rm obs}|\vec{\theta}) =\frac{\exp\left[-\frac{1}{2}\vec{R}_{\rm obs}^{T}C^{-1}\vec{R}_{\rm obs}  \right]}
{\sqrt{(2\pi)\det(C)}},\qquad \textrm{with} \quad {C} = \Delta f \times \vec{\mathcal{F}}\left( \left[S_{R,IJ}\right]+ \sum_n \left[S_n\right] \right)\vec{\mathcal{F}}^{-1}.
\end{equation}
The quantity $\left[S_{R,IJ}\right]$ is the covariance matrix of the SGWB signal in Eq.~\eqref{eq:S_R_IJ_S_h} expanded over the $k$ sampling frequencies $f_k=k \Delta f$ with $\Delta f =1/T$ the frequency spacing and $T$ the observation time. The quantity $\mathcal{F}$ is a vector of sine and cosine converting objects from Fourier representation to time representation. In order to remain schematic, we have omitted the deterministic component $\vec{R}_{\rm det}$. We refer to \cite{NANOGrav:2015qfw,NANOGrav:2015aud,NANOGrav:2023ctt} for the precise likelihood definition, and \cite{Taylor:2021yjx} for a review of the statistical analysis procedure. 
The likelihood $\mathcal{P}(\vec{R}_{\rm obs}|\vec{\theta})$ in Eq.~\eqref{eq:likelihood} can be calculated using the code ${\tt ENTERPRISE}$~\cite{enterprise} and ${\tt enterprise\_extensions}$~\cite{enterprise_ext}.

\underline{Posterior distribution}  ---
 Given observed data, the goal of the GW cosmologist or astronomer is to calculate the probability distribution for certain model parameters, e.g the DW network or SMBH binary parameters. This is encoded in the posterior distribution, $\mathcal{P}(\vec{\theta}|\vec{R}_{\rm obs})$, which illustrates the probability distribution of model parameters $\vec{\theta}$ given the observed data $\vec{R}_{\rm obs}$. It is related to the likelihood function via Bayes's theorem
 \begin{equation}
 \label{eq:posterior_Bayes}
     \mathcal{P}(\vec{\theta}|\vec{R}_{\rm obs}) = \frac{\mathcal{P}(\vec{R}_{\rm obs}|\vec{\theta})\mathcal{P}(\vec{\theta})}{\mathcal{P}(\vec{R}_{\rm obs})}.
 \end{equation}
 The term $\mathcal{P}(\vec{\theta})$ denotes the prior distribution, which reflects our initial knowledge of the parameters before observing the data. $\mathcal{P}(\vec{R}_{\rm obs})$ represents the marginal likelihood or evidence, serving as a normalizing factor to guarantee that the integration of the posterior distribution equals 1.
 The posterior distribution $ \mathcal{P}(\vec{\theta}|\vec{R}_{\rm mes})$ in Eq.~\eqref{eq:posterior_Bayes} can be reconstituted by sampling over the model parameters $\vec{\theta}$ using a Markov Chain Monte-Carlo (MCMC) code. For this purpose, the code ${\tt enterprise\_extensions}$~\cite{enterprise_ext} embeds the parallel-tempering MCMC code ${\tt PTMCMC}$~\cite{justin_ellis_2017_1037579}.  Since we are not interested in the posterior dependence on noise parameters, we marginalize over them by integrating Eq.~\eqref{eq:posterior_Bayes} over all of them. This can be trivially accomplished by visualizing only the relevant entries from the MCMC chains.

\begin{figure*}[t!]
\centering
\begin{adjustbox}{max width=1\linewidth,center}
\raisebox{0cm}{\makebox{\includegraphics[ width=0.49\textwidth, scale=1]{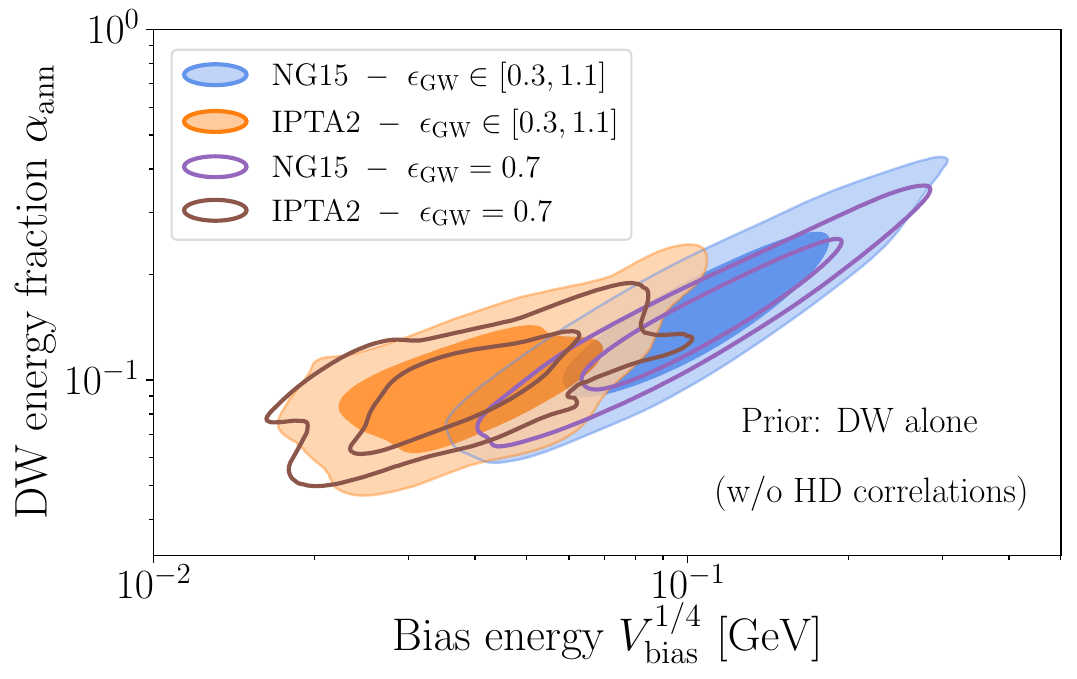}}}
\raisebox{0cm}{\makebox{\includegraphics[ width=0.49\textwidth, scale=1]{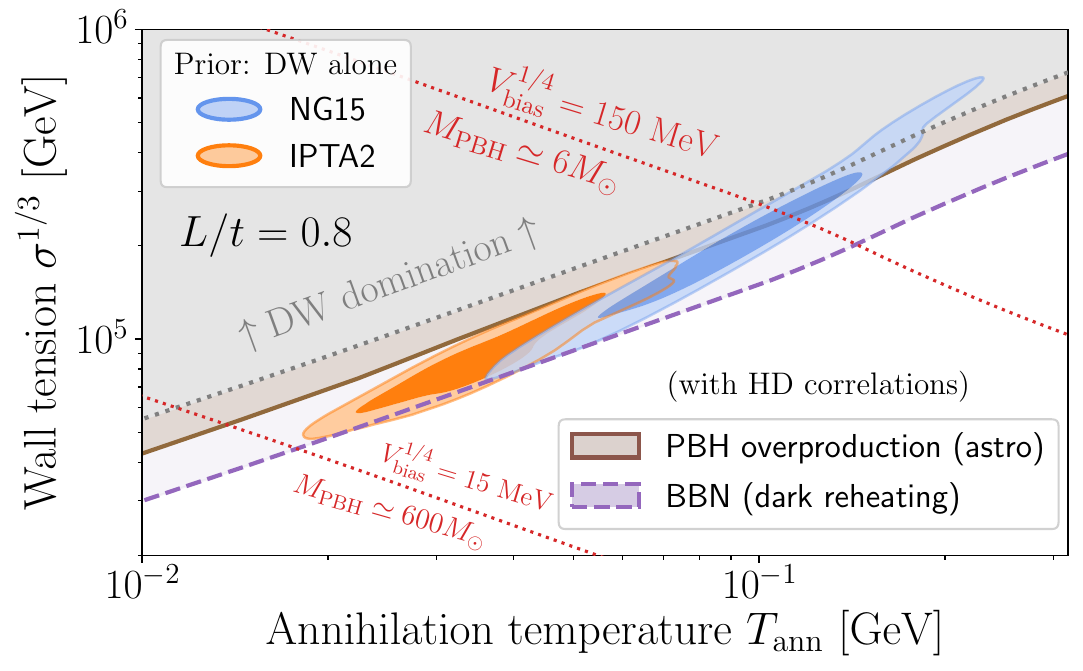}}}
\end{adjustbox}
\caption{\small \label{fig:DW_vary_epsilonGW_vs_fixed} LEFT: the quantity $\epsilon_{\rm GW}$ determines the amplitude of the SGWB from DW networks, see Eq.~\eqref{eq:_Omega_GW_0_DW_app}. Lattice simulations suggest $\epsilon_{\mathsmaller{\rm GW}} = 0.7 \pm 0.4$ \cite{Hiramatsu:2013qaa}. We compare the posteriors with this quantity being fixed to the central value $\epsilon_{\rm GW}=0.7$ or marginalized over the interval $\epsilon_{\mathsmaller{\rm GW}} \in [0.3,1.1]$. We adopt the latter choice in the rest of this work.
RIGHT: we present the posterior distribution in the plane $T_{\rm ann}-\sigma^{1/3}$ instead of $V_{\rm bias}^{1/4}-\alpha_{\rm ann}$, under the assumption of SM reheating (BBN prior evaded), no PBH prior (omitted), while varying $\epsilon_{\rm GW}$ within its error interval and including HD correlations, like in Fig.~\ref{fig:DW_GWspectrum_NANO_IPTA}-right in the main text. }
\end{figure*}

 \underline{PTArcade} --- All the steps described beforehand can be realised at once using the wrapper ${\tt PTArcade}$ \cite{Mitridate:2023oar}.  We considered the first $14$ frequency bins of NG15 \cite{NANOGrav:2020spf} and the first 13 frequency bins of IPTA2 \cite{Antoniadis:2022pcn}. We used ${\tt GetDist}$ tool~\cite{Lewis:2019xzd} to visualize the posteriors. We included $5\times 10^5$ steps and discarded 25$\%$ of each MCMC chain as burn-in. 
 We use the mode ``enterprise'' of ${\tt PTArcade}$, which accounts for the full likelihood in Eq.~\eqref{eq:likelihood}. We included the HD correlation defined in Eq.~\eqref{eq:ORF}. Each Bayesian analysis takes around $300$~hours per CPU. We estimated that omitting the HD correlation speeds up the calculation by a factor $5$ while the mode ``ceffyl'' which uses the free spectrum templates from $\tt ceffyl$ \cite{Lamb:2023jls,ceffylgit} allows to calculate posteriors within a few minutes on a single CPU. We compare the posteriors distribution for the different modes in Fig.~\ref{fig:DW_ceffyl_HD}. Despite the time-saving alternatives, all the analyses in this study were conducted using the most precise modes.  The violin-shaped posteriors of the PTA signals shown in Fig.~\ref{fig:DW_GWspectrum_NANO_IPTA}-left were not computed in this work but were imported directly from \href{https://zenodo.org/record/8060824}{NG15} and
\href{https://zenodo.org/record/5787557}{IPTA2}.

\subsection{Bayes factor.}
To evaluate how well the observed PTA data $\vec{R}_{\rm obs}$ supports one theoretical model $X$ over another $Y$, we can rely on the Bayesian factor:
\begin{equation}
    \mathcal{B}_{Y,X} \equiv \frac{\mathcal{P}(\vec{R}_{\rm obs}|Y)}{\mathcal{P}(\vec{R}_{\rm obs}|X)}.
\end{equation}
The software ${\tt enterprise\_extensions}$~\cite{enterprise_ext} can be used to calculate $\mathcal{B}_{Y,X}$ by sampling the two models $X$ and $Y$ at the same time, with a new parameter $m\in \vec{\theta}$ labelling the two models, e.g. $m_1$ and $m_2$. Then the ratio of likelihoods turns into a ratio of posteriors  \cite{Taylor:2021yjx}:
\begin{equation}
    \mathcal{B}_{Y,X} = \frac{\mathcal{P}(m_1|\vec{R}_{\rm obs})}{\mathcal{P}(m_2|\vec{R}_{\rm obs})}.
\end{equation} 
According to Jeffrey's scale \cite{1939thpr.book.....J,kass1995bayes},  evidence for or against an interpretation is barely worth mentioning for $|{\log}_{10}{\mathcal{B}_{Y,X}}| < 0.5$, becomes substantial for $>0.5$, strong for $>1$, very strong for $>1.5$ and decisive for $>2$.

\subsection{Bayesian analysis conducted in this work.}
\begin{figure*}[t]
\centering
\begin{adjustbox}{max width=1\linewidth,center}
\raisebox{0cm}{\makebox{\includegraphics[ width=0.95\textwidth, scale=1]{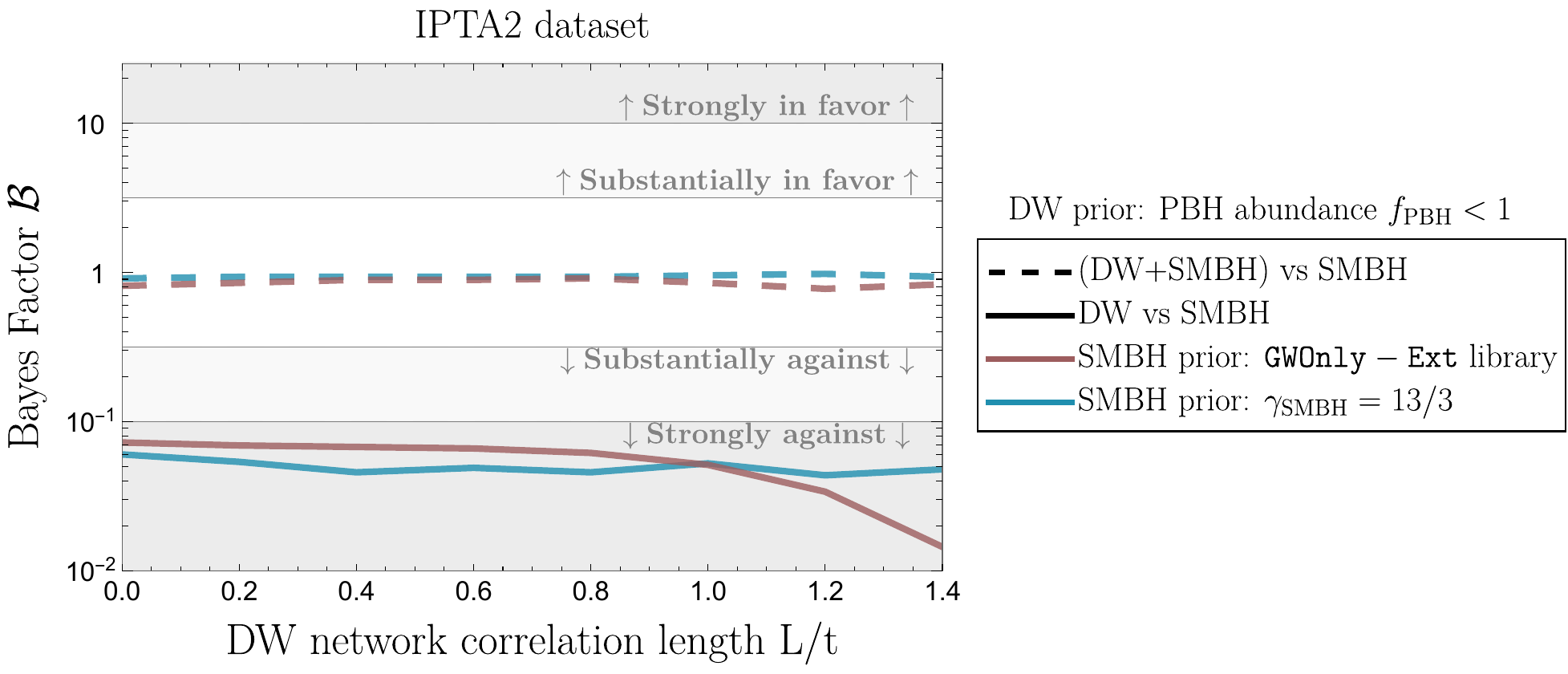}}}
\end{adjustbox}
\caption{  \label{fig:BayesFactor_IPTA2} Similar to Fig.~\ref{fig:f_PBH_vs_alpha} in the main text, but for the IPTA2 dataset, we analyze the Bayes factor which compares the predictive power of the DW and DW+SMBH interpretations of the PTA signal against the sole SMBH binary interpretation. Two distinct models of the SGWB originating from SMBH binaries are considered, as detailed in App.~\ref{app:DW_SMBH_interp}. A greater correlation length $L$ implies a higher abundance of PBHs and stricter constraints due to potential PBH overproduction.  The influence of PBH abundance on the IPTA2 fit is less pronounced than in the NG15 dataset. We refer to Tab.~\ref{tab:BF} for values of Bayes factors in different scenarios.}
\end{figure*}

 In this work, we perform various types of Bayesian analysis of PTA datasets, including either SGWB from DW networks, from SMBH binaries or a combination of them, with various priors for both signals: SM or dark reheating, PBH overproduction bound for different values of the correlation length included or not, and astrophysical priors on SMBH binaries included or not. We compare the different models using the Bayes factor in Table~\ref{tab:BF} and provide the the mean values of the posteriors for model parameters in Table~\ref{tab:meanPoseterior}.
The impact of the PBH abundance on the Bayes factor, through the DW network correlation length, can be appreciated in Figs.~\ref{fig:f_PBH_vs_alpha} of the main text for NG15 and Fig.~\ref{fig:BayesFactor_IPTA2} of this SuM for IPTA2.
We acknowledge that all the Bayesian analysis presented in this paper were generated using the incorrect expression $\mathcal{F}(r_{\rm ann}^{\rm PBH}) \simeq p^{s_{\rm ball}(r_{\rm ann}^{\rm PBH})}$ of a previous version of the manuscript rather than the updated expression $\mathcal{F}(r_{\rm ann}^{\rm PBH}) \simeq s_{\rm ball}(r_{\rm ann}^{\rm PBH}) \times p^{s_{\rm ball}(r_{\rm ann}^{\rm PBH})}$ as specified in Eq.~11 in the main text. This only leads to a slight underestimation of the abundance of PBHs, effectively similar to underestimating the DW network correlation length $L/t$ by approximately 5$\%$.

\underline{NG15 vs IPTA2} ---
The impact of PBH on the IPTA2 analysis is more moderate due to the fact that the DW-only interpretation is disfavoured over the SMBH-only interpretation. This is an important difference between the Bayesian analysis of IPTA2 and NG15. This difference should be imputed to the inclusion in IPTA2 of ``legacy" data collected between 1994 and 2005 on a single frequency bandwidth and a low number of pulsars (seven) \cite{Antoniadis:2022pcn}. In spite that it allows to reach lower frequencies, see Fig.~\ref{fig:DW_GWspectrum_NANO_IPTA}-left in the main text, this part of the dataset is much noisier. In particular the dispersion between frequency channels, induced by the ionized interstellar medium, is removed only incoherently, loosing phase information, and limiting timing precision \cite{EPTA:2023sfo}. Instead, NG15 \cite{NANOGrav:2023ctt} and also the latest EPTA \cite{EPTA:2023sfo} and PPTA datasets \cite{Zic:2023gta} rely on next-generation coherent dedispersion systems, which offer better pulse recovery and a higher timing precision. This could explain why no evidence (Bayes factor of 0.5) for Hellings-Downs quadrupolar correlations was found in IPTA2. Instead, NG15 has a Bayes factor of 226 in favor of HD correlations. Also in EPTA, the Bayes factor supporting the HD correlations decreases from 60 to 4 when the ``legacy" data is incorporated in the analysis \cite{EPTA:2023fyk}.
Moreover, the GW amplitude posterior for the lowest frequency bins of PPTA \cite{Zic:2023gta} and EPTA without ``legacy" data \cite{EPTA:2023sfo} agree well with NG15 \cite{NANOGrav:2023ctt} and not really with IPTA2. This seems to suggest for the presence of systematic effects in the ``legacy'' data from IPTA2. Less important but worth mentioning: NG15 has more pulsars, it contains 67 millisecond pulsars with more than 3 years of observation, while IPTA2 has only 53 of those (for information, EPTA and PPTA have 25 and 29 pulsars with more than 3 years of observation). Because of the above reasons, we have judged the IPTA2 dataset less reliable than NG15 and we have reported its Bayesian analysis to this SuM (Fig.~\ref{fig:BayesFactor_IPTA2}). 

 \section{DW interpretation}
\label{app:DW_interp}
\subsection{SGWB signal.}
The GW power spectrum today produced by long-lived DWs annihilating at a temperature $T_{\rm ann}$ can be expressed as \cite{Hiramatsu:2010yz,Kawasaki:2011vv,Hiramatsu:2013qaa}:
 \begin{align}
\label{eq:_Omega_GW_0_DW_app}
\Omega_{\rm GW}h^2 =\mathcal{D}\frac{3}{32\pi}\epsilon_{\mathsmaller{\rm GW}}\alpha_{\rm ann}^2 S(f),
\end{align}
where $\mathcal{D}$ is the redshift factor:
\begin{equation}
    \mathcal{D} = \Omega_{\rm rad}^{0}h^2\left( \frac{g_{*}(T_{\rm ann})}{g_\star(T_0)}\right)\left( \frac{g_{s,*}(T_0)}{g_{s,*}(T_{\rm ann})}\right)^{4/3}\simeq 3.49 \times 10^{-5}\left( \frac{g_{*}(T_{\rm ann})}{10.75}\right)\left( \frac{10.75}{g_{s,*}(T_{\rm ann})}\right)^{4/3}.
\end{equation}
where $\Omega_{\rm rad}^{0}h^2 \simeq 4.18\times 10^{-5}$ is the radiation relic density \cite{ParticleDataGroup:2022pth}.
The dimensionless factor $\epsilon_{\mathsmaller{\rm GW}}$ encodes deviation from the quadrupole formula \cite{Saikawa:2017hiv} with lattice simulations suggesting \cite{Hiramatsu:2013qaa}:
\begin{equation}
\label{eq:epsilong_GW}
  \epsilon_{\mathsmaller{\rm GW}} = 0.7 \pm 0.4.
\end{equation}
We model the spectral function:
\begin{equation}
S(f) = \frac{2}{\left(f/f_{\rm peak}\right)+\left(f_{\rm peak}/f\right)^3},
\end{equation}
where the IR slope is $\Omega_{\rm GW}\propto f^{3}$ to respect causality \cite{Durrer:2003ja,Caprini:2009fx,Cai:2019cdl,Hook:2020phx} and the UV slope is $\Omega_{\rm GW}\propto f^{-1}$ as suggested by lattice simulations results \cite{Hiramatsu:2013qaa} (though more complicated spectra are possible, see e.g.~\cite{Gelmini:2020bqg}). The peak frequency today is given by 
\begin{equation}
f_{\rm peak} = \frac{a(t_{\rm ann})}{a(t_0)} H(t_{\rm ann}) \simeq 1.08~{\rm nHz}\left( \frac{g_\star(T_{\rm ann})}{10.75} \right)^{1/2}
\times\left( \frac{10.75}{g_{*s}(T_{\rm ann})}\right)^{1/3}\left(\frac{T_{\rm ann}}{10~\rm MeV} \right).
\end{equation}

\begin{figure*}[t!]
\centering
\begin{adjustbox}{max width=1\linewidth,center}
\raisebox{0cm}{\makebox{\includegraphics[ width=0.7\textwidth, scale=1]{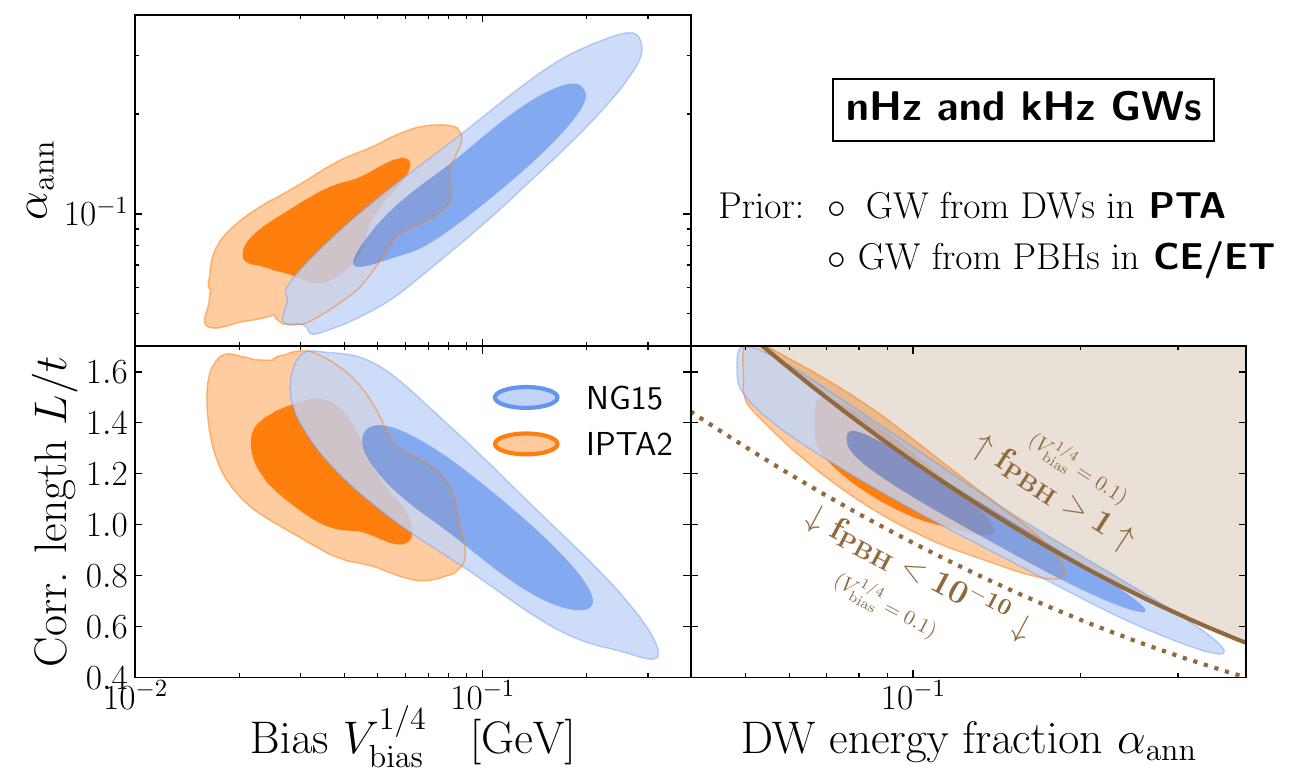}}}
\end{adjustbox}
\caption{\small \label{fig:DW_BBN_PBH_vary_alpha_ann_left}  The ellipses show the $68\%$ and $90\%$ preferred regions of the DW interpretation of the PTA signal additionally assuming that the produced PBHs emit kilo-Hz SGWB within the reach of future Earth-based interferometer CE/ET assuming the astrophysical foreground can be subtracted \cite{Chen:2019irf,Pujolas:2021yaw}. We simultaneously impose the future detectability in CE/ET and current LVK exclusion constraints as priors \cite{Nakamura:1997sm,Raidal:2018bbj,Kavanagh:2018ggo,LIGOScientific:2019kan,DeLuca:2020qqa}. This explains why the posterior is peaked around $f_{\rm PBH}\lesssim 1$.  In brown, we display a sketch -- as it moderately depends on $V_{\rm bias}$ -- of the PBH overclosure bounds.}
\end{figure*}
\subsection{Priors}
\underline{Prior on ${\epsilon}_{{\rm GW}}$} ---
 In order to account for the uncertainties in the GW modeling, we consider $\epsilon_{\mathsmaller{\rm GW}}$ in Eq.~\eqref{eq:epsilong_GW} as a model parameter with a prior $\epsilon_{\mathsmaller{\rm GW}} \in [0.3,1.1]$, over which we marginalize at the end in all the Bayesian analysis of this work. The impact  on the posterior distribution of this choice with respect to fixing $\epsilon_{\mathsmaller{\rm GW}}=0.7$ can be appreciated in the left panel of Fig.~\eqref{fig:DW_vary_epsilonGW_vs_fixed}.

 \underline{BBN prior} ---
 The presence of extra number $\Delta N_{\rm eff}$ of relativistic degrees of freedom (DoFs) at BBN and CMB would change the expansion rate of the universe and impact the CMB data or the abundance of light elements \cite{Pitrou:2018cgg,Dvorkin:2022jyg}. In \cite{Gouttenoire:2023gbn}, we show that DW network contribute to the number of relativistic DoFs as:
\begin{equation}
\label{eq:Delta_Neff_bound_DW}
    \Delta N_{\rm eff}  =\frac{8}{7}\left( \frac{{g_{*}(T_0)}}{2} \right)\left( \frac{11}{4} \right)^{4/3}\alpha_{\rm DW}(T) \simeq 7.4\,\alpha_{\rm ann}~\lesssim~ 0.4,
\end{equation}
where $g_{*}(T_0) \simeq 3.38$. We choose a conservative bound $\Delta N_{\rm eff}\lesssim 0.4$ \cite{ParticleDataGroup:2022pth}.
Two scenarios must be distinguished according to whether upon annihilating the DW network reheates dark DoFs or Standard Model (SM) DoFs. In the first case, the constraint in Eq.~\eqref{eq:Delta_Neff_bound_DW} must be applied to all the parameter space. In the second case, the bound Eq.~\eqref{eq:Delta_Neff_bound_DW} is relaxed if DW annihilate above the neutrino decoupling temperature $T_{\rm ann} \gtrsim 1~\rm MeV$ \cite{Bai:2021ibt,Kawasaki:2000en,Hasegawa:2019jsa}. We find that only the first scenario is constraining DW interpretation of PTA, see the dashed purple line in Fig.~\ref{fig:DW_vary_epsilonGW_vs_fixed}-right and Fig.~\ref{fig:DW_GWspectrum_NANO_IPTA}-right in the main text.

\underline{DW-domination prior} ---
Another bounds comes from the possibility for DWs to dominate the energy budget of the universe when $t_{\rm ann} \gtrsim \textrm{Min}(t_{\rm dom},t_{\rm dom}^{\rm unbias})$ according to whether the DW network energy is dominated by its volume or surface contribution. For annihilating DW network, we find that $t_{\rm dom}<t_{\rm dom}^{\rm unbias}$ always hold, hence we have the bound:
\begin{equation}
\label{eq:DW_domination}
\alpha_{\rm ann} ~\gtrsim~C_d^{-1}.
\end{equation}
Since it is less constraining that the PBH prior discussed next, we do not include the DW-domination prior in this work. We however indicate this region with a gray dotted line in Fig.~\ref{fig:DW_vary_epsilonGW_vs_fixed}-right of the SuM and Figs.~\ref{fig:DW_GWspectrum_NANO_IPTA}-right of the main text.

 \underline{PBH prior: overclosure bound} ---
 One of the novelty of this work is to be able to include PBH overproduction as a prior in the Bayesian analysis. Since the PBH abundance is very sensitive on the model parameter, we considered sufficient to content with the DM overclosure bound:
 \begin{equation}
 \label{eq:PBH_overclosure_bound}
     f_{\rm PBH} < 1,
 \end{equation}
 with $f_{\rm PBH}$ given in the main text, instead of the more precise astrophysical constraints from LVK and CMB, see main text. The abundance of PBHs is in particular very sensitive to the correlation length $L$, whose precise value remains uncertain (see however the discussion in the main text). Therefore, either we fix it to $L/t=0.8$ or treat it as a free parameter within the range $L/t \in [X,1.5]$, with $X$ varying from $X=0$ to $X=1.4$, see Tab.~\ref{tab:BF}.

  \underline{PBH prior: future detectability} ---
  The previous paragraph treats the PBH abundance as a constraint. We now consider a positivist perspective in which PBHs would be detected by the next-generation of terrestrial interferometers, namely the Einstein Telescope and the Cosmic Explorer (ET/CE) \cite{Chen:2019irf,Pujolas:2021yaw,Gouttenoire:2023gbn}. 
 Fig.~\ref{fig:DW_BBN_PBH_vary_alpha_ann_left} shows the preferred range of DW parameters that simultaneously account for the nano-Hz PTA signal and source kilo-Hz GW within the future detectability range of CE/ET \cite{Chen:2019irf,Pujolas:2021yaw}.  We assume that all astrophysical foregrounds have been successfully removed from CE/ET data.  This scenario becomes feasible as soon as $L/t \gtrsim 0.6$, see also Fig.~\ref{fig:f_PBH_vs_alpha} of the main text.

\underline{Additional effects in PBH production} --- On one hand, the estimate of the PBH abundance in the main text is conservative for several reasons.
Firstly, by assuming that the DWs are spherical, our results provide a lower limit to their mass, which leads to an underestimation of the PBH abundance. Instead, the late-annihilator fraction is independent of the DW shapes because these DWs have been specifically selected for being sufficiently large to enclose a sphere with radius $R_{\text{ann}}^{\text{PBH}}$.
Secondly, we do not consider contributions to PBH abundance from DWs that collapse deep within the horizon. These DWs must maintain nearly spherical shapes until they contract within their Schwarzschild radius. Such possibilities strongly depend on the precise shape of the DWs and the extent to which they deviate from sphericity, which we leave for future research.

\underline{Practical implementation of the priors} ---  
In this work, we implemented the priors using the ``Parameter'' class in ${\tt enterprise}$, utilizing methods such as ``Uniform'' or ``LinearExp''. In ${\tt PTArcade}$, we can equivalently use the ``prior'' function. This allows for proper normalization of the priors. The only exception is the PBH prior $f_{\rm PBH}<1$ which involves a complex function $f_{\rm PBH}(\theta)$ of model parameters $\theta=(\alpha_{\rm ann},V_{\rm bias}^{1/4})$ and therefore cannot be implemented as easily as the others. We implemented this prior by setting the GW spectrum in Eq.~\eqref{eq:_Omega_GW_0_DW_app} to 0 whenever $f_{\rm PBH}>1$ (another -- possibly better -- choice would have been to set the GW spectrum to infinity instead of 0). This approach artificially drives the likelihood to very small values, effectively approximating the PBH prior as a Heaviside function $\Theta(1-f_{\rm PBH}(\theta))$ in Eq.~\eqref{eq:posterior_Bayes}. This approach neglects the normalization factor $N$ of the prior probability distribution $\Theta(1-f_{\rm PBH}(\theta))/N$, which we hope does not significantly impact the calculation of the Bayes factors.

\section{DW + SMBH interpretation}
\label{app:DW_SMBH_interp}
\subsection{SGWB from SMBH binaries}

\underline{Single SMBH binary} --- It is generally accepted that most large galaxies have a supermassive black hole (SMBH) at their center \cite{Kormendy:2013dxa}. These galaxies grow over time by merging with others, leading to scenarios where a single galaxy might host several SMBHs. When these SMBHs are sufficiently close, they could potentially form a binary system.
The GW power radiated by a circular binary system of masses $m_1$ and $m_2$ in the quadrupole approximation reads \cite{Peters:1963ux}:
\begin{equation}
\label{eq:quadrupole_Egw_0}
\frac{dE_{\rm GW}}{dt_r} = \frac{32}{5}\frac{c^5}{G}\left( \frac{\pi G \mathcal{M} f_r}{c^3}\right)^{10/3},
\end{equation}
where $t_r$ and $f_r$ are the time and GW frequency measured in the binary rest frame. They are related to the observed GW frequency by $f_r=f(1+z)$ where $z$ is the source redshift and to the orbital frequency by  $f_r=2f_{\rm orb}$. The quantity $\mathcal{M}=(m_1m_2)^{3/5}/(m_1+m_2)^{1/5}$ is the chirp mass.
Equating Eq.~\eqref{eq:quadrupole_Egw_0} with the derivative of the orbital energy $E_{\rm orb}=-Gm_1m_2/2R$ using Kepler's law $\omega_{\rm orb}^2=G(m_1+m_2)/R^3$, we obtain the so-called ``residence time'' per logarithmic frequency interval:
\begin{equation}
\label{eq:residence_time}
    \frac{dt_r}{d\ln {f_r}} = \frac{5}{96\pi^{8/3}} \left( \frac{c^3}{G\mathcal{M}} \right)^{5/3} f_r^{-8/3}.
\end{equation}
Plugging Eq.~\eqref{eq:residence_time} into Eq.~\eqref{eq:quadrupole_Egw_0}, we deduce the source-frame GW energy per logarithmic frequency interval \cite{Thorne:1987af,Maggiore:2007ulw}:
\begin{equation}
\label{eq:quadrupole_Egw}
    \frac{dE_{\rm gw}}{d\ln{f_r}}  = \frac{dt}{\ln{f_r}} \frac{dE_{\rm gw}}{dt_r} = \frac{\pi^{2/3}}{3G}(G \mathcal{M})^{5/3} f_r^{2/3}.
\end{equation}

\underline{Population of SMBH binaries} ---
We consider a population of SMBH binaries with comoving number density per unit redshift and per unit of chirp mass $d^2n/dz/d\mathcal{M}$. The associated radiated energy spectrum $d\rho_{\rm GW}/d\ln f$ is \cite{Sesana:2008mz}:
\begin{equation}
\label{eq:rho_gw_lnf}
    \frac{d\rho_{\rm GW}}{d\ln f} = \frac{\pi}{4}\frac{c^2}{G}f^2h_c^2(f) =\int_0^{\infty}dz\int_0^{\infty}d\mathcal{M} \frac{d^2n}{dzd\mathcal{M}}\frac{1}{(1+z)} \frac{dt}{\ln{f_r}} \frac{dE_{\rm gw}}{dt_r}\bigg|_{f_r = f(1+z)}.
\end{equation}
where $1/(1+z)$ accounts for the cosmological redshifting of the GW energy.
Plugging Eq.~\eqref{eq:quadrupole_Egw} into Eq.~\eqref{eq:rho_gw_lnf}, we obtain the characteristic GW strain:
\begin{equation}
\label{eq:h_c_dn}
   h_c^2(f) =  \frac{4}{3\pi^{1/3}c^2} \int_0^{\infty}dz\int_0^{\infty}d\mathcal{M}\left[\frac{(G\mathcal{M})^{5}}{(1+z)f^4} \right]^{1/3}\frac{d^2n}{dzd\mathcal{M}}.
\end{equation}
From the previous equation, we conclude that the SGWB power spectrum of a population of GW-driven coalescing circular SMBH binaries is a power-law:
\begin{equation}
\label{eq:hc_SMBH-B}
    h_c(f)= A_{\rm SMBH}\left( \frac{f}{f_{\rm yr}} \right)^{\alpha},\qquad \alpha = -2/3,\quad f_{\rm yr} = 1~\rm yr^{-1}\simeq 31.7~\rm nHz,
\end{equation}
with amplitude:
\begin{align}
\label{eq:ASMBH_dn}
    A_{\rm SMBH} =  5 \times 10^{-15} \left( \frac{\mathcal{M}_0}{10^9~\rm M_{\mathSun}} \right)^{5/3} \left( \frac{n_0}{2\times 10^{-3}~\rm Mpc^{-3}} \right) I^{1/2},
\end{align}
where $I$ is a dimensionless quantity:
\begin{equation}
    I = \int_0^{\infty}\frac{dz}{(1+z)^{1/3}}\int_0^{\infty}d\mathcal{M}\left(\frac{\mathcal{M}}{\mathcal{M}_0} \right)^{5/3}n_0^{-1}\frac{d^2n}{dzd\mathcal{M}}.
\end{equation}
In terms of the fractional energy density, defined in Eq.~\eqref{eq:Omega_GW_def}, it gives
\begin{equation}
\label{eq:ASMBH_Omega_GW}
\Omega_{\rm GW}(f) =\Omega_{\rm SMBH }\left( \frac{f}{f_{\rm yr}} \right)^{n_t},\qquad n_t =2(1+\alpha)=-2/3,
\end{equation}
and in terms of timing residual power spectrum, defined in Eq.~\eqref{eq:S_R_IJ_S_h}, it gives
\begin{equation}
\label{eq:ASMBH_S}
   S_{R,IJ}(f) = \Gamma_{IJ} \frac{A_{\rm SMBH}}{12\pi^2}f_{\rm yr}^{-3} \left( \frac{f}{f_{\rm yr}} \right)^{-\gamma_{\rm SMBH}}, \qquad \gamma_{\rm SMBH}=3 - 2 \alpha = 13/3.
\end{equation}

\begin{figure*}[t!]
\centering
\begin{adjustbox}{max width=1\linewidth,center}
\raisebox{0cm}{\makebox{\includegraphics[ width=1\textwidth, scale=1]{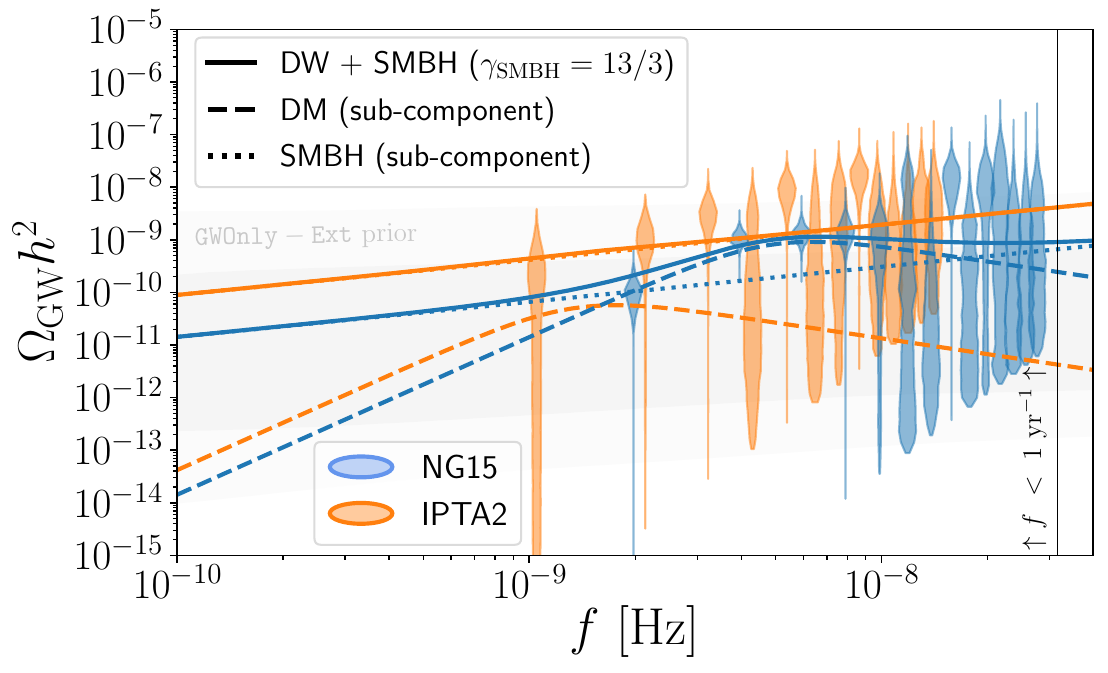}}}
\raisebox{0cm}{\makebox{\includegraphics[ width=1\textwidth, scale=1]{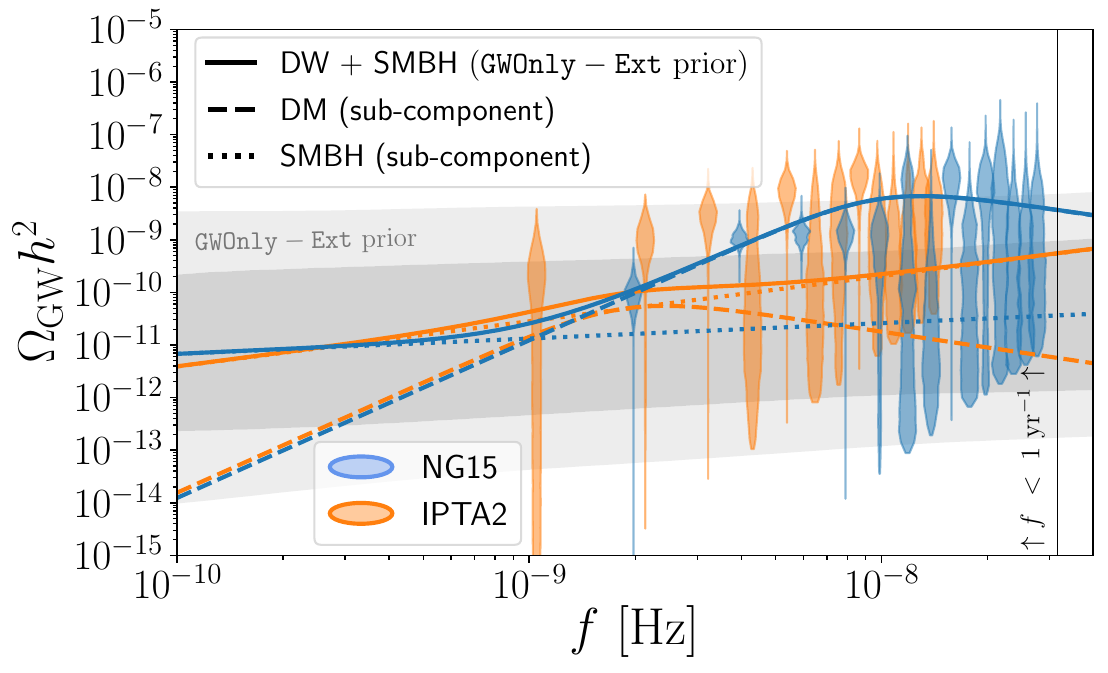}}}
\end{adjustbox}
\caption{\small \label{fig:DW_SMBH_spectra} We show the favored SGWBs from DWs (\textbf{dashed}) and SMBH binaries (\textbf{dotted}), assuming that the PTA signal is a combination of both. Two different models for the SGWB from SMBH binaries are considered. On the \textbf{left}, we use an analytical approach for GW-driven circular binaries, which predicts a power-law SGWB with a spectral tilt of $n_t = 2/3$, equivalent to $\gamma_{\rm SMBH} = 13/3$ in the timing residual power spectrum as indicated by Eq.~\eqref{eq:ASMBH_S}. On the \textbf{right}, we consider the SGWB produced by the population of GW-driven and circular binaries sampled from astrophysical priors, termed as ${\tt GWOnly-Ext}$ library in \cite{NANOGrav:2023hvm}, and accessible in ${\tt PTArcade}$. The $68\%$ and $90\%$ posterior distributions for these SGWBs are depicted using \textbf{gray} bands. We refer to App.~\ref{app:DW_SMBH_interp} for the details.}
\end{figure*}

\begin{figure*}[t!]
\centering
\begin{adjustbox}{max width=1\linewidth,center}
\raisebox{0cm}{\makebox{\includegraphics[ width=0.9\textwidth, scale=1]{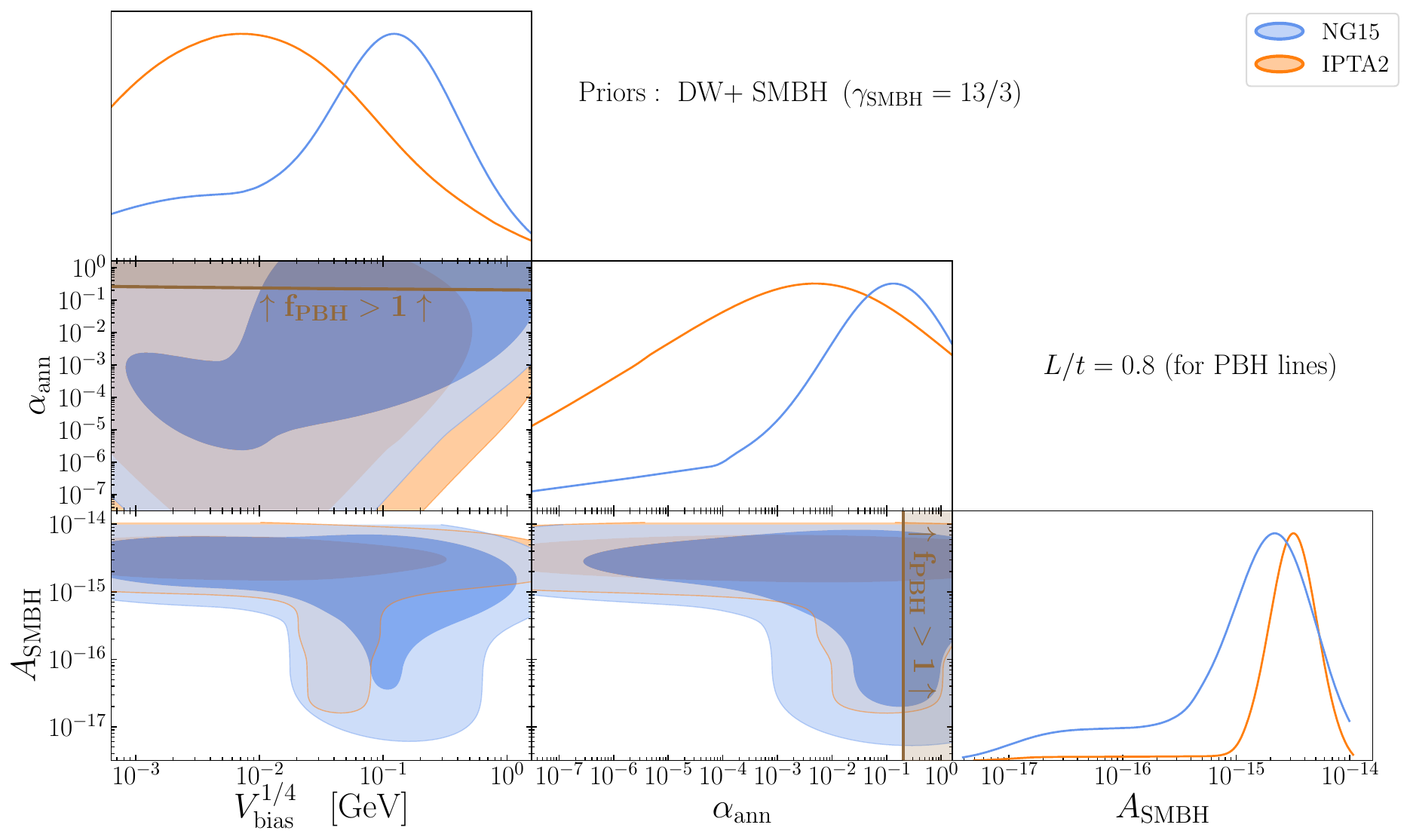}}}
\end{adjustbox}
\begin{adjustbox}{max width=1\linewidth,center}
\raisebox{0cm}{\makebox{\includegraphics[ width=0.9\textwidth, scale=1]{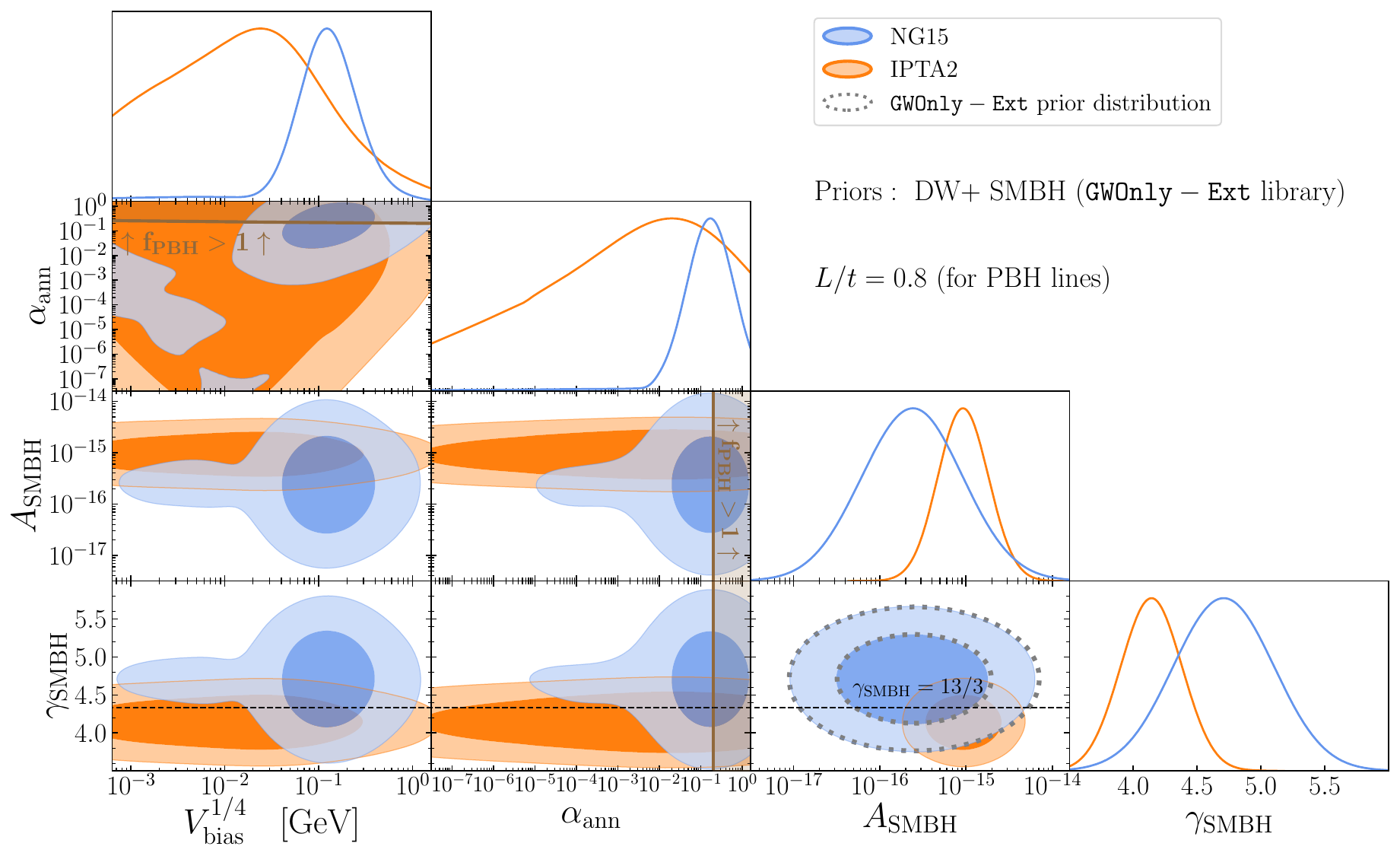}}}
\end{adjustbox}
\caption{\small \label{fig:DW_SMBH_posterior} We show the posterior distribution for the combined DW+SMBH interpretation of the PTA signal. We consider two types of power laws for the SGWB from SMBH binaries. On \textbf{top}, the SGWB is inferred from the purely analytical description of circular and GW-driven binaries, which fixes the spectral tilt to $\gamma_{\rm SMBH}=13/3$ and leaves the amplitude $A_{\rm SMBH}$ as a free parameter. On \textbf{bottom}, the spectral tilt and amplitude $( \gamma_{\rm SMBH},A_{\rm SMBH})$ are drawn from astrophysical priors, still assuming circular and GW-driven binaries, as in \cite{NANOGrav:2023hfp}. The presence of DW networks is favored by the NG15 data with the posteriors overlapping substantially with the region overproducing PBHs shown in \textbf{brown}. In contrast the IPTA2 posterior is rather flat in DW parameters, showing no preference for such interpretation. We refer to Fig.~\ref{fig:BayesFactor_NG15} in main text for a visualisation of the Bayes factors and to App.~\ref{app:DW_SMBH_interp} for more details on the astrophysical priors. }
\end{figure*}
\underline{Discretization} --- The data collected by PTAs is segmented into intervals of length $\Delta f = 1/T$, with $T$ the time period of observation. The number of binary systems emitting at each logarithmic frequency bin, which is given by ${dt_r/d\ln {f_r}}$ in Eq.~\eqref{eq:residence_time}, decreases with frequency. When this number approaches one or less, the continuous and uniform spectrum described in Eq.\eqref{eq:rho_gw_lnf} is accounting for fictitious fractional sources and therefore over-estimate the SWGB signal. The continuous distribution of SMBH, $d^2 n/dz d\mathcal{M}$, must to be replaced with an integer count of SMBH binaries, $N_{ij}$, sorted into bins of redshift $z$ and chirp mass $\mathcal{M}$, and the integral in Eq.~\eqref{eq:rho_gw_lnf} must be replaced by a sum \cite{Sesana:2008mz,McWilliams:2012an,Rosado:2015epa,Kelley:2017lek,Middleton:2020asl}:
\begin{equation}
\label{eq:discretization}
    \int dz d\mathcal{M} \cdots \frac{d^2 n}{dz d\mathcal{M}} \quad\rightarrow\quad \sum_{ij} \cdots N_{ij}.
\end{equation}

\subsection{Priors}

In this work we considered two types of modeling for the SGWB from SMBH binaries, labelled ``$\gamma_{\rm SMBH}=13/3$'' and ``${\tt GWOnly-Ext}$'', respectively.

\underline{Free amplitude} ---
 At first we considered the simple analytical power law in Eq.~\eqref{eq:ASMBH_S} with the amplitude considered as a free parameter:
\begin{equation}
\label{eq:SMBH_prior1}
  \gamma_{\rm SMBH}=13/3\quad \textrm{and}\quad  \log_{\rm 10}(A_{\rm SMBH})\in[-18,~-14]. 
\end{equation}

\underline{Astrophysical prior} ---
Then, we consider the same modeling as in \cite{NANOGrav:2023hvm} where $A_{\rm SMBH}$ and $\gamma_{\rm SMBH}$ are drawn from a Gaussian distribution with mean and covariance matrix:
\begin{equation}
\label{eq:SMBH_prior2}
{\tt GWOnly-Ext:}\qquad \boldsymbol{\mu}_{\rm SMBH} = \begin{pmatrix} -15.6 \\ 4.7 \end{pmatrix} \,, \qquad \boldsymbol{\sigma}_{\rm SMBH} = 10^{-1}\times\begin{pmatrix} 2.8 & -0.026 \\ -0.026 & 1.2 \end{pmatrix} \,.
\end{equation}
Those values are fitted on the SGWBs calculated from SMBH populations taken from the \texttt{GWOnly-Ext} library \cite{NANOGrav:2023hfp,NANOGrav:2023hvm}. This library only contains circular and GW-driven binaries and use a range of model parameters -- like the galaxy stellar mass functions, pair fractions, merger rates, and SMBH-mass versus galaxy-mass relations -- that are narrowly defined \cite{NANOGrav:2023hvm}. The \texttt{GWOnly-Ext} library has been generated using the ${\tt Holodeck}$ program which is still under development \cite{Holodeck}. The astrophysical priors on $(\gamma_{\rm SMBH}, A_{\rm SMBH})$ in Eq.~\eqref{eq:SMBH_prior2} are accessible in ${\tt PTArcade}$ \cite{Mitridate:2023oar}. The reason why Eq.~\eqref{eq:SMBH_prior2} deviates from the $\gamma_{\rm SMBH}=13/3$ while only considering circular GW-driven binaries can be attributed to the summation in Eq.~\eqref{eq:discretization} which reduces the power at higher frequencies \cite{Rosado:2015epa}.

\underline{Results} ---
We conduct a Bayesian analysis to interpret the PTA signal as a combination of SGWBs from DW networks and from SMBH binaries, assuming the two different SMBH priors just presented in Eqs.~\eqref{eq:SMBH_prior1} and \eqref{eq:SMBH_prior2}. The preferred SGWBs are depicted in Fig.~\ref{fig:DW_SMBH_spectra}, while the posterior distributions are illustrated in Fig.~\ref{fig:DW_SMBH_posterior}.

\underline{Environmental effects} ---
In our study, we have focused exclusively on circular GW-driven binaries. The influence of factors such as binary orbital eccentricity and environmental effects, including binary interaction with nearby stars, accretion disk or third black holes, is generally thought to diminish the power at low frequencies \cite{Sesana:2013wja,Kelley:2016gse,Burke-Spolaor:2018bvk,Taylor:2021yjx}. 
Consequently, this could cause an increase in $n_t$ in Eq.~\eqref{eq:ASMBH_Omega_GW} (leading to a redder spectral tilt) and a decrease in $\gamma_{\rm SMBH}$, potentially resembling the contribution from a DW network \cite{NANOGrav:2023hvm}, see Fig.~\ref{fig:DW_GWspectrum_NANO_IPTA}-left in the main text. Both the DW hypothesis and the SMBH model, accounting for a significant amount of eccentricity and environmental effects, receive comparable support from the PTA data \cite{Antoniadis:2023zhi,NANOGrav:2023hfp,NANOGrav:2023hfp,Ellis:2023dgf}. In the future, extending the observation time and increasing the number of detected pulsars could enable the resolution of individual binary sources \cite{Taylor:2020zpk,NANOGrav:2023tcn}. This enhancement would refine the astrophysical priors in Eq.~\eqref{eq:SMBH_prior2} and yield greater precision regarding potential contributions from SGWB originating from primordial sources.

\begin{table}[h!t]
  \begin{center}
    \begin{tblr}{|Q[c,3.1cm]|Q[c,1.7cm]|Q[c,3.8cm]|Q[c,2.2cm]|Q[c,2.0cm]|Q[c,2cm]|}
      \hline
      \SetCell[r=2]{c}{{{\textbf{Model X} }}}
    & \SetCell[r=2]{c}{{{\textbf{Model Y} }}}& \SetCell[r=2,c=2]{c}{{{\textbf{Prior} }}}&& 
\SetCell[c=2]{c}{{{$\log_{10}{\mathcal{B}_{Y,X}}$}}} \\
 & &&&
 \hline
 $\textrm{NG15}$ &
 $\textrm{IPTA2}$\\
        \hline
        \SetCell[r=20]{c=1}{{{SMBH ($\gamma_{\rm SMBH}=13/3$)}}}& \SetCell[r=10]{c=1}{{{DW}}}& \SetCell[c=2]{c=1}{{{dark reheating ($\Delta N_{\rm eff}<0.4$)}}}&& $-2.62 \pm 0.22$ & $-2.71 \pm 0.20$ \\
        \hline
           &&  \SetCell[c=2]{c=1}{{{SM reheating}}}
        & & $0.09 \pm 0.03$ & $-0.98 \pm 0.04$ \\
        \hline
        & &\SetCell[r=8]{c=1}{{{\parbox{3.8cm}{PBH overproduction \\($f_{\rm PBH}<1$)}}}} & \SetCell[r=1]{c}{{{$L/t\in [0.0,1.5]$}}}& $-0.29 \pm 0.03$ & $-1.22 \pm 0.05$ \\
                \hline
                &&& \SetCell[r=1]{c}{{{$L/t\in [0.2,1.5]$}}}& $ -0.33 \pm 0.03$ & $-1.27 \pm 0.05$  \\
                \hline
            &&& \SetCell[r=1]{c}{{{$L/t\in [0.4,1.5]$}}}& $ -0.38 \pm 0.04$ & $-1.34 \pm 0.05$  \\
                \hline
            &&& \SetCell[r=1]{c}{{{$L/t\in [0.6,1.5]$}}}& $-0.47 \pm 0.04$ & $-1.31 \pm 0.06$\\ \hline
            &&& \SetCell[r=1]{c}{{{$L/t\in [0.8,1.5]$}}}& $-0.65 \pm 0.05$ & $ -1.34 \pm 0.07$     \\
             \hline
                &&& \SetCell[r=1]{c}{{{$L/t\in [1.0,1.5]$}}}& $-1.06 \pm 0.09$ & $-1.28 \pm 0.07$\\ \hline
             &&& \SetCell[r=1]{c}{{{$L/t\in [1.2,1.5]$}}}& $-1.65 \pm 0.20$ & $-1.36 \pm 0.10$ \\ \hline
                    &&& \SetCell[r=1]{c}{{{$L/t\in [1.4,1.5]$}}}& $-2.87 \pm 0.59$ & $-1.32 \pm 0.15$ \\ \hline
    \hline        & \SetCell[r=10]{c=1}{{{DW+SMBH}}} &\SetCell[c=2]{c=1}{{{dark reheating ($\Delta N_{\rm eff}<0.4$)}}}&& $-0.01 \pm 0.00$ &  $0.00 \pm 0.01$ \\ \hline
         &  & \SetCell[c=2]{c=1}{{{SM reheating}}}& &  $0.30 \pm 0.01$ & $-0.09 \pm 0.00$ \\   \hline
        & &\SetCell[r=8]{c=1}{{{\parbox{3.8cm}{PBH overproduction \\($f_{\rm PBH}<1$)}}}} & \SetCell[r=1]{c}{{{$L/t\in [0.0,1.5]$}}}& $0.22 \pm 0.00$ &  $-0.04 \pm 0.00$ \\    \hline
                    &&& \SetCell[r=1]{c}{{{$L/t\in [0.2,1.5]$}}}& $0.21 \pm 0.00$ & $ -0.03 \pm 0.00$
                \\  \hline
            &&& \SetCell[r=1]{c}{{{$L/t\in [0.4,1.5]$}}}& $0.18 \pm 0.01$ & $ -0.03 \pm 0.00$
                \\  \hline
            &&& \SetCell[r=1]{c}{{{$L/t\in [0.6,1.5]$}}}& $ 0.13 \pm 0.01$ & $ -0.03 \pm 0.00$
                \\ \hline
            &&& \SetCell[r=1]{c}{{{$L/t\in [0.8,1.5]$}}}& $ 0.08 \pm 0.01$ & $ -0.03 \pm 0.00$
                \\ \hline
                &&& \SetCell[r=1]{c}{{{$L/t\in [1.0,1.5]$}}}& $0.03 \pm 0.01$ & $-0.02 \pm 0.01$\\ \hline
            &&& \SetCell[r=1]{c}{{{$L/t\in [1.2,1.5]$}}}& $-0.02 \pm 0.01$ & $-0.01 \pm 0.01$ \\ \hline
                &&& \SetCell[r=1]{c}{{{$L/t\in [1.4,1.5]$}}}& $-0.07 \pm 0.02$ & $ -0.03 \pm 0.01$ \\ \hline \hline
              \SetCell[r=20]{c=1}{{{SMBH \\
              ($\tt{GWOnly-Ext}\textrm{ prior}$)}}} & \SetCell[r=10]{c=1}{{{DW}}} &\SetCell[c=2]{c=1}{{{dark reheating ($\Delta N_{\rm eff}<0.4$)}}}&&  $-1.69 \pm 0.10$ &  $-2.69 \pm 0.25$ \\ \hline 
     && \SetCell[c=2]{c=1}{{{SM reheating}}}& &  $0.82 \pm 0.05$ &  $-0.94 \pm 0.02$ \\ \hline
        & &\SetCell[r=8]{c=1}{{{\parbox{3.8cm}{PBH overproduction \\($f_{\rm PBH}<1$)}}}} & \SetCell[r=1]{c}{{{$L/t\in [0.0,1.5]$}}}&  $0.34 \pm 0.02$ & $-1.14 \pm 0.02$ \\ \hline
                    &&& \SetCell[r=1]{c}{{{$L/t\in [0.2,1.5]$}}}& $0.30 \pm 0.02$ & $ -1.16 \pm 0.03$  \\   \hline
            &&& \SetCell[r=1]{c}{{{$L/t\in [0.4,1.5]$}}}& $ 0.25 \pm 0.02$ & $-1.17 \pm 0.03$ \\    \hline
            &&& \SetCell[r=1]{c}{{{$L/t\in [0.6,1.5]$}}}& $0.14 \pm 0.02$ & $-1.18 \pm 0.03$ \\    \hline
            &&& \SetCell[r=1]{c}{{{$L/t\in [0.8,1.5]$}}}& $-0.04 \pm 0.02$ & $-1.21 \pm 0.03$ \\   \hline
                &&& \SetCell[r=1]{c}{{{$L/t\in [1.0,1.5]$}}}&  $-0.32 \pm 0.03$ & $-1.29 \pm 0.04$ \\ \hline
            &&& \SetCell[r=1]{c}{{{$L/t\in [1.2,1.5]$}}}& $-0.71 \pm 0.04$ & $-1.47 \pm 0.05$
            \\ \hline
                    &&& \SetCell[r=1]{c}{{{$L/t\in [1.4,1.5]$}}}& $-1.19 \pm 0.09$ & $-1.84 \pm 0.09$ \\ \hline
     \hline
             &  \SetCell[r=10]{c=1}{{{DW+SMBH}}} & \SetCell[c=2]{c=1}{{{dark reheating ($\Delta N_{\rm eff}<0.4$)}}}&&  $-0.08 \pm 0.04$ &  $0.05 \pm 0.03$  \\
        \hline
        &&\SetCell[c=2]{c=1}{{{SM reheating}}}&
        &   $0.98 \pm 0.02$ & $-0.01 \pm 0.01$ \\
        \hline
        & &\SetCell[r=8]{c=1}{{{\parbox{3.8cm}{PBH overproduction \\($f_{\rm PBH}<1$)}}}} & \SetCell[r=1]{c}{{{$L/t\in [0.0,1.5]$}}}& $0.84 \pm 0.05$ & $-0.09 \pm 0.02$ \\\hline
            &&& \SetCell[r=1]{c}{{{$L/t\in [0.2,1.5]$}}}& $0.83 \pm 0.05$ & $-0.07 \pm 0.02$ \\
        \hline
            &&& \SetCell[r=1]{c}{{{$L/t\in [0.4,1.5]$}}}& $0.79 \pm 0.05$ & $-0.05 \pm 0.02$ \\
                \hline
            &&& \SetCell[r=1]{c}{{{$L/t\in [0.6,1.5]$}}}& $0.73 \pm 0.06$ & $-0.05 \pm 0.02$ \\
                \hline
            &&& \SetCell[r=1]{c}{{{$L/t\in [0.8,1.5]$}}}& $0.68 \pm 0.06$ & $-0.04 \pm 0.02$ \\
                \hline
            &&& \SetCell[r=1]{c}{{{$L/t\in [1.0,1.5]$}}}& $0.58 \pm 0.07$ & $-0.07 \pm 0.03$ \\ \hline
        &&& \SetCell[r=1]{c}{{{$L/t\in [1.2,1.5]$}}}& $0.59 \pm 0.08$ & $-0.11 \pm 0.03$ \\ \hline
        &&& \SetCell[r=1]{c}{{{$L/t\in [1.4,1.5]$}}}& $0.47 \pm 0.14$ & $-0.08 \pm 0.05$ \\ \hline
    \end{tblr}
    \caption{\label{tab:BF} This table summarizes the Bayesian analysis conducted in this work. The Bayesian factor $\mathcal{B}_{Y,X}$ indicates whether data favors or disfavors interpretation $Y$ with respect to $X$.  According to Jeffrey's criterion \cite{1939thpr.book.....J,kass1995bayes}, evidence is considered barely worth mentioning if \( |{\log}_{10}{\mathcal{B}_{Y,X}}| < 0.5 \), substantial if it is greater than 0.5, strong if it exceeds 1, very strong if it surpasses 1.5, and decisive if it is above 2. }
  \end{center}
\end{table}

\begin{table}[h!t]
  \begin{center}
    \begin{tblr}{|Q[c,1.8cm]|Q[c,2.8cm]|Q[l,2.5cm]|Q[c,1.9cm]|Q[c,2.3cm]|Q[c,2.3cm]|}
      \hline
      \SetCell[r=2]{c}{{{$\textbf{Model}$ }}}&\SetCell[r=2,c=2]{c}{{{$\textbf{Prior}$ }}}& &\SetCell[r=2]{c}{{{\parbox{1.9cm}{\textbf{Parameters}\\
     $(\log_{10}{X})$}}}}
    & 
\SetCell[c=2]{c}{{{$\textbf{Posterior mean}$}}} \\
 & 
 \hline
 &&& 
 $\textrm{NG15}$ &
 $\textrm{IPTA2}$\\
      \hline \hline 
          \SetCell[r=8]{c}{{{DW}}}& \SetCell[r=4,c=2]{c}{{{SM reheating}}}& & $\alpha_{\rm ann} $ & $ -0.81\pm 0.18$ & $-1.02^{+0.13}_{-0.12}$   \\\hline 
                  && & $V_{\rm bias}^{1/4}$& $ -0.94^{+0.19}_{-0.17}$ &$-1.42\pm 0.14 $   \\\hline 
      &  &  &$\sigma^{1/3}$ & $5.30\pm 0.16$ & $4.94\pm 0.11$   \\\hline 
         &  &  &      $T_{\rm ann}$ & $-1.04^{+0.15}_{-0.11}$ &$-1.45\pm 0.11 $ \\
      \hline \hline 
      
          &\SetCell[r=4]{c}{{{\textrm{PBH overproduction} \\ $(f_{\rm PBH}<1)$}}} & \SetCell[r=2]{c}{{{$L/t\in [0.0,1.5]$}}}& $\alpha_{\rm ann} $ & $-0.84^{+0.13}_{-0.16}$ & $-0.84^{+0.13}_{-0.16}$   \\\hline 
              &    & & $V_{\rm bias}^{1/4}$& $-0.96^{+0.15}_{-0.13}$ &$-1.49^{+0.19}_{-0.16}$  \\ \hline \hline
         & &\SetCell[r=2]{c}{{{$L/t\in [1.0,1.5]$}}}& $\alpha_{\rm ann} $ & $ -0.97^{+0.10}_{-0.05}$ &   $-1.10\pm 0.09$ \\\hline 
          &    & & $V_{\rm bias}^{1/4}$& $-1.07^{+0.14}_{-0.01}$ &$-1.46\pm 0.18 $  \\ 
      \hline \hline 
          \SetCell[r=8]{c}{{{DW+ SMBH}}}& \SetCell[r=4,c=2]{c}{{{\parbox{5cm}{\textrm{DW: SM reheating} \\ \vspace{0.25cm} $\textrm{SMBH: }\gamma_{\rm SMBH}=13/3$}}}}&& $\alpha_{\rm ann} $ & $-1.4^{+2.4}_{-1.8} $ & $ -2.0^{+3.2}_{-4.5}$   \\\hline 
                      & & & $V_{\rm bias}^{1/4}$& $ -1.40^{+1.3}_{-0.50}  $ &$ -2.12^{+0.87}_{-0.97}$   \\\hline 
                 & & & $A_{\rm SMBH}$& $-14.99^{+0.79}_{-0.21}$ &$-14.59^{+0.32}_{-0.13}$  
  \\\hline 
                & &  & $\gamma_{\rm SMBH} $& $ 13/3 $ &$13/3$   \\
         \hline \hline 
          & \SetCell[r=4,c=2]{c}{{{\parbox{5cm}{\textrm{DW: SM reheating}\\ \vspace{0.25cm}
          $\textrm{SMBH:} \tt{GWOnly-Ext}\textrm{ prior distribution}$}}}}&& $\alpha_{\rm ann} $ & $ -0.867^{+0.27}_{-0.090}$ & $ -2.0^{+3.1}_{-3.7}$   \\\hline 
                 & & & $V_{\rm bias}^{1/4}$& $-0.99^{+0.27}_{-0.13}$ &$-1.98^{+1.1}_{-0.93}$   \\\hline 
               &   & & $A_{\rm SMBH}$& $-15.62\pm 0.53 $ &$-15.03\pm 0.26$   \\\hline 
              &  &   & $\gamma_{\rm SMBH} $& $4.71\pm 0.35$ &$ 4.14\pm 0.22$   \\
         \hline \hline 
           \SetCell[r=3]{c}{{{SMBH}}}& \SetCell[r=1,c=2]{c}{{{$\gamma_{\rm SMBH} =13/3$}}}& &$A_{\rm SMBH} $ & $-14.569^{+0.068}_{-0.055}$ & $ -14.488^{+0.079}_{-0.058}$    \\
         \hline \hline 
           & \SetCell[r=2,c=2]{c}{{{\tt{GWOnly-Ext}\textrm{ prior distribution 
           }}}} && $A_{\rm SMBH}$&  $-14.47^{+0.14}_{-0.12}$ & $-14.50^{+0.12}_{-0.10}$    \\
      \hline
      & && $\gamma_{\rm SMBH}$&  $4.08\pm 0.25$ & $4.23\pm 0.19$    \\
      \hline
    \end{tblr}
    \caption{\label{tab:meanPoseterior} This table displays the mean values of parameters and their $68\%$ confidence intervals. It covers various scenarios: the DW interpretation, the SMBH interpretation, and their combination. Scenario are evaluated with and without PBHs constraints and includes two models of the SGWB from SMBH binaries.}
  \end{center}
\end{table}

 \FloatBarrier

 \bibliography{biblio}

\begin{thebibliography}{154}%
\makeatletter
\providecommand \@ifxundefined [1]{%
 \@ifx{#1\undefined}
}%
\providecommand \@ifnum [1]{%
 \ifnum #1\expandafter \@firstoftwo
 \else \expandafter \@secondoftwo
 \fi
}%
\providecommand \@ifx [1]{%
 \ifx #1\expandafter \@firstoftwo
 \else \expandafter \@secondoftwo
 \fi
}%
\providecommand \natexlab [1]{#1}%
\providecommand \enquote  [1]{``#1''}%
\providecommand \bibnamefont  [1]{#1}%
\providecommand \bibfnamefont [1]{#1}%
\providecommand \citenamefont [1]{#1}%
\providecommand \href@noop [0]{\@secondoftwo}%
\providecommand \href [0]{\begingroup \@sanitize@url \@href}%
\providecommand \@href[1]{\@@startlink{#1}\@@href}%
\providecommand \@@href[1]{\endgroup#1\@@endlink}%
\providecommand \@sanitize@url [0]{\catcode `\\12\catcode `\$12\catcode
  `\&12\catcode `\#12\catcode `\^12\catcode `\_12\catcode `\%12\relax}%
\providecommand \@@startlink[1]{}%
\providecommand \@@endlink[0]{}%
\providecommand \url  [0]{\begingroup\@sanitize@url \@url }%
\providecommand \@url [1]{\endgroup\@href {#1}{\urlprefix }}%
\providecommand \urlprefix  [0]{URL }%
\providecommand \Eprint [0]{\href }%
\providecommand \doibase [0]{https://doi.org/}%
\providecommand \selectlanguage [0]{\@gobble}%
\providecommand \bibinfo  [0]{\@secondoftwo}%
\providecommand \bibfield  [0]{\@secondoftwo}%
\providecommand \translation [1]{[#1]}%
\providecommand \BibitemOpen [0]{}%
\providecommand \bibitemStop [0]{}%
\providecommand \bibitemNoStop [0]{.\EOS\space}%
\providecommand \EOS [0]{\spacefactor3000\relax}%
\providecommand \BibitemShut  [1]{\csname bibitem#1\endcsname}%
\let\auto@bib@innerbib\@empty
\bibitem [{\citenamefont {Agazie}\ \emph
  {et~al.}(2023{\natexlab{a}})\citenamefont {Agazie} \emph
  {et~al.}}]{NANOGrav:2023gor}%
  \BibitemOpen
  \bibfield  {author} {\bibinfo {author} {\bibfnamefont {G.}~\bibnamefont
  {Agazie}} \emph {et~al.} (\bibinfo {collaboration} {NANOGrav}),\ }\bibfield
  {title} {\bibinfo {title} {{The NANOGrav 15-year Data Set: Evidence for a
  Gravitational-Wave Background}},\ }\bibfield  {journal} {\bibinfo  {journal}
  {Astrophys. J. Lett.}\ }\textbf {\bibinfo {volume} {951}},\ \href
  {https://doi.org/10.3847/2041-8213/acdac6} {10.3847/2041-8213/acdac6}
  (\bibinfo {year} {2023}{\natexlab{a}}),\ \Eprint
  {https://arxiv.org/abs/2306.16213} {arXiv:2306.16213 [astro-ph.HE]}
  \BibitemShut {NoStop}%
\bibitem [{\citenamefont {Antoniadis}\ \emph
  {et~al.}(2023{\natexlab{a}})\citenamefont {Antoniadis} \emph
  {et~al.}}]{Antoniadis:2023rey}%
  \BibitemOpen
  \bibfield  {author} {\bibinfo {author} {\bibfnamefont {J.}~\bibnamefont
  {Antoniadis}} \emph {et~al.},\ }\bibfield  {title} {\bibinfo {title} {{The
  second data release from the European Pulsar Timing Array III. Search for
  gravitational wave signals}},\ }\href@noop {} {\  (\bibinfo {year}
  {2023}{\natexlab{a}})},\ \Eprint {https://arxiv.org/abs/2306.16214}
  {arXiv:2306.16214 [astro-ph.HE]} \BibitemShut {NoStop}%
\bibitem [{\citenamefont {Reardon}\ \emph {et~al.}(2023)\citenamefont {Reardon}
  \emph {et~al.}}]{Reardon:2023gzh}%
  \BibitemOpen
  \bibfield  {author} {\bibinfo {author} {\bibfnamefont {D.~J.}\ \bibnamefont
  {Reardon}} \emph {et~al.},\ }\bibfield  {title} {\bibinfo {title} {{Search
  for an Isotropic Gravitational-wave Background with the Parkes Pulsar Timing
  Array}},\ }\href {https://doi.org/10.3847/2041-8213/acdd02} {\bibfield
  {journal} {\bibinfo  {journal} {Astrophys. J. Lett.}\ }\textbf {\bibinfo
  {volume} {951}},\ \bibinfo {pages} {L6} (\bibinfo {year} {2023})},\ \Eprint
  {https://arxiv.org/abs/2306.16215} {arXiv:2306.16215 [astro-ph.HE]}
  \BibitemShut {NoStop}%
\bibitem [{\citenamefont {Xu}\ \emph {et~al.}(2023)\citenamefont {Xu} \emph
  {et~al.}}]{Xu:2023wog}%
  \BibitemOpen
  \bibfield  {author} {\bibinfo {author} {\bibfnamefont {H.}~\bibnamefont {Xu}}
  \emph {et~al.},\ }\bibfield  {title} {\bibinfo {title} {{Searching for the
  Nano-Hertz Stochastic Gravitational Wave Background with the Chinese Pulsar
  Timing Array Data Release I}},\ }\href
  {https://doi.org/10.1088/1674-4527/acdfa5} {\bibfield  {journal} {\bibinfo
  {journal} {Res. Astron. Astrophys.}\ }\textbf {\bibinfo {volume} {23}},\
  \bibinfo {pages} {075024} (\bibinfo {year} {2023})},\ \Eprint
  {https://arxiv.org/abs/2306.16216} {arXiv:2306.16216 [astro-ph.HE]}
  \BibitemShut {NoStop}%
\bibitem [{\citenamefont {Agazie}\ \emph
  {et~al.}(2023{\natexlab{b}})\citenamefont {Agazie} \emph
  {et~al.}}]{InternationalPulsarTimingArray:2023mzf}%
  \BibitemOpen
  \bibfield  {author} {\bibinfo {author} {\bibfnamefont {G.}~\bibnamefont
  {Agazie}} \emph {et~al.} (\bibinfo {collaboration} {International Pulsar
  Timing Array}),\ }\bibfield  {title} {\bibinfo {title} {{Comparing recent PTA
  results on the nanohertz stochastic gravitational wave background}},\
  }\href@noop {} {\  (\bibinfo {year} {2023}{\natexlab{b}})},\ \Eprint
  {https://arxiv.org/abs/2309.00693} {arXiv:2309.00693 [astro-ph.HE]}
  \BibitemShut {NoStop}%
\bibitem [{\citenamefont {Pol}\ \emph {et~al.}(2021)\citenamefont {Pol} \emph
  {et~al.}}]{NANOGrav:2020spf}%
  \BibitemOpen
  \bibfield  {author} {\bibinfo {author} {\bibfnamefont {N.~S.}\ \bibnamefont
  {Pol}} \emph {et~al.} (\bibinfo {collaboration} {NANOGrav}),\ }\bibfield
  {title} {\bibinfo {title} {{Astrophysics Milestones for Pulsar Timing Array
  Gravitational-wave Detection}},\ }\href
  {https://doi.org/10.3847/2041-8213/abf2c9} {\bibfield  {journal} {\bibinfo
  {journal} {Astrophys. J. Lett.}\ }\textbf {\bibinfo {volume} {911}},\
  \bibinfo {pages} {L34} (\bibinfo {year} {2021})},\ \Eprint
  {https://arxiv.org/abs/2010.11950} {arXiv:2010.11950 [astro-ph.HE]}
  \BibitemShut {NoStop}%
\bibitem [{\citenamefont {Chen}\ \emph {et~al.}(2021)\citenamefont {Chen} \emph
  {et~al.}}]{Chen:2021rqp}%
  \BibitemOpen
  \bibfield  {author} {\bibinfo {author} {\bibfnamefont {S.}~\bibnamefont
  {Chen}} \emph {et~al.},\ }\bibfield  {title} {\bibinfo {title}
  {{Common-red-signal analysis with 24-yr high-precision timing of the European
  Pulsar Timing Array: Inferences in the stochastic gravitational-wave
  background search}}\ }\href {https://doi.org/10.1093/mnras/stab2833}
  {10.1093/mnras/stab2833} (\bibinfo {year} {2021}),\ \Eprint
  {https://arxiv.org/abs/2110.13184} {arXiv:2110.13184 [astro-ph.HE]}
  \BibitemShut {NoStop}%
\bibitem [{\citenamefont {Goncharov}\ \emph {et~al.}(2021)\citenamefont
  {Goncharov} \emph {et~al.}}]{Goncharov:2021oub}%
  \BibitemOpen
  \bibfield  {author} {\bibinfo {author} {\bibfnamefont {B.}~\bibnamefont
  {Goncharov}} \emph {et~al.},\ }\bibfield  {title} {\bibinfo {title} {{On the
  evidence for a common-spectrum process in the search for the nanohertz
  gravitational-wave background with the Parkes Pulsar Timing Array}}\ }\href
  {https://doi.org/10.3847/2041-8213/ac17f4} {10.3847/2041-8213/ac17f4}
  (\bibinfo {year} {2021}),\ \Eprint {https://arxiv.org/abs/2107.12112}
  {arXiv:2107.12112 [astro-ph.HE]} \BibitemShut {NoStop}%
\bibitem [{\citenamefont {Antoniadis}\ \emph {et~al.}(2022)\citenamefont
  {Antoniadis} \emph {et~al.}}]{Antoniadis:2022pcn}%
  \BibitemOpen
  \bibfield  {author} {\bibinfo {author} {\bibfnamefont {J.}~\bibnamefont
  {Antoniadis}} \emph {et~al.},\ }\bibfield  {title} {\bibinfo {title} {{The
  International Pulsar Timing Array second data release: Search for an
  isotropic gravitational wave background}},\ }\href
  {https://doi.org/10.1093/mnras/stab3418} {\bibfield  {journal} {\bibinfo
  {journal} {Mon. Not. Roy. Astron. Soc.}\ }\textbf {\bibinfo {volume} {510}},\
  \bibinfo {pages} {4873} (\bibinfo {year} {2022})},\ \Eprint
  {https://arxiv.org/abs/2201.03980} {arXiv:2201.03980 [astro-ph.HE]}
  \BibitemShut {NoStop}%
\bibitem [{\citenamefont {Sesana}\ \emph {et~al.}(2004)\citenamefont {Sesana},
  \citenamefont {Haardt}, \citenamefont {Madau},\ and\ \citenamefont
  {Volonteri}}]{Sesana:2004sp}%
  \BibitemOpen
  \bibfield  {author} {\bibinfo {author} {\bibfnamefont {A.}~\bibnamefont
  {Sesana}}, \bibinfo {author} {\bibfnamefont {F.}~\bibnamefont {Haardt}},
  \bibinfo {author} {\bibfnamefont {P.}~\bibnamefont {Madau}},\ and\ \bibinfo
  {author} {\bibfnamefont {M.}~\bibnamefont {Volonteri}},\ }\bibfield  {title}
  {\bibinfo {title} {{Low - frequency gravitational radiation from coalescing
  massive black hole binaries in hierarchical cosmologies}},\ }\href
  {https://doi.org/10.1086/422185} {\bibfield  {journal} {\bibinfo  {journal}
  {Astrophys. J.}\ }\textbf {\bibinfo {volume} {611}},\ \bibinfo {pages} {623}
  (\bibinfo {year} {2004})},\ \Eprint {https://arxiv.org/abs/astro-ph/0401543}
  {arXiv:astro-ph/0401543} \BibitemShut {NoStop}%
\bibitem [{\citenamefont {Burke-Spolaor}\ \emph {et~al.}(2019)\citenamefont
  {Burke-Spolaor} \emph {et~al.}}]{Burke-Spolaor:2018bvk}%
  \BibitemOpen
  \bibfield  {author} {\bibinfo {author} {\bibfnamefont {S.}~\bibnamefont
  {Burke-Spolaor}} \emph {et~al.},\ }\bibfield  {title} {\bibinfo {title} {{The
  Astrophysics of Nanohertz Gravitational Waves}},\ }\href
  {https://doi.org/10.1007/s00159-019-0115-7} {\bibfield  {journal} {\bibinfo
  {journal} {Astron. Astrophys. Rev.}\ }\textbf {\bibinfo {volume} {27}},\
  \bibinfo {pages} {5} (\bibinfo {year} {2019})},\ \Eprint
  {https://arxiv.org/abs/1811.08826} {arXiv:1811.08826 [astro-ph.HE]}
  \BibitemShut {NoStop}%
\bibitem [{\citenamefont {Antoniadis}\ \emph
  {et~al.}(2023{\natexlab{b}})\citenamefont {Antoniadis} \emph
  {et~al.}}]{Antoniadis:2023zhi}%
  \BibitemOpen
  \bibfield  {author} {\bibinfo {author} {\bibfnamefont {J.}~\bibnamefont
  {Antoniadis}} \emph {et~al.},\ }\bibfield  {title} {\bibinfo {title} {{The
  second data release from the European Pulsar Timing Array: V. Implications
  for massive black holes, dark matter and the early Universe}},\ }\href@noop
  {} {\  (\bibinfo {year} {2023}{\natexlab{b}})},\ \Eprint
  {https://arxiv.org/abs/2306.16227} {arXiv:2306.16227 [astro-ph.CO]}
  \BibitemShut {NoStop}%
\bibitem [{\citenamefont {Agazie}\ \emph
  {et~al.}(2023{\natexlab{c}})\citenamefont {Agazie} \emph
  {et~al.}}]{NANOGrav:2023hfp}%
  \BibitemOpen
  \bibfield  {author} {\bibinfo {author} {\bibfnamefont {G.}~\bibnamefont
  {Agazie}} \emph {et~al.} (\bibinfo {collaboration} {NANOGrav}),\ }\bibfield
  {title} {\bibinfo {title} {{The NANOGrav 15-year Data Set: Constraints on
  Supermassive Black Hole Binaries from the Gravitational Wave Background}},\
  }\href@noop {} {\  (\bibinfo {year} {2023}{\natexlab{c}})},\ \Eprint
  {https://arxiv.org/abs/2306.16220} {arXiv:2306.16220 [astro-ph.HE]}
  \BibitemShut {NoStop}%
\bibitem [{\citenamefont {Ellis}\ \emph {et~al.}(2023)\citenamefont {Ellis},
  \citenamefont {Fairbairn}, \citenamefont {H\"utsi}, \citenamefont {Raidal},
  \citenamefont {Urrutia}, \citenamefont {Vaskonen},\ and\ \citenamefont
  {Veerm\"ae}}]{Ellis:2023dgf}%
  \BibitemOpen
  \bibfield  {author} {\bibinfo {author} {\bibfnamefont {J.}~\bibnamefont
  {Ellis}}, \bibinfo {author} {\bibfnamefont {M.}~\bibnamefont {Fairbairn}},
  \bibinfo {author} {\bibfnamefont {G.}~\bibnamefont {H\"utsi}}, \bibinfo
  {author} {\bibfnamefont {J.}~\bibnamefont {Raidal}}, \bibinfo {author}
  {\bibfnamefont {J.}~\bibnamefont {Urrutia}}, \bibinfo {author} {\bibfnamefont
  {V.}~\bibnamefont {Vaskonen}},\ and\ \bibinfo {author} {\bibfnamefont
  {H.}~\bibnamefont {Veerm\"ae}},\ }\bibfield  {title} {\bibinfo {title}
  {{Gravitational Waves from SMBH Binaries in Light of the NANOGrav 15-Year
  Data}},\ }\href@noop {} {\  (\bibinfo {year} {2023})},\ \Eprint
  {https://arxiv.org/abs/2306.17021} {arXiv:2306.17021 [astro-ph.CO]}
  \BibitemShut {NoStop}%
\bibitem [{\citenamefont {Afzal}\ \emph {et~al.}(2023)\citenamefont {Afzal}
  \emph {et~al.}}]{NANOGrav:2023hvm}%
  \BibitemOpen
  \bibfield  {author} {\bibinfo {author} {\bibfnamefont {A.}~\bibnamefont
  {Afzal}} \emph {et~al.} (\bibinfo {collaboration} {NANOGrav}),\ }\bibfield
  {title} {\bibinfo {title} {{The NANOGrav 15-year Data Set: Search for Signals
  from New Physics}},\ }\bibfield  {journal} {\bibinfo  {journal} {Astrophys.
  J. Lett.}\ }\textbf {\bibinfo {volume} {951}},\ \href
  {https://doi.org/10.3847/2041-8213/acdc91} {10.3847/2041-8213/acdc91}
  (\bibinfo {year} {2023}),\ \Eprint {https://arxiv.org/abs/2306.16219}
  {arXiv:2306.16219 [astro-ph.HE]} \BibitemShut {NoStop}%
\bibitem [{\citenamefont {Antoniadis}\ \emph
  {et~al.}(2023{\natexlab{c}})\citenamefont {Antoniadis} \emph
  {et~al.}}]{EPTA:2023xxk}%
  \BibitemOpen
  \bibfield  {author} {\bibinfo {author} {\bibfnamefont {J.}~\bibnamefont
  {Antoniadis}} \emph {et~al.} (\bibinfo {collaboration} {EPTA}),\ }\bibfield
  {title} {\bibinfo {title} {{The second data release from the European Pulsar
  Timing Array: V. Implications for massive black holes, dark matter and the
  early Universe}},\ }\href@noop {} {\  (\bibinfo {year}
  {2023}{\natexlab{c}})},\ \Eprint {https://arxiv.org/abs/2306.16227}
  {arXiv:2306.16227 [astro-ph.CO]} \BibitemShut {NoStop}%
\bibitem [{\citenamefont {Caprini}\ \emph {et~al.}(2010)\citenamefont
  {Caprini}, \citenamefont {Durrer},\ and\ \citenamefont
  {Siemens}}]{Caprini:2010xv}%
  \BibitemOpen
  \bibfield  {author} {\bibinfo {author} {\bibfnamefont {C.}~\bibnamefont
  {Caprini}}, \bibinfo {author} {\bibfnamefont {R.}~\bibnamefont {Durrer}},\
  and\ \bibinfo {author} {\bibfnamefont {X.}~\bibnamefont {Siemens}},\
  }\bibfield  {title} {\bibinfo {title} {{Detection of gravitational waves from
  the QCD phase transition with pulsar timing arrays}},\ }\href
  {https://doi.org/10.1103/PhysRevD.82.063511} {\bibfield  {journal} {\bibinfo
  {journal} {Phys. Rev. D}\ }\textbf {\bibinfo {volume} {82}},\ \bibinfo
  {pages} {063511} (\bibinfo {year} {2010})},\ \Eprint
  {https://arxiv.org/abs/1007.1218} {arXiv:1007.1218 [astro-ph.CO]}
  \BibitemShut {NoStop}%
\bibitem [{\citenamefont {Ratzinger}\ and\ \citenamefont
  {Schwaller}(2021)}]{Ratzinger:2020koh}%
  \BibitemOpen
  \bibfield  {author} {\bibinfo {author} {\bibfnamefont {W.}~\bibnamefont
  {Ratzinger}}\ and\ \bibinfo {author} {\bibfnamefont {P.}~\bibnamefont
  {Schwaller}},\ }\bibfield  {title} {\bibinfo {title} {{Whispers from the dark
  side: Confronting light new physics with NANOGrav data}},\ }\href
  {https://doi.org/10.21468/SciPostPhys.10.2.047} {\bibfield  {journal}
  {\bibinfo  {journal} {SciPost Phys.}\ }\textbf {\bibinfo {volume} {10}},\
  \bibinfo {pages} {047} (\bibinfo {year} {2021})},\ \Eprint
  {https://arxiv.org/abs/2009.11875} {arXiv:2009.11875 [astro-ph.CO]}
  \BibitemShut {NoStop}%
\bibitem [{\citenamefont {Arzoumanian}\ \emph {et~al.}(2021)\citenamefont
  {Arzoumanian} \emph {et~al.}}]{NANOGrav:2021flc}%
  \BibitemOpen
  \bibfield  {author} {\bibinfo {author} {\bibfnamefont {Z.}~\bibnamefont
  {Arzoumanian}} \emph {et~al.} (\bibinfo {collaboration} {NANOGrav}),\
  }\bibfield  {title} {\bibinfo {title} {{Searching for Gravitational Waves
  from Cosmological Phase Transitions with the NANOGrav 12.5-Year Dataset}},\
  }\href {https://doi.org/10.1103/PhysRevLett.127.251302} {\bibfield  {journal}
  {\bibinfo  {journal} {Phys. Rev. Lett.}\ }\textbf {\bibinfo {volume} {127}},\
  \bibinfo {pages} {251302} (\bibinfo {year} {2021})},\ \Eprint
  {https://arxiv.org/abs/2104.13930} {arXiv:2104.13930 [astro-ph.CO]}
  \BibitemShut {NoStop}%
\bibitem [{\citenamefont {Bringmann}\ \emph {et~al.}(2023)\citenamefont
  {Bringmann}, \citenamefont {Depta}, \citenamefont {Konstandin}, \citenamefont
  {Schmidt-Hoberg},\ and\ \citenamefont {Tasillo}}]{Bringmann:2023opz}%
  \BibitemOpen
  \bibfield  {author} {\bibinfo {author} {\bibfnamefont {T.}~\bibnamefont
  {Bringmann}}, \bibinfo {author} {\bibfnamefont {P.~F.}\ \bibnamefont
  {Depta}}, \bibinfo {author} {\bibfnamefont {T.}~\bibnamefont {Konstandin}},
  \bibinfo {author} {\bibfnamefont {K.}~\bibnamefont {Schmidt-Hoberg}},\ and\
  \bibinfo {author} {\bibfnamefont {C.}~\bibnamefont {Tasillo}},\ }\bibfield
  {title} {\bibinfo {title} {{Does NANOGrav observe a dark sector phase
  transition?}},\ }\href@noop {} {\  (\bibinfo {year} {2023})},\ \Eprint
  {https://arxiv.org/abs/2306.09411} {arXiv:2306.09411 [astro-ph.CO]}
  \BibitemShut {NoStop}%
\bibitem [{\citenamefont {Gouttenoire}(2023)}]{Gouttenoire:2023bqy}%
  \BibitemOpen
  \bibfield  {author} {\bibinfo {author} {\bibfnamefont {Y.}~\bibnamefont
  {Gouttenoire}},\ }\bibfield  {title} {\bibinfo {title} {{First-Order Phase
  Transition Interpretation of Pulsar Timing Array Signal Is Consistent with
  Solar-Mass Black Holes}},\ }\href
  {https://doi.org/10.1103/PhysRevLett.131.171404} {\bibfield  {journal}
  {\bibinfo  {journal} {Phys. Rev. Lett.}\ }\textbf {\bibinfo {volume} {131}},\
  \bibinfo {pages} {171404} (\bibinfo {year} {2023})},\ \Eprint
  {https://arxiv.org/abs/2307.04239} {arXiv:2307.04239 [hep-ph]} \BibitemShut
  {NoStop}%
\bibitem [{\citenamefont {Wang}\ \emph {et~al.}(2019)\citenamefont {Wang},
  \citenamefont {Terada},\ and\ \citenamefont {Kohri}}]{Wang:2019kaf}%
  \BibitemOpen
  \bibfield  {author} {\bibinfo {author} {\bibfnamefont {S.}~\bibnamefont
  {Wang}}, \bibinfo {author} {\bibfnamefont {T.}~\bibnamefont {Terada}},\ and\
  \bibinfo {author} {\bibfnamefont {K.}~\bibnamefont {Kohri}},\ }\bibfield
  {title} {\bibinfo {title} {{Prospective constraints on the primordial black
  hole abundance from the stochastic gravitational-wave backgrounds produced by
  coalescing events and curvature perturbations}},\ }\href
  {https://doi.org/10.1103/PhysRevD.99.103531} {\bibfield  {journal} {\bibinfo
  {journal} {Phys. Rev. D}\ }\textbf {\bibinfo {volume} {99}},\ \bibinfo
  {pages} {103531} (\bibinfo {year} {2019})},\ \bibinfo {note} {[Erratum:
  Phys.Rev.D 101, 069901 (2020)]},\ \Eprint {https://arxiv.org/abs/1903.05924}
  {arXiv:1903.05924 [astro-ph.CO]} \BibitemShut {NoStop}%
\bibitem [{\citenamefont {Chen}\ \emph {et~al.}(2020)\citenamefont {Chen},
  \citenamefont {Yuan},\ and\ \citenamefont {Huang}}]{Chen:2019xse}%
  \BibitemOpen
  \bibfield  {author} {\bibinfo {author} {\bibfnamefont {Z.-C.}\ \bibnamefont
  {Chen}}, \bibinfo {author} {\bibfnamefont {C.}~\bibnamefont {Yuan}},\ and\
  \bibinfo {author} {\bibfnamefont {Q.-G.}\ \bibnamefont {Huang}},\ }\bibfield
  {title} {\bibinfo {title} {{Pulsar Timing Array Constraints on Primordial
  Black Holes with NANOGrav 11-Year Dataset}},\ }\href
  {https://doi.org/10.1103/PhysRevLett.124.251101} {\bibfield  {journal}
  {\bibinfo  {journal} {Phys. Rev. Lett.}\ }\textbf {\bibinfo {volume} {124}},\
  \bibinfo {pages} {251101} (\bibinfo {year} {2020})},\ \Eprint
  {https://arxiv.org/abs/1910.12239} {arXiv:1910.12239 [astro-ph.CO]}
  \BibitemShut {NoStop}%
\bibitem [{\citenamefont {De~Luca}\ \emph {et~al.}(2021)\citenamefont
  {De~Luca}, \citenamefont {Franciolini},\ and\ \citenamefont
  {Riotto}}]{DeLuca:2020agl}%
  \BibitemOpen
  \bibfield  {author} {\bibinfo {author} {\bibfnamefont {V.}~\bibnamefont
  {De~Luca}}, \bibinfo {author} {\bibfnamefont {G.}~\bibnamefont
  {Franciolini}},\ and\ \bibinfo {author} {\bibfnamefont {A.}~\bibnamefont
  {Riotto}},\ }\bibfield  {title} {\bibinfo {title} {{NANOGrav Data Hints at
  Primordial Black Holes as Dark Matter}},\ }\href
  {https://doi.org/10.1103/PhysRevLett.126.041303} {\bibfield  {journal}
  {\bibinfo  {journal} {Phys. Rev. Lett.}\ }\textbf {\bibinfo {volume} {126}},\
  \bibinfo {pages} {041303} (\bibinfo {year} {2021})},\ \Eprint
  {https://arxiv.org/abs/2009.08268} {arXiv:2009.08268 [astro-ph.CO]}
  \BibitemShut {NoStop}%
\bibitem [{\citenamefont {Sugiyama}\ \emph {et~al.}(2021)\citenamefont
  {Sugiyama}, \citenamefont {Takhistov}, \citenamefont {Vitagliano},
  \citenamefont {Kusenko}, \citenamefont {Sasaki},\ and\ \citenamefont
  {Takada}}]{Sugiyama:2020roc}%
  \BibitemOpen
  \bibfield  {author} {\bibinfo {author} {\bibfnamefont {S.}~\bibnamefont
  {Sugiyama}}, \bibinfo {author} {\bibfnamefont {V.}~\bibnamefont {Takhistov}},
  \bibinfo {author} {\bibfnamefont {E.}~\bibnamefont {Vitagliano}}, \bibinfo
  {author} {\bibfnamefont {A.}~\bibnamefont {Kusenko}}, \bibinfo {author}
  {\bibfnamefont {M.}~\bibnamefont {Sasaki}},\ and\ \bibinfo {author}
  {\bibfnamefont {M.}~\bibnamefont {Takada}},\ }\bibfield  {title} {\bibinfo
  {title} {{Testing Stochastic Gravitational Wave Signals from Primordial Black
  Holes with Optical Telescopes}},\ }\href
  {https://doi.org/10.1016/j.physletb.2021.136097} {\bibfield  {journal}
  {\bibinfo  {journal} {Phys. Lett. B}\ }\textbf {\bibinfo {volume} {814}},\
  \bibinfo {pages} {136097} (\bibinfo {year} {2021})},\ \Eprint
  {https://arxiv.org/abs/2010.02189} {arXiv:2010.02189 [astro-ph.CO]}
  \BibitemShut {NoStop}%
\bibitem [{\citenamefont {Vaskonen}\ and\ \citenamefont
  {Veerm\"ae}(2021)}]{Vaskonen:2020lbd}%
  \BibitemOpen
  \bibfield  {author} {\bibinfo {author} {\bibfnamefont {V.}~\bibnamefont
  {Vaskonen}}\ and\ \bibinfo {author} {\bibfnamefont {H.}~\bibnamefont
  {Veerm\"ae}},\ }\bibfield  {title} {\bibinfo {title} {{Did NANOGrav see a
  signal from primordial black hole formation?}},\ }\href
  {https://doi.org/10.1103/PhysRevLett.126.051303} {\bibfield  {journal}
  {\bibinfo  {journal} {Phys. Rev. Lett.}\ }\textbf {\bibinfo {volume} {126}},\
  \bibinfo {pages} {051303} (\bibinfo {year} {2021})},\ \Eprint
  {https://arxiv.org/abs/2009.07832} {arXiv:2009.07832 [astro-ph.CO]}
  \BibitemShut {NoStop}%
\bibitem [{\citenamefont {Kohri}\ and\ \citenamefont
  {Terada}(2021)}]{Kohri:2020qqd}%
  \BibitemOpen
  \bibfield  {author} {\bibinfo {author} {\bibfnamefont {K.}~\bibnamefont
  {Kohri}}\ and\ \bibinfo {author} {\bibfnamefont {T.}~\bibnamefont {Terada}},\
  }\bibfield  {title} {\bibinfo {title} {{Solar-Mass Primordial Black Holes
  Explain NANOGrav Hint of Gravitational Waves}},\ }\href
  {https://doi.org/10.1016/j.physletb.2020.136040} {\bibfield  {journal}
  {\bibinfo  {journal} {Phys. Lett. B}\ }\textbf {\bibinfo {volume} {813}},\
  \bibinfo {pages} {136040} (\bibinfo {year} {2021})},\ \Eprint
  {https://arxiv.org/abs/2009.11853} {arXiv:2009.11853 [astro-ph.CO]}
  \BibitemShut {NoStop}%
\bibitem [{\citenamefont {Zhao}\ and\ \citenamefont
  {Wang}(2023)}]{Zhao:2022kvz}%
  \BibitemOpen
  \bibfield  {author} {\bibinfo {author} {\bibfnamefont {Z.-C.}\ \bibnamefont
  {Zhao}}\ and\ \bibinfo {author} {\bibfnamefont {S.}~\bibnamefont {Wang}},\
  }\bibfield  {title} {\bibinfo {title} {{Bayesian Implications for the
  Primordial Black Holes from NANOGrav\textquoteright{}s Pulsar-Timing Data
  Using the Scalar-Induced Gravitational Waves}},\ }\href
  {https://doi.org/10.3390/universe9040157} {\bibfield  {journal} {\bibinfo
  {journal} {Universe}\ }\textbf {\bibinfo {volume} {9}},\ \bibinfo {pages}
  {157} (\bibinfo {year} {2023})},\ \Eprint {https://arxiv.org/abs/2211.09450}
  {arXiv:2211.09450 [astro-ph.CO]} \BibitemShut {NoStop}%
\bibitem [{\citenamefont {Ellis}\ and\ \citenamefont
  {Lewicki}(2021)}]{Ellis:2020ena}%
  \BibitemOpen
  \bibfield  {author} {\bibinfo {author} {\bibfnamefont {J.}~\bibnamefont
  {Ellis}}\ and\ \bibinfo {author} {\bibfnamefont {M.}~\bibnamefont
  {Lewicki}},\ }\bibfield  {title} {\bibinfo {title} {{Cosmic String
  Interpretation of NANOGrav Pulsar Timing Data}},\ }\href
  {https://doi.org/10.1103/PhysRevLett.126.041304} {\bibfield  {journal}
  {\bibinfo  {journal} {Phys. Rev. Lett.}\ }\textbf {\bibinfo {volume} {126}},\
  \bibinfo {pages} {041304} (\bibinfo {year} {2021})},\ \Eprint
  {https://arxiv.org/abs/2009.06555} {arXiv:2009.06555 [astro-ph.CO]}
  \BibitemShut {NoStop}%
\bibitem [{\citenamefont {Blasi}\ \emph {et~al.}(2021)\citenamefont {Blasi},
  \citenamefont {Brdar},\ and\ \citenamefont {Schmitz}}]{Blasi:2020mfx}%
  \BibitemOpen
  \bibfield  {author} {\bibinfo {author} {\bibfnamefont {S.}~\bibnamefont
  {Blasi}}, \bibinfo {author} {\bibfnamefont {V.}~\bibnamefont {Brdar}},\ and\
  \bibinfo {author} {\bibfnamefont {K.}~\bibnamefont {Schmitz}},\ }\bibfield
  {title} {\bibinfo {title} {{Has NANOGrav found first evidence for cosmic
  strings?}},\ }\href {https://doi.org/10.1103/PhysRevLett.126.041305}
  {\bibfield  {journal} {\bibinfo  {journal} {Phys. Rev. Lett.}\ }\textbf
  {\bibinfo {volume} {126}},\ \bibinfo {pages} {041305} (\bibinfo {year}
  {2021})},\ \Eprint {https://arxiv.org/abs/2009.06607} {arXiv:2009.06607
  [astro-ph.CO]} \BibitemShut {NoStop}%
\bibitem [{\citenamefont {Buchmuller}\ \emph {et~al.}(2020)\citenamefont
  {Buchmuller}, \citenamefont {Domcke},\ and\ \citenamefont
  {Schmitz}}]{Buchmuller:2020lbh}%
  \BibitemOpen
  \bibfield  {author} {\bibinfo {author} {\bibfnamefont {W.}~\bibnamefont
  {Buchmuller}}, \bibinfo {author} {\bibfnamefont {V.}~\bibnamefont {Domcke}},\
  and\ \bibinfo {author} {\bibfnamefont {K.}~\bibnamefont {Schmitz}},\
  }\bibfield  {title} {\bibinfo {title} {{From NANOGrav to LIGO with metastable
  cosmic strings}},\ }\href {https://doi.org/10.1016/j.physletb.2020.135914}
  {\bibfield  {journal} {\bibinfo  {journal} {Phys. Lett. B}\ }\textbf
  {\bibinfo {volume} {811}},\ \bibinfo {pages} {135914} (\bibinfo {year}
  {2020})},\ \Eprint {https://arxiv.org/abs/2009.10649} {arXiv:2009.10649
  [astro-ph.CO]} \BibitemShut {NoStop}%
\bibitem [{\citenamefont {Bian}\ \emph {et~al.}(2021)\citenamefont {Bian},
  \citenamefont {Cai}, \citenamefont {Liu}, \citenamefont {Yang},\ and\
  \citenamefont {Zhou}}]{Bian:2020urb}%
  \BibitemOpen
  \bibfield  {author} {\bibinfo {author} {\bibfnamefont {L.}~\bibnamefont
  {Bian}}, \bibinfo {author} {\bibfnamefont {R.-G.}\ \bibnamefont {Cai}},
  \bibinfo {author} {\bibfnamefont {J.}~\bibnamefont {Liu}}, \bibinfo {author}
  {\bibfnamefont {X.-Y.}\ \bibnamefont {Yang}},\ and\ \bibinfo {author}
  {\bibfnamefont {R.}~\bibnamefont {Zhou}},\ }\bibfield  {title} {\bibinfo
  {title} {{Evidence for different gravitational-wave sources in the NANOGrav
  dataset}},\ }\href {https://doi.org/10.1103/PhysRevD.103.L081301} {\bibfield
  {journal} {\bibinfo  {journal} {Phys. Rev. D}\ }\textbf {\bibinfo {volume}
  {103}},\ \bibinfo {pages} {L081301} (\bibinfo {year} {2021})},\ \Eprint
  {https://arxiv.org/abs/2009.13893} {arXiv:2009.13893 [astro-ph.CO]}
  \BibitemShut {NoStop}%
\bibitem [{\citenamefont {King}\ \emph {et~al.}(2021)\citenamefont {King},
  \citenamefont {Pascoli}, \citenamefont {Turner},\ and\ \citenamefont
  {Zhou}}]{King:2020hyd}%
  \BibitemOpen
  \bibfield  {author} {\bibinfo {author} {\bibfnamefont {S.~F.}\ \bibnamefont
  {King}}, \bibinfo {author} {\bibfnamefont {S.}~\bibnamefont {Pascoli}},
  \bibinfo {author} {\bibfnamefont {J.}~\bibnamefont {Turner}},\ and\ \bibinfo
  {author} {\bibfnamefont {Y.-L.}\ \bibnamefont {Zhou}},\ }\bibfield  {title}
  {\bibinfo {title} {{Gravitational Waves and Proton Decay: Complementary
  Windows into Grand Unified Theories}},\ }\href
  {https://doi.org/10.1103/PhysRevLett.126.021802} {\bibfield  {journal}
  {\bibinfo  {journal} {Phys. Rev. Lett.}\ }\textbf {\bibinfo {volume} {126}},\
  \bibinfo {pages} {021802} (\bibinfo {year} {2021})},\ \Eprint
  {https://arxiv.org/abs/2005.13549} {arXiv:2005.13549 [hep-ph]} \BibitemShut
  {NoStop}%
\bibitem [{\citenamefont {Madge}\ \emph {et~al.}(2023)\citenamefont {Madge},
  \citenamefont {Morgante}, \citenamefont {Puchades-Ib\'a\~nez}, \citenamefont
  {Ramberg}, \citenamefont {Ratzinger}, \citenamefont {Schenk},\ and\
  \citenamefont {Schwaller}}]{Madge:2023dxc}%
  \BibitemOpen
  \bibfield  {author} {\bibinfo {author} {\bibfnamefont {E.}~\bibnamefont
  {Madge}}, \bibinfo {author} {\bibfnamefont {E.}~\bibnamefont {Morgante}},
  \bibinfo {author} {\bibfnamefont {C.}~\bibnamefont {Puchades-Ib\'a\~nez}},
  \bibinfo {author} {\bibfnamefont {N.}~\bibnamefont {Ramberg}}, \bibinfo
  {author} {\bibfnamefont {W.}~\bibnamefont {Ratzinger}}, \bibinfo {author}
  {\bibfnamefont {S.}~\bibnamefont {Schenk}},\ and\ \bibinfo {author}
  {\bibfnamefont {P.}~\bibnamefont {Schwaller}},\ }\bibfield  {title} {\bibinfo
  {title} {{Primordial gravitational waves in the nano-Hertz regime and PTA
  data \textemdash{} towards solving the GW inverse problem}},\ }\href
  {https://doi.org/10.1007/JHEP10(2023)171} {\bibfield  {journal} {\bibinfo
  {journal} {JHEP}\ }\textbf {\bibinfo {volume} {10}},\ \bibinfo {pages}
  {171}},\ \Eprint {https://arxiv.org/abs/2306.14856} {arXiv:2306.14856
  [hep-ph]} \BibitemShut {NoStop}%
\bibitem [{\citenamefont {Servant}\ and\ \citenamefont
  {Simakachorn}(2023)}]{Servant:2023mwt}%
  \BibitemOpen
  \bibfield  {author} {\bibinfo {author} {\bibfnamefont {G.}~\bibnamefont
  {Servant}}\ and\ \bibinfo {author} {\bibfnamefont {P.}~\bibnamefont
  {Simakachorn}},\ }\bibfield  {title} {\bibinfo {title} {{Constraining
  Post-Inflationary Axions with Pulsar Timing Arrays}},\ }\href@noop {} {\
  (\bibinfo {year} {2023})},\ \Eprint {https://arxiv.org/abs/2307.03121}
  {arXiv:2307.03121 [hep-ph]} \BibitemShut {NoStop}%
\bibitem [{\citenamefont {Murai}\ and\ \citenamefont
  {Yin}(2023)}]{Murai:2023gkv}%
  \BibitemOpen
  \bibfield  {author} {\bibinfo {author} {\bibfnamefont {K.}~\bibnamefont
  {Murai}}\ and\ \bibinfo {author} {\bibfnamefont {W.}~\bibnamefont {Yin}},\
  }\bibfield  {title} {\bibinfo {title} {{A novel probe of supersymmetry in
  light of nanohertz gravitational waves}},\ }\href
  {https://doi.org/10.1007/JHEP10(2023)062} {\bibfield  {journal} {\bibinfo
  {journal} {JHEP}\ }\textbf {\bibinfo {volume} {10}},\ \bibinfo {pages}
  {062}},\ \Eprint {https://arxiv.org/abs/2307.00628} {arXiv:2307.00628
  [hep-ph]} \BibitemShut {NoStop}%
\bibitem [{\citenamefont {Unal}\ \emph {et~al.}(2024)\citenamefont {Unal},
  \citenamefont {Papageorgiou},\ and\ \citenamefont {Obata}}]{Unal:2023srk}%
  \BibitemOpen
  \bibfield  {author} {\bibinfo {author} {\bibfnamefont {C.}~\bibnamefont
  {Unal}}, \bibinfo {author} {\bibfnamefont {A.}~\bibnamefont {Papageorgiou}},\
  and\ \bibinfo {author} {\bibfnamefont {I.}~\bibnamefont {Obata}},\ }\bibfield
   {title} {\bibinfo {title} {{Axion-gauge dynamics during inflation as the
  origin of pulsar timing array signals and primordial black holes}},\ }\href
  {https://doi.org/10.1016/j.physletb.2024.138873} {\bibfield  {journal}
  {\bibinfo  {journal} {Phys. Lett. B}\ }\textbf {\bibinfo {volume} {856}},\
  \bibinfo {pages} {138873} (\bibinfo {year} {2024})},\ \Eprint
  {https://arxiv.org/abs/2307.02322} {arXiv:2307.02322 [astro-ph.CO]}
  \BibitemShut {NoStop}%
\bibitem [{\citenamefont {Geller}\ \emph {et~al.}(2024)\citenamefont {Geller},
  \citenamefont {Ghosh}, \citenamefont {Lu},\ and\ \citenamefont
  {Tsai}}]{Geller:2023shn}%
  \BibitemOpen
  \bibfield  {author} {\bibinfo {author} {\bibfnamefont {M.}~\bibnamefont
  {Geller}}, \bibinfo {author} {\bibfnamefont {S.}~\bibnamefont {Ghosh}},
  \bibinfo {author} {\bibfnamefont {S.}~\bibnamefont {Lu}},\ and\ \bibinfo
  {author} {\bibfnamefont {Y.}~\bibnamefont {Tsai}},\ }\bibfield  {title}
  {\bibinfo {title} {{Challenges in interpreting the NANOGrav 15-year dataset
  as early Universe gravitational waves produced by an ALP induced
  instability}},\ }\href {https://doi.org/10.1103/PhysRevD.109.063537}
  {\bibfield  {journal} {\bibinfo  {journal} {Phys. Rev. D}\ }\textbf {\bibinfo
  {volume} {109}},\ \bibinfo {pages} {063537} (\bibinfo {year} {2024})},\
  \Eprint {https://arxiv.org/abs/2307.03724} {arXiv:2307.03724 [hep-ph]}
  \BibitemShut {NoStop}%
\bibitem [{\citenamefont {Depta}\ \emph {et~al.}(2023)\citenamefont {Depta},
  \citenamefont {Schmidt-Hoberg},\ and\ \citenamefont
  {Tasillo}}]{Depta:2023qst}%
  \BibitemOpen
  \bibfield  {author} {\bibinfo {author} {\bibfnamefont {P.~F.}\ \bibnamefont
  {Depta}}, \bibinfo {author} {\bibfnamefont {K.}~\bibnamefont
  {Schmidt-Hoberg}},\ and\ \bibinfo {author} {\bibfnamefont {C.}~\bibnamefont
  {Tasillo}},\ }\bibfield  {title} {\bibinfo {title} {{Do pulsar timing arrays
  observe merging primordial black holes?}},\ }\href@noop {} {\  (\bibinfo
  {year} {2023})},\ \Eprint {https://arxiv.org/abs/2306.17836}
  {arXiv:2306.17836 [astro-ph.CO]} \BibitemShut {NoStop}%
\bibitem [{\citenamefont {Gouttenoire}\ \emph {et~al.}(2023)\citenamefont
  {Gouttenoire}, \citenamefont {Trifinopoulos}, \citenamefont {Valogiannis},\
  and\ \citenamefont {Vanvlasselaer}}]{Gouttenoire:2023nzr}%
  \BibitemOpen
  \bibfield  {author} {\bibinfo {author} {\bibfnamefont {Y.}~\bibnamefont
  {Gouttenoire}}, \bibinfo {author} {\bibfnamefont {S.}~\bibnamefont
  {Trifinopoulos}}, \bibinfo {author} {\bibfnamefont {G.}~\bibnamefont
  {Valogiannis}},\ and\ \bibinfo {author} {\bibfnamefont {M.}~\bibnamefont
  {Vanvlasselaer}},\ }\bibfield  {title} {\bibinfo {title} {{Scrutinizing the
  Primordial Black Holes Interpretation of PTA Gravitational Waves and JWST
  Early Galaxies}},\ }\href@noop {} {\  (\bibinfo {year} {2023})},\ \Eprint
  {https://arxiv.org/abs/2307.01457} {arXiv:2307.01457 [astro-ph.CO]}
  \BibitemShut {NoStop}%
\bibitem [{\citenamefont {Vilenkin}(1981)}]{Vilenkin:1981zs}%
  \BibitemOpen
  \bibfield  {author} {\bibinfo {author} {\bibfnamefont {A.}~\bibnamefont
  {Vilenkin}},\ }\bibfield  {title} {\bibinfo {title} {{Gravitational Field of
  Vacuum Domain Walls and Strings}},\ }\href
  {https://doi.org/10.1103/PhysRevD.23.852} {\bibfield  {journal} {\bibinfo
  {journal} {Phys. Rev. D}\ }\textbf {\bibinfo {volume} {23}},\ \bibinfo
  {pages} {852} (\bibinfo {year} {1981})}\BibitemShut {NoStop}%
\bibitem [{\citenamefont {Preskill}\ \emph {et~al.}(1991)\citenamefont
  {Preskill}, \citenamefont {Trivedi}, \citenamefont {Wilczek},\ and\
  \citenamefont {Wise}}]{Preskill:1991kd}%
  \BibitemOpen
  \bibfield  {author} {\bibinfo {author} {\bibfnamefont {J.}~\bibnamefont
  {Preskill}}, \bibinfo {author} {\bibfnamefont {S.~P.}\ \bibnamefont
  {Trivedi}}, \bibinfo {author} {\bibfnamefont {F.}~\bibnamefont {Wilczek}},\
  and\ \bibinfo {author} {\bibfnamefont {M.~B.}\ \bibnamefont {Wise}},\
  }\bibfield  {title} {\bibinfo {title} {{Cosmology and broken discrete
  symmetry}},\ }\href {https://doi.org/10.1016/0550-3213(91)90241-O} {\bibfield
   {journal} {\bibinfo  {journal} {Nucl. Phys. B}\ }\textbf {\bibinfo {volume}
  {363}},\ \bibinfo {pages} {207} (\bibinfo {year} {1991})}\BibitemShut
  {NoStop}%
\bibitem [{\citenamefont {Saikawa}(2017)}]{Saikawa:2017hiv}%
  \BibitemOpen
  \bibfield  {author} {\bibinfo {author} {\bibfnamefont {K.}~\bibnamefont
  {Saikawa}},\ }\bibfield  {title} {\bibinfo {title} {{A review of
  gravitational waves from cosmic domain walls}},\ }\href
  {https://doi.org/10.3390/universe3020040} {\bibfield  {journal} {\bibinfo
  {journal} {Universe}\ }\textbf {\bibinfo {volume} {3}},\ \bibinfo {pages}
  {40} (\bibinfo {year} {2017})},\ \Eprint {https://arxiv.org/abs/1703.02576}
  {arXiv:1703.02576 [hep-ph]} \BibitemShut {NoStop}%
\bibitem [{\citenamefont {Gelmini}\ \emph {et~al.}(2021)\citenamefont
  {Gelmini}, \citenamefont {Pascoli}, \citenamefont {Vitagliano},\ and\
  \citenamefont {Zhou}}]{Gelmini:2020bqg}%
  \BibitemOpen
  \bibfield  {author} {\bibinfo {author} {\bibfnamefont {G.~B.}\ \bibnamefont
  {Gelmini}}, \bibinfo {author} {\bibfnamefont {S.}~\bibnamefont {Pascoli}},
  \bibinfo {author} {\bibfnamefont {E.}~\bibnamefont {Vitagliano}},\ and\
  \bibinfo {author} {\bibfnamefont {Y.-L.}\ \bibnamefont {Zhou}},\ }\bibfield
  {title} {\bibinfo {title} {{Gravitational wave signatures from discrete
  flavor symmetries}},\ }\href {https://doi.org/10.1088/1475-7516/2021/02/032}
  {\bibfield  {journal} {\bibinfo  {journal} {JCAP}\ }\textbf {\bibinfo
  {volume} {02}},\ \bibinfo {pages} {032}},\ \Eprint
  {https://arxiv.org/abs/2009.01903} {arXiv:2009.01903 [hep-ph]} \BibitemShut
  {NoStop}%
\bibitem [{\citenamefont {Zeldovich}\ \emph {et~al.}(1974)\citenamefont
  {Zeldovich}, \citenamefont {Kobzarev},\ and\ \citenamefont
  {Okun}}]{Zeldovich:1974uw}%
  \BibitemOpen
  \bibfield  {author} {\bibinfo {author} {\bibfnamefont {Y.~B.}\ \bibnamefont
  {Zeldovich}}, \bibinfo {author} {\bibfnamefont {I.~Y.}\ \bibnamefont
  {Kobzarev}},\ and\ \bibinfo {author} {\bibfnamefont {L.~B.}\ \bibnamefont
  {Okun}},\ }\bibfield  {title} {\bibinfo {title} {{Cosmological Consequences
  of the Spontaneous Breakdown of Discrete Symmetry}},\ }\href@noop {}
  {\bibfield  {journal} {\bibinfo  {journal} {Zh. Eksp. Teor. Fiz.}\ }\textbf
  {\bibinfo {volume} {67}},\ \bibinfo {pages} {3} (\bibinfo {year}
  {1974})}\BibitemShut {NoStop}%
\bibitem [{\citenamefont {Vilenkin}\ and\ \citenamefont
  {Shellard}(2000)}]{Vilenkin:2000jqa}%
  \BibitemOpen
  \bibfield  {author} {\bibinfo {author} {\bibfnamefont {A.}~\bibnamefont
  {Vilenkin}}\ and\ \bibinfo {author} {\bibfnamefont {E.~P.~S.}\ \bibnamefont
  {Shellard}},\ }\href@noop {} {\emph {\bibinfo {title} {{Cosmic Strings and
  Other Topological Defects}}}}\ (\bibinfo  {publisher} {Cambridge University
  Press},\ \bibinfo {year} {2000})\BibitemShut {NoStop}%
\bibitem [{\citenamefont {Gleiser}\ and\ \citenamefont
  {Roberts}(1998)}]{Gleiser:1998na}%
  \BibitemOpen
  \bibfield  {author} {\bibinfo {author} {\bibfnamefont {M.}~\bibnamefont
  {Gleiser}}\ and\ \bibinfo {author} {\bibfnamefont {R.}~\bibnamefont
  {Roberts}},\ }\bibfield  {title} {\bibinfo {title} {{Gravitational waves from
  collapsing vacuum domains}},\ }\href
  {https://doi.org/10.1103/PhysRevLett.81.5497} {\bibfield  {journal} {\bibinfo
   {journal} {Phys. Rev. Lett.}\ }\textbf {\bibinfo {volume} {81}},\ \bibinfo
  {pages} {5497} (\bibinfo {year} {1998})},\ \Eprint
  {https://arxiv.org/abs/astro-ph/9807260} {arXiv:astro-ph/9807260}
  \BibitemShut {NoStop}%
\bibitem [{\citenamefont {Hiramatsu}\ \emph {et~al.}(2010)\citenamefont
  {Hiramatsu}, \citenamefont {Kawasaki},\ and\ \citenamefont
  {Saikawa}}]{Hiramatsu:2010yz}%
  \BibitemOpen
  \bibfield  {author} {\bibinfo {author} {\bibfnamefont {T.}~\bibnamefont
  {Hiramatsu}}, \bibinfo {author} {\bibfnamefont {M.}~\bibnamefont
  {Kawasaki}},\ and\ \bibinfo {author} {\bibfnamefont {K.}~\bibnamefont
  {Saikawa}},\ }\bibfield  {title} {\bibinfo {title} {{Gravitational Waves from
  Collapsing Domain Walls}},\ }\href
  {https://doi.org/10.1088/1475-7516/2010/05/032} {\bibfield  {journal}
  {\bibinfo  {journal} {JCAP}\ }\textbf {\bibinfo {volume} {05}},\ \bibinfo
  {pages} {032}},\ \Eprint {https://arxiv.org/abs/1002.1555} {arXiv:1002.1555
  [astro-ph.CO]} \BibitemShut {NoStop}%
\bibitem [{\citenamefont {Kawasaki}\ and\ \citenamefont
  {Saikawa}(2011)}]{Kawasaki:2011vv}%
  \BibitemOpen
  \bibfield  {author} {\bibinfo {author} {\bibfnamefont {M.}~\bibnamefont
  {Kawasaki}}\ and\ \bibinfo {author} {\bibfnamefont {K.}~\bibnamefont
  {Saikawa}},\ }\bibfield  {title} {\bibinfo {title} {{Study of gravitational
  radiation from cosmic domain walls}},\ }\href
  {https://doi.org/10.1088/1475-7516/2011/09/008} {\bibfield  {journal}
  {\bibinfo  {journal} {JCAP}\ }\textbf {\bibinfo {volume} {09}},\ \bibinfo
  {pages} {008}},\ \Eprint {https://arxiv.org/abs/1102.5628} {arXiv:1102.5628
  [astro-ph.CO]} \BibitemShut {NoStop}%
\bibitem [{\citenamefont {Hiramatsu}\ \emph {et~al.}(2014)\citenamefont
  {Hiramatsu}, \citenamefont {Kawasaki},\ and\ \citenamefont
  {Saikawa}}]{Hiramatsu:2013qaa}%
  \BibitemOpen
  \bibfield  {author} {\bibinfo {author} {\bibfnamefont {T.}~\bibnamefont
  {Hiramatsu}}, \bibinfo {author} {\bibfnamefont {M.}~\bibnamefont
  {Kawasaki}},\ and\ \bibinfo {author} {\bibfnamefont {K.}~\bibnamefont
  {Saikawa}},\ }\bibfield  {title} {\bibinfo {title} {{On the estimation of
  gravitational wave spectrum from cosmic domain walls}},\ }\href
  {https://doi.org/10.1088/1475-7516/2014/02/031} {\bibfield  {journal}
  {\bibinfo  {journal} {JCAP}\ }\textbf {\bibinfo {volume} {02}},\ \bibinfo
  {pages} {031}},\ \Eprint {https://arxiv.org/abs/1309.5001} {arXiv:1309.5001
  [astro-ph.CO]} \BibitemShut {NoStop}%
\bibitem [{\citenamefont {Ferreira}\ \emph {et~al.}(2023)\citenamefont
  {Ferreira}, \citenamefont {Notari}, \citenamefont {Pujolas},\ and\
  \citenamefont {Rompineve}}]{Ferreira:2022zzo}%
  \BibitemOpen
  \bibfield  {author} {\bibinfo {author} {\bibfnamefont {R.~Z.}\ \bibnamefont
  {Ferreira}}, \bibinfo {author} {\bibfnamefont {A.}~\bibnamefont {Notari}},
  \bibinfo {author} {\bibfnamefont {O.}~\bibnamefont {Pujolas}},\ and\ \bibinfo
  {author} {\bibfnamefont {F.}~\bibnamefont {Rompineve}},\ }\bibfield  {title}
  {\bibinfo {title} {{Gravitational waves from domain walls in Pulsar Timing
  Array datasets}},\ }\href {https://doi.org/10.1088/1475-7516/2023/02/001}
  {\bibfield  {journal} {\bibinfo  {journal} {JCAP}\ }\textbf {\bibinfo
  {volume} {02}},\ \bibinfo {pages} {001}},\ \Eprint
  {https://arxiv.org/abs/2204.04228} {arXiv:2204.04228 [astro-ph.CO]}
  \BibitemShut {NoStop}%
\bibitem [{\citenamefont {Ferrer}\ \emph {et~al.}(2019)\citenamefont {Ferrer},
  \citenamefont {Masso}, \citenamefont {Panico}, \citenamefont {Pujolas},\ and\
  \citenamefont {Rompineve}}]{Ferrer:2018uiu}%
  \BibitemOpen
  \bibfield  {author} {\bibinfo {author} {\bibfnamefont {F.}~\bibnamefont
  {Ferrer}}, \bibinfo {author} {\bibfnamefont {E.}~\bibnamefont {Masso}},
  \bibinfo {author} {\bibfnamefont {G.}~\bibnamefont {Panico}}, \bibinfo
  {author} {\bibfnamefont {O.}~\bibnamefont {Pujolas}},\ and\ \bibinfo {author}
  {\bibfnamefont {F.}~\bibnamefont {Rompineve}},\ }\bibfield  {title} {\bibinfo
  {title} {{Primordial Black Holes from the QCD axion}},\ }\href
  {https://doi.org/10.1103/PhysRevLett.122.101301} {\bibfield  {journal}
  {\bibinfo  {journal} {Phys. Rev. Lett.}\ }\textbf {\bibinfo {volume} {122}},\
  \bibinfo {pages} {101301} (\bibinfo {year} {2019})},\ \Eprint
  {https://arxiv.org/abs/1807.01707} {arXiv:1807.01707 [hep-ph]} \BibitemShut
  {NoStop}%
\bibitem [{\citenamefont {Ge}(2020)}]{Ge:2019ihf}%
  \BibitemOpen
  \bibfield  {author} {\bibinfo {author} {\bibfnamefont {S.}~\bibnamefont
  {Ge}},\ }\bibfield  {title} {\bibinfo {title} {{Sublunar-Mass Primordial
  Black Holes from Closed Axion Domain Walls}},\ }\href
  {https://doi.org/10.1016/j.dark.2019.100440} {\bibfield  {journal} {\bibinfo
  {journal} {Phys. Dark Univ.}\ }\textbf {\bibinfo {volume} {27}},\ \bibinfo
  {pages} {100440} (\bibinfo {year} {2020})},\ \Eprint
  {https://arxiv.org/abs/1905.12182} {arXiv:1905.12182 [hep-ph]} \BibitemShut
  {NoStop}%
\bibitem [{\citenamefont {Gouttenoire}\ and\ \citenamefont
  {Vitagliano}(2023)}]{Gouttenoire:2023gbn}%
  \BibitemOpen
  \bibfield  {author} {\bibinfo {author} {\bibfnamefont {Y.}~\bibnamefont
  {Gouttenoire}}\ and\ \bibinfo {author} {\bibfnamefont {E.}~\bibnamefont
  {Vitagliano}},\ }\bibfield  {title} {\bibinfo {title} {{Primordial Black
  Holes and Wormholes from Domain Wall Networks}},\ }\href@noop {} {\
  (\bibinfo {year} {2023})},\ \Eprint {https://arxiv.org/abs/2311.07670}
  {arXiv:2311.07670 [hep-ph]} \BibitemShut {NoStop}%
\bibitem [{\citenamefont {Gelmini}\ \emph
  {et~al.}(2023{\natexlab{a}})\citenamefont {Gelmini}, \citenamefont
  {Simpson},\ and\ \citenamefont {Vitagliano}}]{Gelmini:2022nim}%
  \BibitemOpen
  \bibfield  {author} {\bibinfo {author} {\bibfnamefont {G.~B.}\ \bibnamefont
  {Gelmini}}, \bibinfo {author} {\bibfnamefont {A.}~\bibnamefont {Simpson}},\
  and\ \bibinfo {author} {\bibfnamefont {E.}~\bibnamefont {Vitagliano}},\
  }\bibfield  {title} {\bibinfo {title} {{Catastrogenesis: DM, GWs, and PBHs
  from ALP string-wall networks}},\ }\href
  {https://doi.org/10.1088/1475-7516/2023/02/031} {\bibfield  {journal}
  {\bibinfo  {journal} {JCAP}\ }\textbf {\bibinfo {volume} {02}},\ \bibinfo
  {pages} {031}},\ \Eprint {https://arxiv.org/abs/2207.07126} {arXiv:2207.07126
  [hep-ph]} \BibitemShut {NoStop}%
\bibitem [{\citenamefont {Gelmini}\ \emph
  {et~al.}(2023{\natexlab{b}})\citenamefont {Gelmini}, \citenamefont {Hyman},
  \citenamefont {Simpson},\ and\ \citenamefont {Vitagliano}}]{Gelmini:2023ngs}%
  \BibitemOpen
  \bibfield  {author} {\bibinfo {author} {\bibfnamefont {G.~B.}\ \bibnamefont
  {Gelmini}}, \bibinfo {author} {\bibfnamefont {J.}~\bibnamefont {Hyman}},
  \bibinfo {author} {\bibfnamefont {A.}~\bibnamefont {Simpson}},\ and\ \bibinfo
  {author} {\bibfnamefont {E.}~\bibnamefont {Vitagliano}},\ }\bibfield  {title}
  {\bibinfo {title} {{Primordial black hole dark matter from catastrogenesis
  with unstable pseudo-Goldstone bosons}},\ }\href@noop {} {\  (\bibinfo {year}
  {2023}{\natexlab{b}})},\ \Eprint {https://arxiv.org/abs/2303.14107}
  {arXiv:2303.14107 [hep-ph]} \BibitemShut {NoStop}%
\bibitem [{\citenamefont {Kitajima}\ \emph {et~al.}(2024)\citenamefont
  {Kitajima}, \citenamefont {Lee}, \citenamefont {Murai}, \citenamefont
  {Takahashi},\ and\ \citenamefont {Yin}}]{Kitajima:2023cek}%
  \BibitemOpen
  \bibfield  {author} {\bibinfo {author} {\bibfnamefont {N.}~\bibnamefont
  {Kitajima}}, \bibinfo {author} {\bibfnamefont {J.}~\bibnamefont {Lee}},
  \bibinfo {author} {\bibfnamefont {K.}~\bibnamefont {Murai}}, \bibinfo
  {author} {\bibfnamefont {F.}~\bibnamefont {Takahashi}},\ and\ \bibinfo
  {author} {\bibfnamefont {W.}~\bibnamefont {Yin}},\ }\bibfield  {title}
  {\bibinfo {title} {{Gravitational waves from domain wall collapse, and
  application to nanohertz signals with QCD-coupled axions}},\ }\href
  {https://doi.org/10.1016/j.physletb.2024.138586} {\bibfield  {journal}
  {\bibinfo  {journal} {Phys. Lett. B}\ }\textbf {\bibinfo {volume} {851}},\
  \bibinfo {pages} {138586} (\bibinfo {year} {2024})},\ \Eprint
  {https://arxiv.org/abs/2306.17146} {arXiv:2306.17146 [hep-ph]} \BibitemShut
  {NoStop}%
\bibitem [{\citenamefont {Nakamura}\ \emph {et~al.}(1997)\citenamefont
  {Nakamura}, \citenamefont {Sasaki}, \citenamefont {Tanaka},\ and\
  \citenamefont {Thorne}}]{Nakamura:1997sm}%
  \BibitemOpen
  \bibfield  {author} {\bibinfo {author} {\bibfnamefont {T.}~\bibnamefont
  {Nakamura}}, \bibinfo {author} {\bibfnamefont {M.}~\bibnamefont {Sasaki}},
  \bibinfo {author} {\bibfnamefont {T.}~\bibnamefont {Tanaka}},\ and\ \bibinfo
  {author} {\bibfnamefont {K.~S.}\ \bibnamefont {Thorne}},\ }\bibfield  {title}
  {\bibinfo {title} {{Gravitational waves from coalescing black hole MACHO
  binaries}},\ }\href {https://doi.org/10.1086/310886} {\bibfield  {journal}
  {\bibinfo  {journal} {Astrophys. J. Lett.}\ }\textbf {\bibinfo {volume}
  {487}},\ \bibinfo {pages} {L139} (\bibinfo {year} {1997})},\ \Eprint
  {https://arxiv.org/abs/astro-ph/9708060} {arXiv:astro-ph/9708060}
  \BibitemShut {NoStop}%
\bibitem [{\citenamefont {Raidal}\ \emph {et~al.}(2019)\citenamefont {Raidal},
  \citenamefont {Spethmann}, \citenamefont {Vaskonen},\ and\ \citenamefont
  {Veerm\"ae}}]{Raidal:2018bbj}%
  \BibitemOpen
  \bibfield  {author} {\bibinfo {author} {\bibfnamefont {M.}~\bibnamefont
  {Raidal}}, \bibinfo {author} {\bibfnamefont {C.}~\bibnamefont {Spethmann}},
  \bibinfo {author} {\bibfnamefont {V.}~\bibnamefont {Vaskonen}},\ and\
  \bibinfo {author} {\bibfnamefont {H.}~\bibnamefont {Veerm\"ae}},\ }\bibfield
  {title} {\bibinfo {title} {{Formation and Evolution of Primordial Black Hole
  Binaries in the Early Universe}},\ }\href
  {https://doi.org/10.1088/1475-7516/2019/02/018} {\bibfield  {journal}
  {\bibinfo  {journal} {JCAP}\ }\textbf {\bibinfo {volume} {02}},\ \bibinfo
  {pages} {018}},\ \Eprint {https://arxiv.org/abs/1812.01930} {arXiv:1812.01930
  [astro-ph.CO]} \BibitemShut {NoStop}%
\bibitem [{\citenamefont {Kavanagh}\ \emph {et~al.}(2018)\citenamefont
  {Kavanagh}, \citenamefont {Gaggero},\ and\ \citenamefont
  {Bertone}}]{Kavanagh:2018ggo}%
  \BibitemOpen
  \bibfield  {author} {\bibinfo {author} {\bibfnamefont {B.~J.}\ \bibnamefont
  {Kavanagh}}, \bibinfo {author} {\bibfnamefont {D.}~\bibnamefont {Gaggero}},\
  and\ \bibinfo {author} {\bibfnamefont {G.}~\bibnamefont {Bertone}},\
  }\bibfield  {title} {\bibinfo {title} {{Merger rate of a subdominant
  population of primordial black holes}},\ }\href
  {https://doi.org/10.1103/PhysRevD.98.023536} {\bibfield  {journal} {\bibinfo
  {journal} {Phys. Rev. D}\ }\textbf {\bibinfo {volume} {98}},\ \bibinfo
  {pages} {023536} (\bibinfo {year} {2018})},\ \Eprint
  {https://arxiv.org/abs/1805.09034} {arXiv:1805.09034 [astro-ph.CO]}
  \BibitemShut {NoStop}%
\bibitem [{\citenamefont {Abbott}\ \emph {et~al.}(2019)\citenamefont {Abbott}
  \emph {et~al.}}]{LIGOScientific:2019kan}%
  \BibitemOpen
  \bibfield  {author} {\bibinfo {author} {\bibfnamefont {B.~P.}\ \bibnamefont
  {Abbott}} \emph {et~al.} (\bibinfo {collaboration} {LIGO Scientific,
  Virgo}),\ }\bibfield  {title} {\bibinfo {title} {{Search for Subsolar Mass
  Ultracompact Binaries in Advanced LIGO\textquoteright{}s Second Observing
  Run}},\ }\href {https://doi.org/10.1103/PhysRevLett.123.161102} {\bibfield
  {journal} {\bibinfo  {journal} {Phys. Rev. Lett.}\ }\textbf {\bibinfo
  {volume} {123}},\ \bibinfo {pages} {161102} (\bibinfo {year} {2019})},\
  \Eprint {https://arxiv.org/abs/1904.08976} {arXiv:1904.08976 [astro-ph.CO]}
  \BibitemShut {NoStop}%
\bibitem [{\citenamefont {De~Luca}\ \emph {et~al.}(2020)\citenamefont
  {De~Luca}, \citenamefont {Franciolini}, \citenamefont {Pani},\ and\
  \citenamefont {Riotto}}]{DeLuca:2020qqa}%
  \BibitemOpen
  \bibfield  {author} {\bibinfo {author} {\bibfnamefont {V.}~\bibnamefont
  {De~Luca}}, \bibinfo {author} {\bibfnamefont {G.}~\bibnamefont
  {Franciolini}}, \bibinfo {author} {\bibfnamefont {P.}~\bibnamefont {Pani}},\
  and\ \bibinfo {author} {\bibfnamefont {A.}~\bibnamefont {Riotto}},\
  }\bibfield  {title} {\bibinfo {title} {{Primordial Black Holes Confront
  LIGO/Virgo data: Current situation}},\ }\href
  {https://doi.org/10.1088/1475-7516/2020/06/044} {\bibfield  {journal}
  {\bibinfo  {journal} {JCAP}\ }\textbf {\bibinfo {volume} {06}},\ \bibinfo
  {pages} {044}},\ \Eprint {https://arxiv.org/abs/2005.05641} {arXiv:2005.05641
  [astro-ph.CO]} \BibitemShut {NoStop}%
\bibitem [{\citenamefont {Ali-Ha\"\i{}moud}\ and\ \citenamefont
  {Kamionkowski}(2017)}]{Ali-Haimoud:2016mbv}%
  \BibitemOpen
  \bibfield  {author} {\bibinfo {author} {\bibfnamefont {Y.}~\bibnamefont
  {Ali-Ha\"\i{}moud}}\ and\ \bibinfo {author} {\bibfnamefont {M.}~\bibnamefont
  {Kamionkowski}},\ }\bibfield  {title} {\bibinfo {title} {{Cosmic microwave
  background limits on accreting primordial black holes}},\ }\href
  {https://doi.org/10.1103/PhysRevD.95.043534} {\bibfield  {journal} {\bibinfo
  {journal} {Phys. Rev. D}\ }\textbf {\bibinfo {volume} {95}},\ \bibinfo
  {pages} {043534} (\bibinfo {year} {2017})},\ \Eprint
  {https://arxiv.org/abs/1612.05644} {arXiv:1612.05644 [astro-ph.CO]}
  \BibitemShut {NoStop}%
\bibitem [{\citenamefont {Poulin}\ \emph {et~al.}(2017)\citenamefont {Poulin},
  \citenamefont {Serpico}, \citenamefont {Calore}, \citenamefont {Clesse},\
  and\ \citenamefont {Kohri}}]{Poulin:2017bwe}%
  \BibitemOpen
  \bibfield  {author} {\bibinfo {author} {\bibfnamefont {V.}~\bibnamefont
  {Poulin}}, \bibinfo {author} {\bibfnamefont {P.~D.}\ \bibnamefont {Serpico}},
  \bibinfo {author} {\bibfnamefont {F.}~\bibnamefont {Calore}}, \bibinfo
  {author} {\bibfnamefont {S.}~\bibnamefont {Clesse}},\ and\ \bibinfo {author}
  {\bibfnamefont {K.}~\bibnamefont {Kohri}},\ }\bibfield  {title} {\bibinfo
  {title} {{Cmb Bounds on Disk-Accreting Massive Primordial Black Holes}},\
  }\href {https://doi.org/10.1103/PhysRevD.96.083524} {\bibfield  {journal}
  {\bibinfo  {journal} {Phys. Rev. D}\ }\textbf {\bibinfo {volume} {96}},\
  \bibinfo {pages} {083524} (\bibinfo {year} {2017})},\ \Eprint
  {https://arxiv.org/abs/1707.04206} {arXiv:1707.04206 [astro-ph.CO]}
  \BibitemShut {NoStop}%
\bibitem [{\citenamefont {Serpico}\ \emph {et~al.}(2020)\citenamefont
  {Serpico}, \citenamefont {Poulin}, \citenamefont {Inman},\ and\ \citenamefont
  {Kohri}}]{Serpico:2020ehh}%
  \BibitemOpen
  \bibfield  {author} {\bibinfo {author} {\bibfnamefont {P.~D.}\ \bibnamefont
  {Serpico}}, \bibinfo {author} {\bibfnamefont {V.}~\bibnamefont {Poulin}},
  \bibinfo {author} {\bibfnamefont {D.}~\bibnamefont {Inman}},\ and\ \bibinfo
  {author} {\bibfnamefont {K.}~\bibnamefont {Kohri}},\ }\bibfield  {title}
  {\bibinfo {title} {{Cosmic microwave background bounds on primordial black
  holes including dark matter halo accretion}},\ }\href
  {https://doi.org/10.1103/PhysRevResearch.2.023204} {\bibfield  {journal}
  {\bibinfo  {journal} {Phys. Rev. Res.}\ }\textbf {\bibinfo {volume} {2}},\
  \bibinfo {pages} {023204} (\bibinfo {year} {2020})},\ \Eprint
  {https://arxiv.org/abs/2002.10771} {arXiv:2002.10771 [astro-ph.CO]}
  \BibitemShut {NoStop}%
\bibitem [{\citenamefont {Blasi}\ \emph
  {et~al.}(2023{\natexlab{a}})\citenamefont {Blasi}, \citenamefont {Mariotti},
  \citenamefont {Rase}, \citenamefont {Sevrin},\ and\ \citenamefont
  {Turbang}}]{Blasi:2022ayo}%
  \BibitemOpen
  \bibfield  {author} {\bibinfo {author} {\bibfnamefont {S.}~\bibnamefont
  {Blasi}}, \bibinfo {author} {\bibfnamefont {A.}~\bibnamefont {Mariotti}},
  \bibinfo {author} {\bibfnamefont {A.}~\bibnamefont {Rase}}, \bibinfo {author}
  {\bibfnamefont {A.}~\bibnamefont {Sevrin}},\ and\ \bibinfo {author}
  {\bibfnamefont {K.}~\bibnamefont {Turbang}},\ }\bibfield  {title} {\bibinfo
  {title} {{Friction on ALP domain walls and gravitational waves}},\ }\href
  {https://doi.org/10.1088/1475-7516/2023/04/008} {\bibfield  {journal}
  {\bibinfo  {journal} {JCAP}\ }\textbf {\bibinfo {volume} {04}},\ \bibinfo
  {pages} {008}},\ \Eprint {https://arxiv.org/abs/2210.14246} {arXiv:2210.14246
  [hep-ph]} \BibitemShut {NoStop}%
\bibitem [{\citenamefont {Martins}\ \emph {et~al.}(2016)\citenamefont
  {Martins}, \citenamefont {Rybak}, \citenamefont {Avgoustidis},\ and\
  \citenamefont {Shellard}}]{Martins:2016ois}%
  \BibitemOpen
  \bibfield  {author} {\bibinfo {author} {\bibfnamefont {C.~J. A.~P.}\
  \bibnamefont {Martins}}, \bibinfo {author} {\bibfnamefont {I.~Y.}\
  \bibnamefont {Rybak}}, \bibinfo {author} {\bibfnamefont {A.}~\bibnamefont
  {Avgoustidis}},\ and\ \bibinfo {author} {\bibfnamefont {E.~P.~S.}\
  \bibnamefont {Shellard}},\ }\bibfield  {title} {\bibinfo {title} {{Extending
  the Velocity-Dependent One-Scale Model for Domain Walls}},\ }\href
  {https://doi.org/10.1103/PhysRevD.93.043534} {\bibfield  {journal} {\bibinfo
  {journal} {Phys. Rev. D}\ }\textbf {\bibinfo {volume} {93}},\ \bibinfo
  {pages} {043534} (\bibinfo {year} {2016})},\ \Eprint
  {https://arxiv.org/abs/1602.01322} {arXiv:1602.01322 [hep-ph]} \BibitemShut
  {NoStop}%
\bibitem [{\citenamefont {Kibble}(1976)}]{Kibble:1976sj}%
  \BibitemOpen
  \bibfield  {author} {\bibinfo {author} {\bibfnamefont {T.~W.~B.}\
  \bibnamefont {Kibble}},\ }\bibfield  {title} {\bibinfo {title} {{Topology of
  Cosmic Domains and Strings}},\ }\href
  {https://doi.org/10.1088/0305-4470/9/8/029} {\bibfield  {journal} {\bibinfo
  {journal} {J. Phys. A}\ }\textbf {\bibinfo {volume} {9}},\ \bibinfo {pages}
  {1387} (\bibinfo {year} {1976})}\BibitemShut {NoStop}%
\bibitem [{\citenamefont {Sikivie}(1982)}]{Sikivie:1982qv}%
  \BibitemOpen
  \bibfield  {author} {\bibinfo {author} {\bibfnamefont {P.}~\bibnamefont
  {Sikivie}},\ }\bibfield  {title} {\bibinfo {title} {{Of Axions, Domain Walls
  and the Early Universe}},\ }\href
  {https://doi.org/10.1103/PhysRevLett.48.1156} {\bibfield  {journal} {\bibinfo
   {journal} {Phys. Rev. Lett.}\ }\textbf {\bibinfo {volume} {48}},\ \bibinfo
  {pages} {1156} (\bibinfo {year} {1982})}\BibitemShut {NoStop}%
\bibitem [{\citenamefont {Gelmini}\ \emph {et~al.}(1989)\citenamefont
  {Gelmini}, \citenamefont {Gleiser},\ and\ \citenamefont
  {Kolb}}]{Gelmini:1988sf}%
  \BibitemOpen
  \bibfield  {author} {\bibinfo {author} {\bibfnamefont {G.~B.}\ \bibnamefont
  {Gelmini}}, \bibinfo {author} {\bibfnamefont {M.}~\bibnamefont {Gleiser}},\
  and\ \bibinfo {author} {\bibfnamefont {E.~W.}\ \bibnamefont {Kolb}},\
  }\bibfield  {title} {\bibinfo {title} {{Cosmology of Biased Discrete Symmetry
  Breaking}},\ }\href {https://doi.org/10.1103/PhysRevD.39.1558} {\bibfield
  {journal} {\bibinfo  {journal} {Phys. Rev. D}\ }\textbf {\bibinfo {volume}
  {39}},\ \bibinfo {pages} {1558} (\bibinfo {year} {1989})}\BibitemShut
  {NoStop}%
\bibitem [{\citenamefont {Kawasaki}\ \emph {et~al.}(2015)\citenamefont
  {Kawasaki}, \citenamefont {Saikawa},\ and\ \citenamefont
  {Sekiguchi}}]{Kawasaki:2014sqa}%
  \BibitemOpen
  \bibfield  {author} {\bibinfo {author} {\bibfnamefont {M.}~\bibnamefont
  {Kawasaki}}, \bibinfo {author} {\bibfnamefont {K.}~\bibnamefont {Saikawa}},\
  and\ \bibinfo {author} {\bibfnamefont {T.}~\bibnamefont {Sekiguchi}},\
  }\bibfield  {title} {\bibinfo {title} {{Axion dark matter from topological
  defects}},\ }\href {https://doi.org/10.1103/PhysRevD.91.065014} {\bibfield
  {journal} {\bibinfo  {journal} {Phys. Rev. D}\ }\textbf {\bibinfo {volume}
  {91}},\ \bibinfo {pages} {065014} (\bibinfo {year} {2015})},\ \Eprint
  {https://arxiv.org/abs/1412.0789} {arXiv:1412.0789 [hep-ph]} \BibitemShut
  {NoStop}%
\bibitem [{\citenamefont {Chang}\ \emph {et~al.}(1999)\citenamefont {Chang},
  \citenamefont {Hagmann},\ and\ \citenamefont {Sikivie}}]{Chang:1998tb}%
  \BibitemOpen
  \bibfield  {author} {\bibinfo {author} {\bibfnamefont {S.}~\bibnamefont
  {Chang}}, \bibinfo {author} {\bibfnamefont {C.}~\bibnamefont {Hagmann}},\
  and\ \bibinfo {author} {\bibfnamefont {P.}~\bibnamefont {Sikivie}},\
  }\bibfield  {title} {\bibinfo {title} {{Studies of the motion and decay of
  axion walls bounded by strings}},\ }\href
  {https://doi.org/10.1103/PhysRevD.59.023505} {\bibfield  {journal} {\bibinfo
  {journal} {Phys. Rev. D}\ }\textbf {\bibinfo {volume} {59}},\ \bibinfo
  {pages} {023505} (\bibinfo {year} {1999})},\ \Eprint
  {https://arxiv.org/abs/hep-ph/9807374} {arXiv:hep-ph/9807374} \BibitemShut
  {NoStop}%
\bibitem [{\citenamefont {Mitridate}\ \emph {et~al.}(2023)\citenamefont
  {Mitridate}, \citenamefont {Wright}, \citenamefont {von Eckardstein},
  \citenamefont {Schr\"oder}, \citenamefont {Nay}, \citenamefont {Olum},
  \citenamefont {Schmitz},\ and\ \citenamefont {Trickle}}]{Mitridate:2023oar}%
  \BibitemOpen
  \bibfield  {author} {\bibinfo {author} {\bibfnamefont {A.}~\bibnamefont
  {Mitridate}}, \bibinfo {author} {\bibfnamefont {D.}~\bibnamefont {Wright}},
  \bibinfo {author} {\bibfnamefont {R.}~\bibnamefont {von Eckardstein}},
  \bibinfo {author} {\bibfnamefont {T.}~\bibnamefont {Schr\"oder}}, \bibinfo
  {author} {\bibfnamefont {J.}~\bibnamefont {Nay}}, \bibinfo {author}
  {\bibfnamefont {K.}~\bibnamefont {Olum}}, \bibinfo {author} {\bibfnamefont
  {K.}~\bibnamefont {Schmitz}},\ and\ \bibinfo {author} {\bibfnamefont
  {T.}~\bibnamefont {Trickle}},\ }\bibfield  {title} {\bibinfo {title}
  {{PTArcade}},\ }\href@noop {} {\  (\bibinfo {year} {2023})},\ \Eprint
  {https://arxiv.org/abs/2306.16377} {arXiv:2306.16377 [hep-ph]} \BibitemShut
  {NoStop}%
\bibitem [{\citenamefont {Deng}\ \emph {et~al.}(2017)\citenamefont {Deng},
  \citenamefont {Garriga},\ and\ \citenamefont {Vilenkin}}]{Deng:2016vzb}%
  \BibitemOpen
  \bibfield  {author} {\bibinfo {author} {\bibfnamefont {H.}~\bibnamefont
  {Deng}}, \bibinfo {author} {\bibfnamefont {J.}~\bibnamefont {Garriga}},\ and\
  \bibinfo {author} {\bibfnamefont {A.}~\bibnamefont {Vilenkin}},\ }\bibfield
  {title} {\bibinfo {title} {{Primordial black hole and wormhole formation by
  domain walls}},\ }\href {https://doi.org/10.1088/1475-7516/2017/04/050}
  {\bibfield  {journal} {\bibinfo  {journal} {JCAP}\ }\textbf {\bibinfo
  {volume} {04}},\ \bibinfo {pages} {050}},\ \Eprint
  {https://arxiv.org/abs/1612.03753} {arXiv:1612.03753 [gr-qc]} \BibitemShut
  {NoStop}%
\bibitem [{\citenamefont {Deng}\ and\ \citenamefont
  {Vilenkin}(2017)}]{Deng:2017uwc}%
  \BibitemOpen
  \bibfield  {author} {\bibinfo {author} {\bibfnamefont {H.}~\bibnamefont
  {Deng}}\ and\ \bibinfo {author} {\bibfnamefont {A.}~\bibnamefont
  {Vilenkin}},\ }\bibfield  {title} {\bibinfo {title} {{Primordial black hole
  formation by vacuum bubbles}},\ }\href
  {https://doi.org/10.1088/1475-7516/2017/12/044} {\bibfield  {journal}
  {\bibinfo  {journal} {JCAP}\ }\textbf {\bibinfo {volume} {12}},\ \bibinfo
  {pages} {044}},\ \Eprint {https://arxiv.org/abs/1710.02865} {arXiv:1710.02865
  [gr-qc]} \BibitemShut {NoStop}%
\bibitem [{\citenamefont {Deng}(2020)}]{Deng:2020mds}%
  \BibitemOpen
  \bibfield  {author} {\bibinfo {author} {\bibfnamefont {H.}~\bibnamefont
  {Deng}},\ }\bibfield  {title} {\bibinfo {title} {{Primordial black hole
  formation by vacuum bubbles. Part II}},\ }\href
  {https://doi.org/10.1088/1475-7516/2020/09/023} {\bibfield  {journal}
  {\bibinfo  {journal} {JCAP}\ }\textbf {\bibinfo {volume} {09}},\ \bibinfo
  {pages} {023}},\ \Eprint {https://arxiv.org/abs/2006.11907} {arXiv:2006.11907
  [astro-ph.CO]} \BibitemShut {NoStop}%
\bibitem [{\citenamefont {Berezin}\ \emph {et~al.}(1983)\citenamefont
  {Berezin}, \citenamefont {Kuzmin},\ and\ \citenamefont
  {Tkachev}}]{Berezin:1982ur}%
  \BibitemOpen
  \bibfield  {author} {\bibinfo {author} {\bibfnamefont {V.~A.}\ \bibnamefont
  {Berezin}}, \bibinfo {author} {\bibfnamefont {V.~A.}\ \bibnamefont
  {Kuzmin}},\ and\ \bibinfo {author} {\bibfnamefont {I.~I.}\ \bibnamefont
  {Tkachev}},\ }\bibfield  {title} {\bibinfo {title} {{THIN WALL VACUUM DOMAINS
  EVOLUTION}},\ }\href {https://doi.org/10.1016/0370-2693(83)90630-5}
  {\bibfield  {journal} {\bibinfo  {journal} {Phys. Lett. B}\ }\textbf
  {\bibinfo {volume} {120}},\ \bibinfo {pages} {91} (\bibinfo {year}
  {1983})}\BibitemShut {NoStop}%
\bibitem [{\citenamefont {Berezin}\ \emph {et~al.}(1987)\citenamefont
  {Berezin}, \citenamefont {Kuzmin},\ and\ \citenamefont
  {Tkachev}}]{Berezin:1987bc}%
  \BibitemOpen
  \bibfield  {author} {\bibinfo {author} {\bibfnamefont {V.~A.}\ \bibnamefont
  {Berezin}}, \bibinfo {author} {\bibfnamefont {V.~A.}\ \bibnamefont
  {Kuzmin}},\ and\ \bibinfo {author} {\bibfnamefont {I.~I.}\ \bibnamefont
  {Tkachev}},\ }\bibfield  {title} {\bibinfo {title} {{Dynamics of Bubbles in
  General Relativity}},\ }\href {https://doi.org/10.1103/PhysRevD.36.2919}
  {\bibfield  {journal} {\bibinfo  {journal} {Phys. Rev. D}\ }\textbf {\bibinfo
  {volume} {36}},\ \bibinfo {pages} {2919} (\bibinfo {year}
  {1987})}\BibitemShut {NoStop}%
\bibitem [{\citenamefont {Blau}\ \emph {et~al.}(1987)\citenamefont {Blau},
  \citenamefont {Guendelman},\ and\ \citenamefont {Guth}}]{Blau:1986cw}%
  \BibitemOpen
  \bibfield  {author} {\bibinfo {author} {\bibfnamefont {S.~K.}\ \bibnamefont
  {Blau}}, \bibinfo {author} {\bibfnamefont {E.~I.}\ \bibnamefont
  {Guendelman}},\ and\ \bibinfo {author} {\bibfnamefont {A.~H.}\ \bibnamefont
  {Guth}},\ }\bibfield  {title} {\bibinfo {title} {{The Dynamics of False
  Vacuum Bubbles}},\ }\href {https://doi.org/10.1103/PhysRevD.35.1747}
  {\bibfield  {journal} {\bibinfo  {journal} {Phys. Rev. D}\ }\textbf {\bibinfo
  {volume} {35}},\ \bibinfo {pages} {1747} (\bibinfo {year}
  {1987})}\BibitemShut {NoStop}%
\bibitem [{\citenamefont {Maeda}(1986)}]{Maeda:1985ye}%
  \BibitemOpen
  \bibfield  {author} {\bibinfo {author} {\bibfnamefont {K.-i.}\ \bibnamefont
  {Maeda}},\ }\bibfield  {title} {\bibinfo {title} {{Bubble dynamics in the
  expanding universe}},\ }\href@noop {} {\bibfield  {journal} {\bibinfo
  {journal} {Gen. Rel. Grav.}\ }\textbf {\bibinfo {volume} {18}},\ \bibinfo
  {pages} {931} (\bibinfo {year} {1986})}\BibitemShut {NoStop}%
\bibitem [{\citenamefont {Tanahashi}\ and\ \citenamefont
  {Yoo}(2015)}]{Tanahashi:2014sma}%
  \BibitemOpen
  \bibfield  {author} {\bibinfo {author} {\bibfnamefont {N.}~\bibnamefont
  {Tanahashi}}\ and\ \bibinfo {author} {\bibfnamefont {C.-M.}\ \bibnamefont
  {Yoo}},\ }\bibfield  {title} {\bibinfo {title} {{Spherical Domain Wall
  Collapse in a Dust Universe}},\ }\href
  {https://doi.org/10.1088/0264-9381/32/15/155003} {\bibfield  {journal}
  {\bibinfo  {journal} {Class. Quant. Grav.}\ }\textbf {\bibinfo {volume}
  {32}},\ \bibinfo {pages} {155003} (\bibinfo {year} {2015})},\ \Eprint
  {https://arxiv.org/abs/1411.7479} {arXiv:1411.7479 [gr-qc]} \BibitemShut
  {NoStop}%
\bibitem [{\citenamefont {Stauffer}(1979)}]{Stauffer:1978kr}%
  \BibitemOpen
  \bibfield  {author} {\bibinfo {author} {\bibfnamefont {D.}~\bibnamefont
  {Stauffer}},\ }\bibfield  {title} {\bibinfo {title} {{Scaling theory of
  percolation clusters}},\ }\href
  {https://doi.org/10.1016/0370-1573(79)90060-7} {\bibfield  {journal}
  {\bibinfo  {journal} {Phys. Rept.}\ }\textbf {\bibinfo {volume} {54}},\
  \bibinfo {pages} {1} (\bibinfo {year} {1979})}\BibitemShut {NoStop}%
\bibitem [{\citenamefont {Essam}(1980)}]{Essam_1980}%
  \BibitemOpen
  \bibfield  {author} {\bibinfo {author} {\bibfnamefont {J.~W.}\ \bibnamefont
  {Essam}},\ }\bibfield  {title} {\bibinfo {title} {Percolation theory},\
  }\href {https://doi.org/10.1088/0034-4885/43/7/001} {\bibfield  {journal}
  {\bibinfo  {journal} {Reports on Progress in Physics}\ }\textbf {\bibinfo
  {volume} {43}},\ \bibinfo {pages} {833} (\bibinfo {year} {1980})}\BibitemShut
  {NoStop}%
\bibitem [{\citenamefont {Lalak}\ \emph {et~al.}(1995)\citenamefont {Lalak},
  \citenamefont {Ovrut},\ and\ \citenamefont {Thomas}}]{Lalak:1993bp}%
  \BibitemOpen
  \bibfield  {author} {\bibinfo {author} {\bibfnamefont {Z.}~\bibnamefont
  {Lalak}}, \bibinfo {author} {\bibfnamefont {B.~A.}\ \bibnamefont {Ovrut}},\
  and\ \bibinfo {author} {\bibfnamefont {S.}~\bibnamefont {Thomas}},\
  }\bibfield  {title} {\bibinfo {title} {{Large scale structure as a critical
  phenomenon}},\ }\href {https://doi.org/10.1103/PhysRevD.51.5456} {\bibfield
  {journal} {\bibinfo  {journal} {Phys. Rev. D}\ }\textbf {\bibinfo {volume}
  {51}},\ \bibinfo {pages} {5456} (\bibinfo {year} {1995})}\BibitemShut
  {NoStop}%
\bibitem [{\citenamefont {Chen}\ and\ \citenamefont
  {Huang}(2020)}]{Chen:2019irf}%
  \BibitemOpen
  \bibfield  {author} {\bibinfo {author} {\bibfnamefont {Z.-C.}\ \bibnamefont
  {Chen}}\ and\ \bibinfo {author} {\bibfnamefont {Q.-G.}\ \bibnamefont
  {Huang}},\ }\bibfield  {title} {\bibinfo {title} {{Distinguishing Primordial
  Black Holes from Astrophysical Black Holes by Einstein Telescope and Cosmic
  Explorer}},\ }\href {https://doi.org/10.1088/1475-7516/2020/08/039}
  {\bibfield  {journal} {\bibinfo  {journal} {JCAP}\ }\textbf {\bibinfo
  {volume} {08}},\ \bibinfo {pages} {039}},\ \Eprint
  {https://arxiv.org/abs/1904.02396} {arXiv:1904.02396 [astro-ph.CO]}
  \BibitemShut {NoStop}%
\bibitem [{\citenamefont {Pujolas}\ \emph {et~al.}(2021)\citenamefont
  {Pujolas}, \citenamefont {Vaskonen},\ and\ \citenamefont
  {Veerm\"ae}}]{Pujolas:2021yaw}%
  \BibitemOpen
  \bibfield  {author} {\bibinfo {author} {\bibfnamefont {O.}~\bibnamefont
  {Pujolas}}, \bibinfo {author} {\bibfnamefont {V.}~\bibnamefont {Vaskonen}},\
  and\ \bibinfo {author} {\bibfnamefont {H.}~\bibnamefont {Veerm\"ae}},\
  }\bibfield  {title} {\bibinfo {title} {{Prospects for probing gravitational
  waves from primordial black hole binaries}},\ }\href
  {https://doi.org/10.1103/PhysRevD.104.083521} {\bibfield  {journal} {\bibinfo
   {journal} {Phys. Rev. D}\ }\textbf {\bibinfo {volume} {104}},\ \bibinfo
  {pages} {083521} (\bibinfo {year} {2021})},\ \Eprint
  {https://arxiv.org/abs/2107.03379} {arXiv:2107.03379 [astro-ph.CO]}
  \BibitemShut {NoStop}%
\bibitem [{\citenamefont {Mena}\ \emph {et~al.}(2019)\citenamefont {Mena},
  \citenamefont {Palomares-Ruiz}, \citenamefont {Villanueva-Domingo},\ and\
  \citenamefont {Witte}}]{Mena:2019nhm}%
  \BibitemOpen
  \bibfield  {author} {\bibinfo {author} {\bibfnamefont {O.}~\bibnamefont
  {Mena}}, \bibinfo {author} {\bibfnamefont {S.}~\bibnamefont
  {Palomares-Ruiz}}, \bibinfo {author} {\bibfnamefont {P.}~\bibnamefont
  {Villanueva-Domingo}},\ and\ \bibinfo {author} {\bibfnamefont {S.~J.}\
  \bibnamefont {Witte}},\ }\bibfield  {title} {\bibinfo {title} {{Constraining
  the primordial black hole abundance with 21-cm cosmology}},\ }\href
  {https://doi.org/10.1103/PhysRevD.100.043540} {\bibfield  {journal} {\bibinfo
   {journal} {Phys. Rev. D}\ }\textbf {\bibinfo {volume} {100}},\ \bibinfo
  {pages} {043540} (\bibinfo {year} {2019})},\ \Eprint
  {https://arxiv.org/abs/1906.07735} {arXiv:1906.07735 [astro-ph.CO]}
  \BibitemShut {NoStop}%
\bibitem [{\citenamefont {Villanueva-Domingo}\ and\ \citenamefont
  {Ichiki}(2021)}]{Villanueva-Domingo:2021cgh}%
  \BibitemOpen
  \bibfield  {author} {\bibinfo {author} {\bibfnamefont {P.}~\bibnamefont
  {Villanueva-Domingo}}\ and\ \bibinfo {author} {\bibfnamefont
  {K.}~\bibnamefont {Ichiki}},\ }\bibfield  {title} {\bibinfo {title} {{21 cm
  Forest Constraints on Primordial Black Holes}}\ }\href
  {https://doi.org/10.1093/pasj/psab119} {10.1093/pasj/psab119} (\bibinfo
  {year} {2021}),\ \Eprint {https://arxiv.org/abs/2104.10695} {arXiv:2104.10695
  [astro-ph.CO]} \BibitemShut {NoStop}%
\bibitem [{\citenamefont {Villanueva-Domingo}\ \emph
  {et~al.}(2021)\citenamefont {Villanueva-Domingo}, \citenamefont {Mena},\ and\
  \citenamefont {Palomares-Ruiz}}]{Villanueva-Domingo:2021spv}%
  \BibitemOpen
  \bibfield  {author} {\bibinfo {author} {\bibfnamefont {P.}~\bibnamefont
  {Villanueva-Domingo}}, \bibinfo {author} {\bibfnamefont {O.}~\bibnamefont
  {Mena}},\ and\ \bibinfo {author} {\bibfnamefont {S.}~\bibnamefont
  {Palomares-Ruiz}},\ }\bibfield  {title} {\bibinfo {title} {{A brief review on
  primordial black holes as dark matter}},\ }\href
  {https://doi.org/10.3389/fspas.2021.681084} {\bibfield  {journal} {\bibinfo
  {journal} {Front. Astron. Space Sci.}\ }\textbf {\bibinfo {volume} {8}},\
  \bibinfo {pages} {87} (\bibinfo {year} {2021})},\ \Eprint
  {https://arxiv.org/abs/2103.12087} {arXiv:2103.12087 [astro-ph.CO]}
  \BibitemShut {NoStop}%
\bibitem [{\citenamefont {Chen}\ \emph {et~al.}(2023)\citenamefont {Chen},
  \citenamefont {Kongsore},\ and\ \citenamefont {Van~Tilburg}}]{Chen:2023xyj}%
  \BibitemOpen
  \bibfield  {author} {\bibinfo {author} {\bibfnamefont {I.-K.}\ \bibnamefont
  {Chen}}, \bibinfo {author} {\bibfnamefont {M.}~\bibnamefont {Kongsore}},\
  and\ \bibinfo {author} {\bibfnamefont {K.}~\bibnamefont {Van~Tilburg}},\
  }\bibfield  {title} {\bibinfo {title} {{Detecting Dark Compact Objects in
  Gaia DR4: A Data Analysis Pipeline for Transient Astrometric Lensing
  Searches}},\ }\href@noop {} {\  (\bibinfo {year} {2023})},\ \Eprint
  {https://arxiv.org/abs/2301.00822} {arXiv:2301.00822 [astro-ph.GA]}
  \BibitemShut {NoStop}%
\bibitem [{\citenamefont {Van~Tilburg}\ \emph {et~al.}(2018)\citenamefont
  {Van~Tilburg}, \citenamefont {Taki},\ and\ \citenamefont
  {Weiner}}]{VanTilburg:2018ykj}%
  \BibitemOpen
  \bibfield  {author} {\bibinfo {author} {\bibfnamefont {K.}~\bibnamefont
  {Van~Tilburg}}, \bibinfo {author} {\bibfnamefont {A.-M.}\ \bibnamefont
  {Taki}},\ and\ \bibinfo {author} {\bibfnamefont {N.}~\bibnamefont {Weiner}},\
  }\bibfield  {title} {\bibinfo {title} {{Halometry from Astrometry}},\ }\href
  {https://doi.org/10.1088/1475-7516/2018/07/041} {\bibfield  {journal}
  {\bibinfo  {journal} {JCAP}\ }\textbf {\bibinfo {volume} {07}},\ \bibinfo
  {pages} {041}},\ \Eprint {https://arxiv.org/abs/1804.01991} {arXiv:1804.01991
  [astro-ph.CO]} \BibitemShut {NoStop}%
\bibitem [{\citenamefont {Verma}\ and\ \citenamefont
  {Rentala}(2022)}]{Verma:2022pym}%
  \BibitemOpen
  \bibfield  {author} {\bibinfo {author} {\bibfnamefont {H.}~\bibnamefont
  {Verma}}\ and\ \bibinfo {author} {\bibfnamefont {V.}~\bibnamefont
  {Rentala}},\ }\bibfield  {title} {\bibinfo {title} {{Astrometric Microlensing
  of Primordial Black Holes with Gaia}},\ }\href@noop {} {\  (\bibinfo {year}
  {2022})},\ \Eprint {https://arxiv.org/abs/2208.14460} {arXiv:2208.14460
  [astro-ph.GA]} \BibitemShut {NoStop}%
\bibitem [{\citenamefont {Blasi}\ \emph
  {et~al.}(2023{\natexlab{b}})\citenamefont {Blasi}, \citenamefont {Mariotti},
  \citenamefont {Rase},\ and\ \citenamefont {Sevrin}}]{Blasi:2023sej}%
  \BibitemOpen
  \bibfield  {author} {\bibinfo {author} {\bibfnamefont {S.}~\bibnamefont
  {Blasi}}, \bibinfo {author} {\bibfnamefont {A.}~\bibnamefont {Mariotti}},
  \bibinfo {author} {\bibfnamefont {A.}~\bibnamefont {Rase}},\ and\ \bibinfo
  {author} {\bibfnamefont {A.}~\bibnamefont {Sevrin}},\ }\bibfield  {title}
  {\bibinfo {title} {{Axionic domain walls at Pulsar Timing Arrays: QCD bias
  and particle friction}},\ }\href@noop {} {\  (\bibinfo {year}
  {2023}{\natexlab{b}})},\ \Eprint {https://arxiv.org/abs/2306.17830}
  {arXiv:2306.17830 [hep-ph]} \BibitemShut {NoStop}%
\bibitem [{\citenamefont {Coulson}\ \emph {et~al.}(1996)\citenamefont
  {Coulson}, \citenamefont {Lalak},\ and\ \citenamefont
  {Ovrut}}]{Coulson:1995nv}%
  \BibitemOpen
  \bibfield  {author} {\bibinfo {author} {\bibfnamefont {D.}~\bibnamefont
  {Coulson}}, \bibinfo {author} {\bibfnamefont {Z.}~\bibnamefont {Lalak}},\
  and\ \bibinfo {author} {\bibfnamefont {B.~A.}\ \bibnamefont {Ovrut}},\
  }\bibfield  {title} {\bibinfo {title} {{Biased domain walls}},\ }\href
  {https://doi.org/10.1103/PhysRevD.53.4237} {\bibfield  {journal} {\bibinfo
  {journal} {Phys. Rev. D}\ }\textbf {\bibinfo {volume} {53}},\ \bibinfo
  {pages} {4237} (\bibinfo {year} {1996})}\BibitemShut {NoStop}%
\bibitem [{\citenamefont {Hindmarsh}(1996)}]{Hindmarsh:1996xv}%
  \BibitemOpen
  \bibfield  {author} {\bibinfo {author} {\bibfnamefont {M.}~\bibnamefont
  {Hindmarsh}},\ }\bibfield  {title} {\bibinfo {title} {{Analytic scaling
  solutions for cosmic domain walls}},\ }\href
  {https://doi.org/10.1103/PhysRevLett.77.4495} {\bibfield  {journal} {\bibinfo
   {journal} {Phys. Rev. Lett.}\ }\textbf {\bibinfo {volume} {77}},\ \bibinfo
  {pages} {4495} (\bibinfo {year} {1996})},\ \Eprint
  {https://arxiv.org/abs/hep-ph/9605332} {arXiv:hep-ph/9605332} \BibitemShut
  {NoStop}%
\bibitem [{\citenamefont {Pujolas}\ and\ \citenamefont
  {Zahariade}(2023)}]{Pujolas:2022qvs}%
  \BibitemOpen
  \bibfield  {author} {\bibinfo {author} {\bibfnamefont {O.}~\bibnamefont
  {Pujolas}}\ and\ \bibinfo {author} {\bibfnamefont {G.}~\bibnamefont
  {Zahariade}},\ }\bibfield  {title} {\bibinfo {title} {{Domain wall
  annihilation: A QFT perspective}},\ }\href
  {https://doi.org/10.1103/PhysRevD.107.123527} {\bibfield  {journal} {\bibinfo
   {journal} {Phys. Rev. D}\ }\textbf {\bibinfo {volume} {107}},\ \bibinfo
  {pages} {123527} (\bibinfo {year} {2023})},\ \Eprint
  {https://arxiv.org/abs/2212.11204} {arXiv:2212.11204 [hep-th]} \BibitemShut
  {NoStop}%
\bibitem [{\citenamefont {Babichev}\ \emph {et~al.}(2022)\citenamefont
  {Babichev}, \citenamefont {Gorbunov}, \citenamefont {Ramazanov},\ and\
  \citenamefont {Vikman}}]{Babichev:2021uvl}%
  \BibitemOpen
  \bibfield  {author} {\bibinfo {author} {\bibfnamefont {E.}~\bibnamefont
  {Babichev}}, \bibinfo {author} {\bibfnamefont {D.}~\bibnamefont {Gorbunov}},
  \bibinfo {author} {\bibfnamefont {S.}~\bibnamefont {Ramazanov}},\ and\
  \bibinfo {author} {\bibfnamefont {A.}~\bibnamefont {Vikman}},\ }\bibfield
  {title} {\bibinfo {title} {{Gravitational shine of dark domain walls}},\
  }\href {https://doi.org/10.1088/1475-7516/2022/04/028} {\bibfield  {journal}
  {\bibinfo  {journal} {JCAP}\ }\textbf {\bibinfo {volume} {04}}\bibfield
  {number} {\bibinfo  {number} { (04)},\ \bibinfo {pages} {028}},\ }\Eprint
  {https://arxiv.org/abs/2112.12608} {arXiv:2112.12608 [hep-ph]} \BibitemShut
  {NoStop}%
\bibitem [{\citenamefont {Ramazanov}\ \emph {et~al.}(2022)\citenamefont
  {Ramazanov}, \citenamefont {Babichev}, \citenamefont {Gorbunov},\ and\
  \citenamefont {Vikman}}]{Ramazanov:2021eya}%
  \BibitemOpen
  \bibfield  {author} {\bibinfo {author} {\bibfnamefont {S.}~\bibnamefont
  {Ramazanov}}, \bibinfo {author} {\bibfnamefont {E.}~\bibnamefont {Babichev}},
  \bibinfo {author} {\bibfnamefont {D.}~\bibnamefont {Gorbunov}},\ and\
  \bibinfo {author} {\bibfnamefont {A.}~\bibnamefont {Vikman}},\ }\bibfield
  {title} {\bibinfo {title} {{Beyond freeze-in: Dark matter via inverse phase
  transition and gravitational wave signal}},\ }\href
  {https://doi.org/10.1103/PhysRevD.105.063530} {\bibfield  {journal} {\bibinfo
   {journal} {Phys. Rev. D}\ }\textbf {\bibinfo {volume} {105}},\ \bibinfo
  {pages} {063530} (\bibinfo {year} {2022})},\ \Eprint
  {https://arxiv.org/abs/2104.13722} {arXiv:2104.13722 [hep-ph]} \BibitemShut
  {NoStop}%
\bibitem [{\citenamefont {Babichev}\ \emph {et~al.}(2023)\citenamefont
  {Babichev}, \citenamefont {Gorbunov}, \citenamefont {Ramazanov},
  \citenamefont {Samanta},\ and\ \citenamefont {Vikman}}]{Babichev:2023pbf}%
  \BibitemOpen
  \bibfield  {author} {\bibinfo {author} {\bibfnamefont {E.}~\bibnamefont
  {Babichev}}, \bibinfo {author} {\bibfnamefont {D.}~\bibnamefont {Gorbunov}},
  \bibinfo {author} {\bibfnamefont {S.}~\bibnamefont {Ramazanov}}, \bibinfo
  {author} {\bibfnamefont {R.}~\bibnamefont {Samanta}},\ and\ \bibinfo {author}
  {\bibfnamefont {A.}~\bibnamefont {Vikman}},\ }\bibfield  {title} {\bibinfo
  {title} {{NANOGrav spectral index $\gamma=3$ from melting domain walls}},\
  }\href@noop {} {\  (\bibinfo {year} {2023})},\ \Eprint
  {https://arxiv.org/abs/2307.04582} {arXiv:2307.04582 [hep-ph]} \BibitemShut
  {NoStop}%
\bibitem [{\citenamefont {Agazie}\ \emph
  {et~al.}(2023{\natexlab{d}})\citenamefont {Agazie} \emph
  {et~al.}}]{NANOGrav:2023ctt}%
  \BibitemOpen
  \bibfield  {author} {\bibinfo {author} {\bibfnamefont {G.}~\bibnamefont
  {Agazie}} \emph {et~al.} (\bibinfo {collaboration} {NANOGrav}),\ }\bibfield
  {title} {\bibinfo {title} {{The NANOGrav 15 yr Data Set: Detector
  Characterization and Noise Budget}},\ }\href
  {https://doi.org/10.3847/2041-8213/acda88} {\bibfield  {journal} {\bibinfo
  {journal} {Astrophys. J. Lett.}\ }\textbf {\bibinfo {volume} {951}},\
  \bibinfo {pages} {L10} (\bibinfo {year} {2023}{\natexlab{d}})},\ \Eprint
  {https://arxiv.org/abs/2306.16218} {arXiv:2306.16218 [astro-ph.HE]}
  \BibitemShut {NoStop}%
\bibitem [{\citenamefont {Antoniadis}\ \emph
  {et~al.}(2023{\natexlab{d}})\citenamefont {Antoniadis} \emph
  {et~al.}}]{EPTA:2023sfo}%
  \BibitemOpen
  \bibfield  {author} {\bibinfo {author} {\bibfnamefont {J.}~\bibnamefont
  {Antoniadis}} \emph {et~al.} (\bibinfo {collaboration} {EPTA}),\ }\bibfield
  {title} {\bibinfo {title} {{The second data release from the European Pulsar
  Timing Array - I. The dataset and timing analysis}},\ }\href
  {https://doi.org/10.1051/0004-6361/202346841} {\bibfield  {journal} {\bibinfo
   {journal} {Astron. Astrophys.}\ }\textbf {\bibinfo {volume} {678}},\
  \bibinfo {pages} {A48} (\bibinfo {year} {2023}{\natexlab{d}})},\ \Eprint
  {https://arxiv.org/abs/2306.16224} {arXiv:2306.16224 [astro-ph.HE]}
  \BibitemShut {NoStop}%
\bibitem [{\citenamefont {Zic}\ \emph {et~al.}(2023)\citenamefont {Zic} \emph
  {et~al.}}]{Zic:2023gta}%
  \BibitemOpen
  \bibfield  {author} {\bibinfo {author} {\bibfnamefont {A.}~\bibnamefont
  {Zic}} \emph {et~al.},\ }\bibfield  {title} {\bibinfo {title} {{The Parkes
  Pulsar Timing Array third data release}},\ }\href
  {https://doi.org/10.1017/pasa.2023.36} {\bibfield  {journal} {\bibinfo
  {journal} {Publ. Astron. Soc. Austral.}\ }\textbf {\bibinfo {volume} {40}},\
  \bibinfo {pages} {e049} (\bibinfo {year} {2023})},\ \Eprint
  {https://arxiv.org/abs/2306.16230} {arXiv:2306.16230 [astro-ph.HE]}
  \BibitemShut {NoStop}%
\bibitem [{\citenamefont {Sakharov}\ \emph {et~al.}(2021)\citenamefont
  {Sakharov}, \citenamefont {Eroshenko},\ and\ \citenamefont
  {Rubin}}]{Sakharov:2021dim}%
  \BibitemOpen
  \bibfield  {author} {\bibinfo {author} {\bibfnamefont {A.~S.}\ \bibnamefont
  {Sakharov}}, \bibinfo {author} {\bibfnamefont {Y.~N.}\ \bibnamefont
  {Eroshenko}},\ and\ \bibinfo {author} {\bibfnamefont {S.~G.}\ \bibnamefont
  {Rubin}},\ }\bibfield  {title} {\bibinfo {title} {{Looking at the NANOGrav
  signal through the anthropic window of axionlike particles}},\ }\href
  {https://doi.org/10.1103/PhysRevD.104.043005} {\bibfield  {journal} {\bibinfo
   {journal} {Phys. Rev. D}\ }\textbf {\bibinfo {volume} {104}},\ \bibinfo
  {pages} {043005} (\bibinfo {year} {2021})},\ \Eprint
  {https://arxiv.org/abs/2104.08750} {arXiv:2104.08750 [hep-ph]} \BibitemShut
  {NoStop}%
\bibitem [{\citenamefont {Zambujal~Ferreira}\ \emph {et~al.}(2022)\citenamefont
  {Zambujal~Ferreira}, \citenamefont {Notari}, \citenamefont {Pujol\`as},\ and\
  \citenamefont {Rompineve}}]{ZambujalFerreira:2021cte}%
  \BibitemOpen
  \bibfield  {author} {\bibinfo {author} {\bibfnamefont {R.}~\bibnamefont
  {Zambujal~Ferreira}}, \bibinfo {author} {\bibfnamefont {A.}~\bibnamefont
  {Notari}}, \bibinfo {author} {\bibfnamefont {O.}~\bibnamefont {Pujol\`as}},\
  and\ \bibinfo {author} {\bibfnamefont {F.}~\bibnamefont {Rompineve}},\
  }\bibfield  {title} {\bibinfo {title} {{High Quality QCD Axion at
  Gravitational Wave Observatories}},\ }\href
  {https://doi.org/10.1103/PhysRevLett.128.141101} {\bibfield  {journal}
  {\bibinfo  {journal} {Phys. Rev. Lett.}\ }\textbf {\bibinfo {volume} {128}},\
  \bibinfo {pages} {141101} (\bibinfo {year} {2022})},\ \Eprint
  {https://arxiv.org/abs/2107.07542} {arXiv:2107.07542 [hep-ph]} \BibitemShut
  {NoStop}%
\bibitem [{\citenamefont {Bai}\ \emph {et~al.}(2023)\citenamefont {Bai},
  \citenamefont {Chen},\ and\ \citenamefont {Korwar}}]{Bai:2023cqj}%
  \BibitemOpen
  \bibfield  {author} {\bibinfo {author} {\bibfnamefont {Y.}~\bibnamefont
  {Bai}}, \bibinfo {author} {\bibfnamefont {T.-K.}\ \bibnamefont {Chen}},\ and\
  \bibinfo {author} {\bibfnamefont {M.}~\bibnamefont {Korwar}},\ }\bibfield
  {title} {\bibinfo {title} {{QCD-Collapsed Domain Walls: QCD Phase Transition
  and Gravitational Wave Spectroscopy}},\ }\href@noop {} {\  (\bibinfo {year}
  {2023})},\ \Eprint {https://arxiv.org/abs/2306.17160} {arXiv:2306.17160
  [hep-ph]} \BibitemShut {NoStop}%
\bibitem [{\citenamefont {Lu}\ and\ \citenamefont {Chiang}(2023)}]{Lu:2023mcz}%
  \BibitemOpen
  \bibfield  {author} {\bibinfo {author} {\bibfnamefont {B.-Q.}\ \bibnamefont
  {Lu}}\ and\ \bibinfo {author} {\bibfnamefont {C.-W.}\ \bibnamefont
  {Chiang}},\ }\bibfield  {title} {\bibinfo {title} {{Nano-Hertz stochastic
  gravitational wave background from domain wall annihilation}},\ }\href@noop
  {} {\  (\bibinfo {year} {2023})},\ \Eprint {https://arxiv.org/abs/2307.00746}
  {arXiv:2307.00746 [hep-ph]} \BibitemShut {NoStop}%
\bibitem [{\citenamefont {Li}(2023)}]{Li:2023tdx}%
  \BibitemOpen
  \bibfield  {author} {\bibinfo {author} {\bibfnamefont {X.-F.}\ \bibnamefont
  {Li}},\ }\bibfield  {title} {\bibinfo {title} {{Probing the high temperature
  symmetry breaking with gravitational waves from domain walls}},\ }\href@noop
  {} {\  (\bibinfo {year} {2023})},\ \Eprint {https://arxiv.org/abs/2307.03163}
  {arXiv:2307.03163 [hep-ph]} \BibitemShut {NoStop}%
\bibitem [{\citenamefont {Guo}\ \emph {et~al.}(2023)\citenamefont {Guo},
  \citenamefont {Khlopov}, \citenamefont {Liu}, \citenamefont {Wu},
  \citenamefont {Wu},\ and\ \citenamefont {Zhu}}]{Guo:2023hyp}%
  \BibitemOpen
  \bibfield  {author} {\bibinfo {author} {\bibfnamefont {S.-Y.}\ \bibnamefont
  {Guo}}, \bibinfo {author} {\bibfnamefont {M.}~\bibnamefont {Khlopov}},
  \bibinfo {author} {\bibfnamefont {X.}~\bibnamefont {Liu}}, \bibinfo {author}
  {\bibfnamefont {L.}~\bibnamefont {Wu}}, \bibinfo {author} {\bibfnamefont
  {Y.}~\bibnamefont {Wu}},\ and\ \bibinfo {author} {\bibfnamefont
  {B.}~\bibnamefont {Zhu}},\ }\bibfield  {title} {\bibinfo {title} {{Footprints
  of Axion-Like Particle in Pulsar Timing Array Data and JWST Observations}},\
  }\href@noop {} {\  (\bibinfo {year} {2023})},\ \Eprint
  {https://arxiv.org/abs/2306.17022} {arXiv:2306.17022 [hep-ph]} \BibitemShut
  {NoStop}%
\bibitem [{\citenamefont {King}\ \emph {et~al.}(2023)\citenamefont {King},
  \citenamefont {Marfatia},\ and\ \citenamefont {Rahat}}]{King:2023cgv}%
  \BibitemOpen
  \bibfield  {author} {\bibinfo {author} {\bibfnamefont {S.~F.}\ \bibnamefont
  {King}}, \bibinfo {author} {\bibfnamefont {D.}~\bibnamefont {Marfatia}},\
  and\ \bibinfo {author} {\bibfnamefont {M.~H.}\ \bibnamefont {Rahat}},\
  }\bibfield  {title} {\bibinfo {title} {{Towards distinguishing Dirac from
  Majorana neutrino mass with gravitational waves}},\ }\href@noop {} {\
  (\bibinfo {year} {2023})},\ \Eprint {https://arxiv.org/abs/2306.05389}
  {arXiv:2306.05389 [hep-ph]} \BibitemShut {NoStop}%
\bibitem [{\citenamefont {Allen}\ and\ \citenamefont
  {Romano}(1999)}]{Allen:1997ad}%
  \BibitemOpen
  \bibfield  {author} {\bibinfo {author} {\bibfnamefont {B.}~\bibnamefont
  {Allen}}\ and\ \bibinfo {author} {\bibfnamefont {J.~D.}\ \bibnamefont
  {Romano}},\ }\bibfield  {title} {\bibinfo {title} {{Detecting a stochastic
  background of gravitational radiation: Signal processing strategies and
  sensitivities}},\ }\href {https://doi.org/10.1103/PhysRevD.59.102001}
  {\bibfield  {journal} {\bibinfo  {journal} {Phys. Rev. D}\ }\textbf {\bibinfo
  {volume} {59}},\ \bibinfo {pages} {102001} (\bibinfo {year} {1999})},\
  \Eprint {https://arxiv.org/abs/gr-qc/9710117} {arXiv:gr-qc/9710117}
  \BibitemShut {NoStop}%
\bibitem [{\citenamefont {Anholm}\ \emph {et~al.}(2009)\citenamefont {Anholm},
  \citenamefont {Ballmer}, \citenamefont {Creighton}, \citenamefont {Price},\
  and\ \citenamefont {Siemens}}]{Anholm:2008wy}%
  \BibitemOpen
  \bibfield  {author} {\bibinfo {author} {\bibfnamefont {M.}~\bibnamefont
  {Anholm}}, \bibinfo {author} {\bibfnamefont {S.}~\bibnamefont {Ballmer}},
  \bibinfo {author} {\bibfnamefont {J.~D.~E.}\ \bibnamefont {Creighton}},
  \bibinfo {author} {\bibfnamefont {L.~R.}\ \bibnamefont {Price}},\ and\
  \bibinfo {author} {\bibfnamefont {X.}~\bibnamefont {Siemens}},\ }\bibfield
  {title} {\bibinfo {title} {{Optimal strategies for gravitational wave
  stochastic background searches in pulsar timing data}},\ }\href
  {https://doi.org/10.1103/PhysRevD.79.084030} {\bibfield  {journal} {\bibinfo
  {journal} {Phys. Rev. D}\ }\textbf {\bibinfo {volume} {79}},\ \bibinfo
  {pages} {084030} (\bibinfo {year} {2009})},\ \Eprint
  {https://arxiv.org/abs/0809.0701} {arXiv:0809.0701 [gr-qc]} \BibitemShut
  {NoStop}%
\bibitem [{\citenamefont {Maggiore}(2018)}]{Maggiore:2018sht}%
  \BibitemOpen
  \bibfield  {author} {\bibinfo {author} {\bibfnamefont {M.}~\bibnamefont
  {Maggiore}},\ }\href@noop {} {\emph {\bibinfo {title} {{Gravitational Waves.
  Vol. 2: Astrophysics and Cosmology}}}}\ (\bibinfo  {publisher} {Oxford
  University Press},\ \bibinfo {year} {2018})\BibitemShut {NoStop}%
\bibitem [{\citenamefont {Hellings}\ and\ \citenamefont
  {Downs}(1983)}]{Hellings:1983fr}%
  \BibitemOpen
  \bibfield  {author} {\bibinfo {author} {\bibfnamefont {R.~w.}\ \bibnamefont
  {Hellings}}\ and\ \bibinfo {author} {\bibfnamefont {G.~s.}\ \bibnamefont
  {Downs}},\ }\bibfield  {title} {\bibinfo {title} {{UPPER LIMITS ON THE
  ISOTROPIC GRAVITATIONAL RADIATION BACKGROUND FROM PULSAR TIMING ANALYSIS}},\
  }\href {https://doi.org/10.1086/183954} {\bibfield  {journal} {\bibinfo
  {journal} {Astrophys. J. Lett.}\ }\textbf {\bibinfo {volume} {265}},\
  \bibinfo {pages} {L39} (\bibinfo {year} {1983})}\BibitemShut {NoStop}%
\bibitem [{\citenamefont {Hobbs}\ \emph {et~al.}(2006)\citenamefont {Hobbs},
  \citenamefont {Edwards},\ and\ \citenamefont {Manchester}}]{Hobbs:2006cd}%
  \BibitemOpen
  \bibfield  {author} {\bibinfo {author} {\bibfnamefont {G.}~\bibnamefont
  {Hobbs}}, \bibinfo {author} {\bibfnamefont {R.}~\bibnamefont {Edwards}},\
  and\ \bibinfo {author} {\bibfnamefont {R.}~\bibnamefont {Manchester}},\
  }\bibfield  {title} {\bibinfo {title} {{Tempo2, a new pulsar timing package.
  1. overview}},\ }\href {https://doi.org/10.1111/j.1365-2966.2006.10302.x}
  {\bibfield  {journal} {\bibinfo  {journal} {Mon. Not. Roy. Astron. Soc.}\
  }\textbf {\bibinfo {volume} {369}},\ \bibinfo {pages} {655} (\bibinfo {year}
  {2006})},\ \Eprint {https://arxiv.org/abs/astro-ph/0603381}
  {arXiv:astro-ph/0603381} \BibitemShut {NoStop}%
\bibitem [{\citenamefont {Edwards}\ \emph {et~al.}(2006)\citenamefont
  {Edwards}, \citenamefont {Hobbs},\ and\ \citenamefont
  {Manchester}}]{Edwards:2006zg}%
  \BibitemOpen
  \bibfield  {author} {\bibinfo {author} {\bibfnamefont {R.~T.}\ \bibnamefont
  {Edwards}}, \bibinfo {author} {\bibfnamefont {G.~B.}\ \bibnamefont {Hobbs}},\
  and\ \bibinfo {author} {\bibfnamefont {R.~N.}\ \bibnamefont {Manchester}},\
  }\bibfield  {title} {\bibinfo {title} {{Tempo2, a new pulsar timing package.
  2. The timing model and precision estimates}},\ }\href
  {https://doi.org/10.1111/j.1365-2966.2006.10870.x} {\bibfield  {journal}
  {\bibinfo  {journal} {Mon. Not. Roy. Astron. Soc.}\ }\textbf {\bibinfo
  {volume} {372}},\ \bibinfo {pages} {1549} (\bibinfo {year} {2006})},\ \Eprint
  {https://arxiv.org/abs/astro-ph/0607664} {arXiv:astro-ph/0607664}
  \BibitemShut {NoStop}%
\bibitem [{\citenamefont {Hobbs}\ \emph {et~al.}(2009)\citenamefont {Hobbs},
  \citenamefont {Jenet}, \citenamefont {Lee}, \citenamefont {Verbiest},
  \citenamefont {Yardley}, \citenamefont {Manchester}, \citenamefont {Lommen},
  \citenamefont {Coles}, \citenamefont {Edwards},\ and\ \citenamefont
  {Shettigara}}]{Hobbs:2009yn}%
  \BibitemOpen
  \bibfield  {author} {\bibinfo {author} {\bibfnamefont {G.}~\bibnamefont
  {Hobbs}}, \bibinfo {author} {\bibfnamefont {F.}~\bibnamefont {Jenet}},
  \bibinfo {author} {\bibfnamefont {K.~J.}\ \bibnamefont {Lee}}, \bibinfo
  {author} {\bibfnamefont {J.~P.~W.}\ \bibnamefont {Verbiest}}, \bibinfo
  {author} {\bibfnamefont {D.}~\bibnamefont {Yardley}}, \bibinfo {author}
  {\bibfnamefont {R.}~\bibnamefont {Manchester}}, \bibinfo {author}
  {\bibfnamefont {A.}~\bibnamefont {Lommen}}, \bibinfo {author} {\bibfnamefont
  {W.}~\bibnamefont {Coles}}, \bibinfo {author} {\bibfnamefont
  {R.}~\bibnamefont {Edwards}},\ and\ \bibinfo {author} {\bibfnamefont
  {C.}~\bibnamefont {Shettigara}},\ }\bibfield  {title} {\bibinfo {title}
  {{TEMPO2, a new pulsar timing package. III: Gravitational wave simulation}},\
  }\href {https://doi.org/10.1111/j.1365-2966.2009.14391.x} {\bibfield
  {journal} {\bibinfo  {journal} {Mon. Not. Roy. Astron. Soc.}\ }\textbf
  {\bibinfo {volume} {394}},\ \bibinfo {pages} {1945} (\bibinfo {year}
  {2009})},\ \Eprint {https://arxiv.org/abs/0901.0592} {arXiv:0901.0592
  [astro-ph.SR]} \BibitemShut {NoStop}%
\bibitem [{\citenamefont {{Hobbs}}\ and\ \citenamefont
  {{Edwards}}(2012)}]{2012ascl.soft10015H}%
  \BibitemOpen
  \bibfield  {author} {\bibinfo {author} {\bibfnamefont {G.}~\bibnamefont
  {{Hobbs}}}\ and\ \bibinfo {author} {\bibfnamefont {R.}~\bibnamefont
  {{Edwards}}},\ }\href@noop {} {\bibinfo {title} {{Tempo2: Pulsar Timing
  Package}}},\ \bibinfo {howpublished} {Astrophysics Source Code Library,
  record ascl:1210.015} (\bibinfo {year} {2012}),\ \Eprint
  {https://arxiv.org/abs/1210.015} {ascl:1210.015} \BibitemShut {NoStop}%
\bibitem [{\citenamefont {Luo}\ \emph {et~al.}(2021)\citenamefont {Luo} \emph
  {et~al.}}]{Luo:2020ksx}%
  \BibitemOpen
  \bibfield  {author} {\bibinfo {author} {\bibfnamefont {J.}~\bibnamefont
  {Luo}} \emph {et~al.},\ }\bibfield  {title} {\bibinfo {title} {{PINT: A
  Modern Software Package for Pulsar Timing}},\ }\href
  {https://doi.org/10.3847/1538-4357/abe62f} {\bibfield  {journal} {\bibinfo
  {journal} {Astrophys. J.}\ }\textbf {\bibinfo {volume} {911}},\ \bibinfo
  {pages} {45} (\bibinfo {year} {2021})},\ \Eprint
  {https://arxiv.org/abs/2012.00074} {arXiv:2012.00074 [astro-ph.IM]}
  \BibitemShut {NoStop}%
\bibitem [{\citenamefont {Ellis}\ \emph {et~al.}(2020)\citenamefont {Ellis},
  \citenamefont {Vallisneri}, \citenamefont {Taylor},\ and\ \citenamefont
  {Baker}}]{enterprise}%
  \BibitemOpen
  \bibfield  {author} {\bibinfo {author} {\bibfnamefont {J.~A.}\ \bibnamefont
  {Ellis}}, \bibinfo {author} {\bibfnamefont {M.}~\bibnamefont {Vallisneri}},
  \bibinfo {author} {\bibfnamefont {S.~R.}\ \bibnamefont {Taylor}},\ and\
  \bibinfo {author} {\bibfnamefont {P.~T.}\ \bibnamefont {Baker}},\ }\href
  {https://doi.org/10.5281/zenodo.4059815} {\bibinfo {title} {Enterprise:
  Enhanced numerical toolbox enabling a robust pulsar inference suite}},\
  \bibinfo {howpublished} {Zenodo} (\bibinfo {year} {2020})\BibitemShut
  {NoStop}%
\bibitem [{\citenamefont {Arzoumanian}\ \emph {et~al.}(2015)\citenamefont
  {Arzoumanian} \emph {et~al.}}]{NANOGrav:2015qfw}%
  \BibitemOpen
  \bibfield  {author} {\bibinfo {author} {\bibfnamefont {Z.}~\bibnamefont
  {Arzoumanian}} \emph {et~al.} (\bibinfo {collaboration} {NANOGrav}),\
  }\bibfield  {title} {\bibinfo {title} {{The NANOGrav Nine-year Data Set:
  Observations, Arrival Time Measurements, and Analysis of 37 Millisecond
  Pulsars}},\ }\href {https://doi.org/10.1088/0004-637X/813/1/65} {\bibfield
  {journal} {\bibinfo  {journal} {Astrophys. J.}\ }\textbf {\bibinfo {volume}
  {813}},\ \bibinfo {pages} {65} (\bibinfo {year} {2015})},\ \Eprint
  {https://arxiv.org/abs/1505.07540} {arXiv:1505.07540 [astro-ph.IM]}
  \BibitemShut {NoStop}%
\bibitem [{\citenamefont {Arzoumanian}\ \emph {et~al.}(2016)\citenamefont
  {Arzoumanian} \emph {et~al.}}]{NANOGrav:2015aud}%
  \BibitemOpen
  \bibfield  {author} {\bibinfo {author} {\bibfnamefont {Z.}~\bibnamefont
  {Arzoumanian}} \emph {et~al.} (\bibinfo {collaboration} {NANOGrav}),\
  }\bibfield  {title} {\bibinfo {title} {{The NANOGrav Nine-year Data Set:
  Limits on the Isotropic Stochastic Gravitational Wave Background}},\ }\href
  {https://doi.org/10.3847/0004-637X/821/1/13} {\bibfield  {journal} {\bibinfo
  {journal} {Astrophys. J.}\ }\textbf {\bibinfo {volume} {821}},\ \bibinfo
  {pages} {13} (\bibinfo {year} {2016})},\ \Eprint
  {https://arxiv.org/abs/1508.03024} {arXiv:1508.03024 [astro-ph.GA]}
  \BibitemShut {NoStop}%
\bibitem [{\citenamefont {Taylor}(2021)}]{Taylor:2021yjx}%
  \BibitemOpen
  \bibfield  {author} {\bibinfo {author} {\bibfnamefont {S.~R.}\ \bibnamefont
  {Taylor}},\ }\bibfield  {title} {\bibinfo {title} {{The Nanohertz
  Gravitational Wave Astronomer}},\ }\href@noop {} {\  (\bibinfo {year}
  {2021})},\ \Eprint {https://arxiv.org/abs/2105.13270} {arXiv:2105.13270
  [astro-ph.HE]} \BibitemShut {NoStop}%
\bibitem [{\citenamefont {Taylor}\ \emph {et~al.}(2021)\citenamefont {Taylor},
  \citenamefont {Baker}, \citenamefont {Hazboun}, \citenamefont {Simon},\ and\
  \citenamefont {Vigeland}}]{enterprise_ext}%
  \BibitemOpen
  \bibfield  {author} {\bibinfo {author} {\bibfnamefont {S.~R.}\ \bibnamefont
  {Taylor}}, \bibinfo {author} {\bibfnamefont {P.~T.}\ \bibnamefont {Baker}},
  \bibinfo {author} {\bibfnamefont {J.~S.}\ \bibnamefont {Hazboun}}, \bibinfo
  {author} {\bibfnamefont {J.}~\bibnamefont {Simon}},\ and\ \bibinfo {author}
  {\bibfnamefont {S.~J.}\ \bibnamefont {Vigeland}},\ }\href
  {https://github.com/nanograv/enterprise_extensions} {\bibinfo {title}
  {enterprise\_extensions}} (\bibinfo {year} {2021}),\ \bibinfo {note}
  {v2.3.3}\BibitemShut {NoStop}%
\bibitem [{\citenamefont {Ellis}\ and\ \citenamefont {van
  Haasteren}(2017)}]{justin_ellis_2017_1037579}%
  \BibitemOpen
  \bibfield  {author} {\bibinfo {author} {\bibfnamefont {J.}~\bibnamefont
  {Ellis}}\ and\ \bibinfo {author} {\bibfnamefont {R.}~\bibnamefont {van
  Haasteren}},\ }\href {https://doi.org/10.5281/zenodo.1037579} {\bibinfo
  {title} {jellis18/ptmcmcsampler: Official release}} (\bibinfo {year}
  {2017})\BibitemShut {NoStop}%
\bibitem [{\citenamefont {Lewis}(2019)}]{Lewis:2019xzd}%
  \BibitemOpen
  \bibfield  {author} {\bibinfo {author} {\bibfnamefont {A.}~\bibnamefont
  {Lewis}},\ }\bibfield  {title} {\bibinfo {title} {{GetDist: a Python package
  for analysing Monte Carlo samples}},\ }\href {https://getdist.readthedocs.io}
  {\  (\bibinfo {year} {2019})},\ \Eprint {https://arxiv.org/abs/1910.13970}
  {arXiv:1910.13970 [astro-ph.IM]} \BibitemShut {NoStop}%
\bibitem [{\citenamefont {Lamb}\ \emph {et~al.}(2023)\citenamefont {Lamb},
  \citenamefont {Taylor},\ and\ \citenamefont {van Haasteren}}]{Lamb:2023jls}%
  \BibitemOpen
  \bibfield  {author} {\bibinfo {author} {\bibfnamefont {W.~G.}\ \bibnamefont
  {Lamb}}, \bibinfo {author} {\bibfnamefont {S.~R.}\ \bibnamefont {Taylor}},\
  and\ \bibinfo {author} {\bibfnamefont {R.}~\bibnamefont {van Haasteren}},\
  }\bibfield  {title} {\bibinfo {title} {{The Need For Speed: Rapid Refitting
  Techniques for Bayesian Spectral Characterization of the Gravitational Wave
  Background Using PTAs}},\ }\href@noop {} {\  (\bibinfo {year} {2023})},\
  \Eprint {https://arxiv.org/abs/2303.15442} {arXiv:2303.15442 [astro-ph.HE]}
  \BibitemShut {NoStop}%
\bibitem [{\citenamefont {Lamb}\ and\ \citenamefont {al.}()}]{ceffylgit}%
  \BibitemOpen
  \bibfield  {author} {\bibinfo {author} {\bibfnamefont {W.~G.}\ \bibnamefont
  {Lamb}}\ and\ \bibinfo {author} {\bibnamefont {al.}},\ }\href@noop {}
  {\bibinfo  {journal} {\href{https://github.com/astrolamb/ceffyl}{\tt
  https://github.com/astrolamb/ceffyl}}\ }\BibitemShut {NoStop}%
\bibitem [{\citenamefont {{Jeffreys}}(1939)}]{1939thpr.book.....J}%
  \BibitemOpen
\bibfield  {journal} {  }\bibfield  {author} {\bibinfo {author} {\bibfnamefont
  {H.}~\bibnamefont {{Jeffreys}}},\ }\href@noop {} {\emph {\bibinfo {title}
  {{Theory of Probability}}}}\ (\bibinfo {year} {1939})\BibitemShut {NoStop}%
\bibitem [{\citenamefont {Kass}\ and\ \citenamefont
  {Raftery}(1995)}]{kass1995bayes}%
  \BibitemOpen
  \bibfield  {author} {\bibinfo {author} {\bibfnamefont {R.~E.}\ \bibnamefont
  {Kass}}\ and\ \bibinfo {author} {\bibfnamefont {A.~E.}\ \bibnamefont
  {Raftery}},\ }\bibfield  {title} {\bibinfo {title} {Bayes factors},\
  }\href@noop {} {\bibfield  {journal} {\bibinfo  {journal} {Journal of the
  american statistical association}\ }\textbf {\bibinfo {volume} {90}},\
  \bibinfo {pages} {773} (\bibinfo {year} {1995})}\BibitemShut {NoStop}%
\bibitem [{\citenamefont {Antoniadis}\ \emph
  {et~al.}(2023{\natexlab{e}})\citenamefont {Antoniadis} \emph
  {et~al.}}]{EPTA:2023fyk}%
  \BibitemOpen
  \bibfield  {author} {\bibinfo {author} {\bibfnamefont {J.}~\bibnamefont
  {Antoniadis}} \emph {et~al.} (\bibinfo {collaboration} {EPTA, InPTA:}),\
  }\bibfield  {title} {\bibinfo {title} {{The second data release from the
  European Pulsar Timing Array - III. Search for gravitational wave signals}},\
  }\href {https://doi.org/10.1051/0004-6361/202346844} {\bibfield  {journal}
  {\bibinfo  {journal} {Astron. Astrophys.}\ }\textbf {\bibinfo {volume}
  {678}},\ \bibinfo {pages} {A50} (\bibinfo {year} {2023}{\natexlab{e}})},\
  \Eprint {https://arxiv.org/abs/2306.16214} {arXiv:2306.16214 [astro-ph.HE]}
  \BibitemShut {NoStop}%
\bibitem [{\citenamefont {Workman}\ \emph {et~al.}(2022)\citenamefont {Workman}
  \emph {et~al.}}]{ParticleDataGroup:2022pth}%
  \BibitemOpen
  \bibfield  {author} {\bibinfo {author} {\bibfnamefont {R.~L.}\ \bibnamefont
  {Workman}} \emph {et~al.} (\bibinfo {collaboration} {Particle Data Group}),\
  }\bibfield  {title} {\bibinfo {title} {{Review of Particle Physics}},\ }\href
  {https://doi.org/10.1093/ptep/ptac097} {\bibfield  {journal} {\bibinfo
  {journal} {PTEP}\ }\textbf {\bibinfo {volume} {2022}},\ \bibinfo {pages}
  {083C01} (\bibinfo {year} {2022})}\BibitemShut {NoStop}%
\bibitem [{\citenamefont {Durrer}\ and\ \citenamefont
  {Caprini}(2003)}]{Durrer:2003ja}%
  \BibitemOpen
  \bibfield  {author} {\bibinfo {author} {\bibfnamefont {R.}~\bibnamefont
  {Durrer}}\ and\ \bibinfo {author} {\bibfnamefont {C.}~\bibnamefont
  {Caprini}},\ }\bibfield  {title} {\bibinfo {title} {{Primordial Magnetic
  Fields and Causality}},\ }\href
  {https://doi.org/10.1088/1475-7516/2003/11/010} {\bibfield  {journal}
  {\bibinfo  {journal} {JCAP}\ }\textbf {\bibinfo {volume} {11}},\ \bibinfo
  {pages} {010}},\ \Eprint {https://arxiv.org/abs/astro-ph/0305059}
  {arXiv:astro-ph/0305059} \BibitemShut {NoStop}%
\bibitem [{\citenamefont {Caprini}\ \emph {et~al.}(2009)\citenamefont
  {Caprini}, \citenamefont {Durrer}, \citenamefont {Konstandin},\ and\
  \citenamefont {Servant}}]{Caprini:2009fx}%
  \BibitemOpen
  \bibfield  {author} {\bibinfo {author} {\bibfnamefont {C.}~\bibnamefont
  {Caprini}}, \bibinfo {author} {\bibfnamefont {R.}~\bibnamefont {Durrer}},
  \bibinfo {author} {\bibfnamefont {T.}~\bibnamefont {Konstandin}},\ and\
  \bibinfo {author} {\bibfnamefont {G.}~\bibnamefont {Servant}},\ }\bibfield
  {title} {\bibinfo {title} {{General Properties of the Gravitational Wave
  Spectrum from Phase Transitions}},\ }\href
  {https://doi.org/10.1103/PhysRevD.79.083519} {\bibfield  {journal} {\bibinfo
  {journal} {Phys. Rev. D}\ }\textbf {\bibinfo {volume} {79}},\ \bibinfo
  {pages} {083519} (\bibinfo {year} {2009})},\ \Eprint
  {https://arxiv.org/abs/0901.1661} {arXiv:0901.1661 [astro-ph.CO]}
  \BibitemShut {NoStop}%
\bibitem [{\citenamefont {Cai}\ \emph {et~al.}(2020)\citenamefont {Cai},
  \citenamefont {Pi},\ and\ \citenamefont {Sasaki}}]{Cai:2019cdl}%
  \BibitemOpen
  \bibfield  {author} {\bibinfo {author} {\bibfnamefont {R.-G.}\ \bibnamefont
  {Cai}}, \bibinfo {author} {\bibfnamefont {S.}~\bibnamefont {Pi}},\ and\
  \bibinfo {author} {\bibfnamefont {M.}~\bibnamefont {Sasaki}},\ }\bibfield
  {title} {\bibinfo {title} {{Universal Infrared Scaling of Gravitational Wave
  Background Spectra}},\ }\href {https://doi.org/10.1103/PhysRevD.102.083528}
  {\bibfield  {journal} {\bibinfo  {journal} {Phys. Rev. D}\ }\textbf {\bibinfo
  {volume} {102}},\ \bibinfo {pages} {083528} (\bibinfo {year} {2020})},\
  \Eprint {https://arxiv.org/abs/1909.13728} {arXiv:1909.13728 [astro-ph.CO]}
  \BibitemShut {NoStop}%
\bibitem [{\citenamefont {Hook}\ \emph {et~al.}(2021)\citenamefont {Hook},
  \citenamefont {Marques-Tavares},\ and\ \citenamefont {Racco}}]{Hook:2020phx}%
  \BibitemOpen
  \bibfield  {author} {\bibinfo {author} {\bibfnamefont {A.}~\bibnamefont
  {Hook}}, \bibinfo {author} {\bibfnamefont {G.}~\bibnamefont
  {Marques-Tavares}},\ and\ \bibinfo {author} {\bibfnamefont {D.}~\bibnamefont
  {Racco}},\ }\bibfield  {title} {\bibinfo {title} {{Causal Gravitational Waves
  as a Probe of Free Streaming Particles and the Expansion of the Universe}},\
  }\href {https://doi.org/10.1007/JHEP02(2021)117} {\bibfield  {journal}
  {\bibinfo  {journal} {JHEP}\ }\textbf {\bibinfo {volume} {02}},\ \bibinfo
  {pages} {117}},\ \Eprint {https://arxiv.org/abs/2010.03568} {arXiv:2010.03568
  [hep-ph]} \BibitemShut {NoStop}%
\bibitem [{\citenamefont {Pitrou}\ \emph {et~al.}(2018)\citenamefont {Pitrou},
  \citenamefont {Coc}, \citenamefont {Uzan},\ and\ \citenamefont
  {Vangioni}}]{Pitrou:2018cgg}%
  \BibitemOpen
  \bibfield  {author} {\bibinfo {author} {\bibfnamefont {C.}~\bibnamefont
  {Pitrou}}, \bibinfo {author} {\bibfnamefont {A.}~\bibnamefont {Coc}},
  \bibinfo {author} {\bibfnamefont {J.-P.}\ \bibnamefont {Uzan}},\ and\
  \bibinfo {author} {\bibfnamefont {E.}~\bibnamefont {Vangioni}},\ }\bibfield
  {title} {\bibinfo {title} {{Precision big bang nucleosynthesis with improved
  Helium-4 predictions}},\ }\href
  {https://doi.org/10.1016/j.physrep.2018.04.005} {\bibfield  {journal}
  {\bibinfo  {journal} {Phys. Rept.}\ }\textbf {\bibinfo {volume} {754}},\
  \bibinfo {pages} {1} (\bibinfo {year} {2018})},\ \Eprint
  {https://arxiv.org/abs/1801.08023} {arXiv:1801.08023 [astro-ph.CO]}
  \BibitemShut {NoStop}%
\bibitem [{\citenamefont {Dvorkin}\ \emph {et~al.}(2022)\citenamefont {Dvorkin}
  \emph {et~al.}}]{Dvorkin:2022jyg}%
  \BibitemOpen
  \bibfield  {author} {\bibinfo {author} {\bibfnamefont {C.}~\bibnamefont
  {Dvorkin}} \emph {et~al.},\ }\bibfield  {title} {\bibinfo {title} {{The
  Physics of Light Relics}},\ }in\ \href@noop {} {\emph {\bibinfo {booktitle}
  {{2022 Snowmass Summer Study}}}}\ (\bibinfo {year} {2022})\ \Eprint
  {https://arxiv.org/abs/2203.07943} {arXiv:2203.07943 [hep-ph]} \BibitemShut
  {NoStop}%
\bibitem [{\citenamefont {Bai}\ and\ \citenamefont
  {Korwar}(2022)}]{Bai:2021ibt}%
  \BibitemOpen
  \bibfield  {author} {\bibinfo {author} {\bibfnamefont {Y.}~\bibnamefont
  {Bai}}\ and\ \bibinfo {author} {\bibfnamefont {M.}~\bibnamefont {Korwar}},\
  }\bibfield  {title} {\bibinfo {title} {{Cosmological constraints on
  first-order phase transitions}},\ }\href
  {https://doi.org/10.1103/PhysRevD.105.095015} {\bibfield  {journal} {\bibinfo
   {journal} {Phys. Rev. D}\ }\textbf {\bibinfo {volume} {105}},\ \bibinfo
  {pages} {095015} (\bibinfo {year} {2022})},\ \Eprint
  {https://arxiv.org/abs/2109.14765} {arXiv:2109.14765 [hep-ph]} \BibitemShut
  {NoStop}%
\bibitem [{\citenamefont {Kawasaki}\ \emph {et~al.}(2000)\citenamefont
  {Kawasaki}, \citenamefont {Kohri},\ and\ \citenamefont
  {Sugiyama}}]{Kawasaki:2000en}%
  \BibitemOpen
  \bibfield  {author} {\bibinfo {author} {\bibfnamefont {M.}~\bibnamefont
  {Kawasaki}}, \bibinfo {author} {\bibfnamefont {K.}~\bibnamefont {Kohri}},\
  and\ \bibinfo {author} {\bibfnamefont {N.}~\bibnamefont {Sugiyama}},\
  }\bibfield  {title} {\bibinfo {title} {{MeV scale reheating temperature and
  thermalization of neutrino background}},\ }\href
  {https://doi.org/10.1103/PhysRevD.62.023506} {\bibfield  {journal} {\bibinfo
  {journal} {Phys. Rev. D}\ }\textbf {\bibinfo {volume} {62}},\ \bibinfo
  {pages} {023506} (\bibinfo {year} {2000})},\ \Eprint
  {https://arxiv.org/abs/astro-ph/0002127} {arXiv:astro-ph/0002127}
  \BibitemShut {NoStop}%
\bibitem [{\citenamefont {Hasegawa}\ \emph {et~al.}(2019)\citenamefont
  {Hasegawa}, \citenamefont {Hiroshima}, \citenamefont {Kohri}, \citenamefont
  {Hansen}, \citenamefont {Tram},\ and\ \citenamefont
  {Hannestad}}]{Hasegawa:2019jsa}%
  \BibitemOpen
  \bibfield  {author} {\bibinfo {author} {\bibfnamefont {T.}~\bibnamefont
  {Hasegawa}}, \bibinfo {author} {\bibfnamefont {N.}~\bibnamefont {Hiroshima}},
  \bibinfo {author} {\bibfnamefont {K.}~\bibnamefont {Kohri}}, \bibinfo
  {author} {\bibfnamefont {R.~S.~L.}\ \bibnamefont {Hansen}}, \bibinfo {author}
  {\bibfnamefont {T.}~\bibnamefont {Tram}},\ and\ \bibinfo {author}
  {\bibfnamefont {S.}~\bibnamefont {Hannestad}},\ }\bibfield  {title} {\bibinfo
  {title} {{MeV-scale reheating temperature and thermalization of oscillating
  neutrinos by radiative and hadronic decays of massive particles}},\ }\href
  {https://doi.org/10.1088/1475-7516/2019/12/012} {\bibfield  {journal}
  {\bibinfo  {journal} {JCAP}\ }\textbf {\bibinfo {volume} {12}},\ \bibinfo
  {pages} {012}},\ \Eprint {https://arxiv.org/abs/1908.10189} {arXiv:1908.10189
  [hep-ph]} \BibitemShut {NoStop}%
\bibitem [{\citenamefont {Kormendy}\ and\ \citenamefont
  {Ho}(2013)}]{Kormendy:2013dxa}%
  \BibitemOpen
  \bibfield  {author} {\bibinfo {author} {\bibfnamefont {J.}~\bibnamefont
  {Kormendy}}\ and\ \bibinfo {author} {\bibfnamefont {L.~C.}\ \bibnamefont
  {Ho}},\ }\bibfield  {title} {\bibinfo {title} {{Coevolution (Or Not) of
  Supermassive Black Holes and Host Galaxies}},\ }\href
  {https://doi.org/10.1146/annurev-astro-082708-101811} {\bibfield  {journal}
  {\bibinfo  {journal} {Ann. Rev. Astron. Astrophys.}\ }\textbf {\bibinfo
  {volume} {51}},\ \bibinfo {pages} {511} (\bibinfo {year} {2013})},\ \Eprint
  {https://arxiv.org/abs/1304.7762} {arXiv:1304.7762 [astro-ph.CO]}
  \BibitemShut {NoStop}%
\bibitem [{\citenamefont {Peters}\ and\ \citenamefont
  {Mathews}(1963)}]{Peters:1963ux}%
  \BibitemOpen
  \bibfield  {author} {\bibinfo {author} {\bibfnamefont {P.~C.}\ \bibnamefont
  {Peters}}\ and\ \bibinfo {author} {\bibfnamefont {J.}~\bibnamefont
  {Mathews}},\ }\bibfield  {title} {\bibinfo {title} {{Gravitational radiation
  from point masses in a Keplerian orbit}},\ }\href
  {https://doi.org/10.1103/PhysRev.131.435} {\bibfield  {journal} {\bibinfo
  {journal} {Phys. Rev.}\ }\textbf {\bibinfo {volume} {131}},\ \bibinfo {pages}
  {435} (\bibinfo {year} {1963})}\BibitemShut {NoStop}%
\bibitem [{\citenamefont {Thorne}(1987)}]{Thorne:1987af}%
  \BibitemOpen
  \bibfield  {author} {\bibinfo {author} {\bibfnamefont {K.~S.}\ \bibnamefont
  {Thorne}},\ }\bibfield  {title} {\bibinfo {title} {{GRAVITATIONAL
  RADIATION}},\ }\href@noop {} {\  (\bibinfo {year} {1987})}\BibitemShut
  {NoStop}%
\bibitem [{\citenamefont {Maggiore}(2007)}]{Maggiore:2007ulw}%
  \BibitemOpen
  \bibfield  {author} {\bibinfo {author} {\bibfnamefont {M.}~\bibnamefont
  {Maggiore}},\ }\href
  {https://doi.org/10.1093/acprof:oso/9780198570745.001.0001} {\emph {\bibinfo
  {title} {{Gravitational Waves. Vol. 1: Theory and Experiments}}}}\ (\bibinfo
  {publisher} {Oxford University Press},\ \bibinfo {year} {2007})\BibitemShut
  {NoStop}%
\bibitem [{\citenamefont {Sesana}\ \emph {et~al.}(2008)\citenamefont {Sesana},
  \citenamefont {Vecchio},\ and\ \citenamefont {Colacino}}]{Sesana:2008mz}%
  \BibitemOpen
  \bibfield  {author} {\bibinfo {author} {\bibfnamefont {A.}~\bibnamefont
  {Sesana}}, \bibinfo {author} {\bibfnamefont {A.}~\bibnamefont {Vecchio}},\
  and\ \bibinfo {author} {\bibfnamefont {C.~N.}\ \bibnamefont {Colacino}},\
  }\bibfield  {title} {\bibinfo {title} {{The stochastic gravitational-wave
  background from massive black hole binary systems: implications for
  observations with Pulsar Timing Arrays}},\ }\href
  {https://doi.org/10.1111/j.1365-2966.2008.13682.x} {\bibfield  {journal}
  {\bibinfo  {journal} {Mon. Not. Roy. Astron. Soc.}\ }\textbf {\bibinfo
  {volume} {390}},\ \bibinfo {pages} {192} (\bibinfo {year} {2008})},\ \Eprint
  {https://arxiv.org/abs/0804.4476} {arXiv:0804.4476 [astro-ph]} \BibitemShut
  {NoStop}%
\bibitem [{\citenamefont {McWilliams}\ \emph {et~al.}(2014)\citenamefont
  {McWilliams}, \citenamefont {Ostriker},\ and\ \citenamefont
  {Pretorius}}]{McWilliams:2012an}%
  \BibitemOpen
  \bibfield  {author} {\bibinfo {author} {\bibfnamefont {S.~T.}\ \bibnamefont
  {McWilliams}}, \bibinfo {author} {\bibfnamefont {J.~P.}\ \bibnamefont
  {Ostriker}},\ and\ \bibinfo {author} {\bibfnamefont {F.}~\bibnamefont
  {Pretorius}},\ }\bibfield  {title} {\bibinfo {title} {{Gravitational waves
  and stalled satellites from massive galaxy mergers at $z \leq 1$}},\ }\href
  {https://doi.org/10.1088/0004-637X/789/2/156} {\bibfield  {journal} {\bibinfo
   {journal} {Astrophys. J.}\ }\textbf {\bibinfo {volume} {789}},\ \bibinfo
  {pages} {156} (\bibinfo {year} {2014})},\ \Eprint
  {https://arxiv.org/abs/1211.5377} {arXiv:1211.5377 [astro-ph.CO]}
  \BibitemShut {NoStop}%
\bibitem [{\citenamefont {Rosado}\ \emph {et~al.}(2015)\citenamefont {Rosado},
  \citenamefont {Sesana},\ and\ \citenamefont {Gair}}]{Rosado:2015epa}%
  \BibitemOpen
  \bibfield  {author} {\bibinfo {author} {\bibfnamefont {P.~A.}\ \bibnamefont
  {Rosado}}, \bibinfo {author} {\bibfnamefont {A.}~\bibnamefont {Sesana}},\
  and\ \bibinfo {author} {\bibfnamefont {J.}~\bibnamefont {Gair}},\ }\bibfield
  {title} {\bibinfo {title} {{Expected properties of the first gravitational
  wave signal detected with pulsar timing arrays}},\ }\href
  {https://doi.org/10.1093/mnras/stv1098} {\bibfield  {journal} {\bibinfo
  {journal} {Mon. Not. Roy. Astron. Soc.}\ }\textbf {\bibinfo {volume} {451}},\
  \bibinfo {pages} {2417} (\bibinfo {year} {2015})},\ \Eprint
  {https://arxiv.org/abs/1503.04803} {arXiv:1503.04803 [astro-ph.HE]}
  \BibitemShut {NoStop}%
\bibitem [{\citenamefont {Kelley}\ \emph
  {et~al.}(2017{\natexlab{a}})\citenamefont {Kelley}, \citenamefont {Blecha},
  \citenamefont {Hernquist}, \citenamefont {Sesana},\ and\ \citenamefont
  {Taylor}}]{Kelley:2017lek}%
  \BibitemOpen
  \bibfield  {author} {\bibinfo {author} {\bibfnamefont {L.~Z.}\ \bibnamefont
  {Kelley}}, \bibinfo {author} {\bibfnamefont {L.}~\bibnamefont {Blecha}},
  \bibinfo {author} {\bibfnamefont {L.}~\bibnamefont {Hernquist}}, \bibinfo
  {author} {\bibfnamefont {A.}~\bibnamefont {Sesana}},\ and\ \bibinfo {author}
  {\bibfnamefont {S.~R.}\ \bibnamefont {Taylor}},\ }\bibfield  {title}
  {\bibinfo {title} {{The Gravitational Wave Background from Massive Black Hole
  Binaries in Illustris: spectral features and time to detection with pulsar
  timing arrays}},\ }\href {https://doi.org/10.1093/mnras/stx1638} {\bibfield
  {journal} {\bibinfo  {journal} {Mon. Not. Roy. Astron. Soc.}\ }\textbf
  {\bibinfo {volume} {471}},\ \bibinfo {pages} {4508} (\bibinfo {year}
  {2017}{\natexlab{a}})},\ \Eprint {https://arxiv.org/abs/1702.02180}
  {arXiv:1702.02180 [astro-ph.HE]} \BibitemShut {NoStop}%
\bibitem [{\citenamefont {Middleton}\ \emph {et~al.}(2021)\citenamefont
  {Middleton}, \citenamefont {Sesana}, \citenamefont {Chen}, \citenamefont
  {Vecchio}, \citenamefont {Del~Pozzo},\ and\ \citenamefont
  {Rosado}}]{Middleton:2020asl}%
  \BibitemOpen
  \bibfield  {author} {\bibinfo {author} {\bibfnamefont {H.}~\bibnamefont
  {Middleton}}, \bibinfo {author} {\bibfnamefont {A.}~\bibnamefont {Sesana}},
  \bibinfo {author} {\bibfnamefont {S.}~\bibnamefont {Chen}}, \bibinfo {author}
  {\bibfnamefont {A.}~\bibnamefont {Vecchio}}, \bibinfo {author} {\bibfnamefont
  {W.}~\bibnamefont {Del~Pozzo}},\ and\ \bibinfo {author} {\bibfnamefont
  {P.~A.}\ \bibnamefont {Rosado}},\ }\bibfield  {title} {\bibinfo {title}
  {{Massive black hole binary systems and the NANOGrav 12.5 yr results}},\
  }\href {https://doi.org/10.1093/mnrasl/slab008} {\bibfield  {journal}
  {\bibinfo  {journal} {Mon. Not. Roy. Astron. Soc.}\ }\textbf {\bibinfo
  {volume} {502}},\ \bibinfo {pages} {L99} (\bibinfo {year} {2021})},\ \Eprint
  {https://arxiv.org/abs/2011.01246} {arXiv:2011.01246 [astro-ph.HE]}
  \BibitemShut {NoStop}%
\bibitem [{\citenamefont {Kelley}\ and\ \citenamefont {al.}()}]{Holodeck}%
  \BibitemOpen
  \bibfield  {author} {\bibinfo {author} {\bibfnamefont {L.~Z.}\ \bibnamefont
  {Kelley}}\ and\ \bibinfo {author} {\bibnamefont {al.}},\ }\href@noop {}
  {\bibinfo  {journal} {\href{https://github.com/nanograv/holodeck}{\tt
  https://github.com/nanograv/holodeck}}\ }\BibitemShut {NoStop}%
\bibitem [{\citenamefont {Sesana}(2013)}]{Sesana:2013wja}%
  \BibitemOpen
\bibfield  {journal} {  }\bibfield  {author} {\bibinfo {author} {\bibfnamefont
  {A.}~\bibnamefont {Sesana}},\ }\bibfield  {title} {\bibinfo {title}
  {{Insights into the astrophysics of supermassive black hole binaries from
  pulsar timing observations}},\ }\href
  {https://doi.org/10.1088/0264-9381/30/22/224014} {\bibfield  {journal}
  {\bibinfo  {journal} {Class. Quant. Grav.}\ }\textbf {\bibinfo {volume}
  {30}},\ \bibinfo {pages} {224014} (\bibinfo {year} {2013})},\ \Eprint
  {https://arxiv.org/abs/1307.2600} {arXiv:1307.2600 [astro-ph.CO]}
  \BibitemShut {NoStop}%
\bibitem [{\citenamefont {Kelley}\ \emph
  {et~al.}(2017{\natexlab{b}})\citenamefont {Kelley}, \citenamefont {Blecha},\
  and\ \citenamefont {Hernquist}}]{Kelley:2016gse}%
  \BibitemOpen
  \bibfield  {author} {\bibinfo {author} {\bibfnamefont {L.~Z.}\ \bibnamefont
  {Kelley}}, \bibinfo {author} {\bibfnamefont {L.}~\bibnamefont {Blecha}},\
  and\ \bibinfo {author} {\bibfnamefont {L.}~\bibnamefont {Hernquist}},\
  }\bibfield  {title} {\bibinfo {title} {{Massive Black Hole Binary Mergers in
  Dynamical Galactic Environments}},\ }\href
  {https://doi.org/10.1093/mnras/stw2452} {\bibfield  {journal} {\bibinfo
  {journal} {Mon. Not. Roy. Astron. Soc.}\ }\textbf {\bibinfo {volume} {464}},\
  \bibinfo {pages} {3131} (\bibinfo {year} {2017}{\natexlab{b}})},\ \Eprint
  {https://arxiv.org/abs/1606.01900} {arXiv:1606.01900 [astro-ph.HE]}
  \BibitemShut {NoStop}%
\bibitem [{\citenamefont {Taylor}\ \emph {et~al.}(2020)\citenamefont {Taylor},
  \citenamefont {van Haasteren},\ and\ \citenamefont
  {Sesana}}]{Taylor:2020zpk}%
  \BibitemOpen
  \bibfield  {author} {\bibinfo {author} {\bibfnamefont {S.~R.}\ \bibnamefont
  {Taylor}}, \bibinfo {author} {\bibfnamefont {R.}~\bibnamefont {van
  Haasteren}},\ and\ \bibinfo {author} {\bibfnamefont {A.}~\bibnamefont
  {Sesana}},\ }\bibfield  {title} {\bibinfo {title} {{From Bright Binaries To
  Bumpy Backgrounds: Mapping Realistic Gravitational Wave Skies With
  Pulsar-Timing Arrays}},\ }\href {https://doi.org/10.1103/PhysRevD.102.084039}
  {\bibfield  {journal} {\bibinfo  {journal} {Phys. Rev. D}\ }\textbf {\bibinfo
  {volume} {102}},\ \bibinfo {pages} {084039} (\bibinfo {year} {2020})},\
  \Eprint {https://arxiv.org/abs/2006.04810} {arXiv:2006.04810 [astro-ph.IM]}
  \BibitemShut {NoStop}%
\bibitem [{\citenamefont {Agazie}\ \emph
  {et~al.}(2023{\natexlab{e}})\citenamefont {Agazie} \emph
  {et~al.}}]{NANOGrav:2023tcn}%
  \BibitemOpen
  \bibfield  {author} {\bibinfo {author} {\bibfnamefont {G.}~\bibnamefont
  {Agazie}} \emph {et~al.} (\bibinfo {collaboration} {NANOGrav}),\ }\bibfield
  {title} {\bibinfo {title} {{The NANOGrav 15 yr Data Set: Search for
  Anisotropy in the Gravitational-wave Background}},\ }\href
  {https://doi.org/10.3847/2041-8213/acf4fd} {\bibfield  {journal} {\bibinfo
  {journal} {Astrophys. J. Lett.}\ }\textbf {\bibinfo {volume} {956}},\
  \bibinfo {pages} {L3} (\bibinfo {year} {2023}{\natexlab{e}})},\ \Eprint
  {https://arxiv.org/abs/2306.16221} {arXiv:2306.16221 [astro-ph.HE]}
  \BibitemShut {NoStop}%
\end{thebibliography}%

\end{document}